\documentclass[12pt,preprint,url]{aastex}
\usepackage{amsbsy}
\usepackage{amsmath}
\newcommand{\be}{\begin{equation}}
\newcommand{\ee}{\end{equation}}

\def\ltsima{$\; \buildrel < \over \sim \;$}

\def\lsim{\lower.5ex\hbox{\ltsima}}
\def\gtsima{$\; \buildrel > \over \sim \;$}
\def\gsim{\lower.5ex\hbox{\gtsima}}

\shorttitle{Glitch detection with a hidden Markov model}
\shortauthors{Melatos et al.}

\begin{document}
\title{Pulsar glitch detection with a hidden Markov model}

\author{A. Melatos\altaffilmark{1,2} 
 and L. M. Dunn \altaffilmark{1,2}
 and S. Suvorova \altaffilmark{1,2,3}
 and W. Moran \altaffilmark{3}
 and R. J. Evans \altaffilmark{2,3}}

\email{amelatos@unimelb.edu.au}

\altaffiltext{1}{School of Physics, University of Melbourne,
 Parkville, VIC 3010, Australia}

\altaffiltext{2}{Australian Research Council Centre of Excellence
 for Gravitational Wave Discovery (OzGrav),
 Parkville, VIC 3010, Australia}

\altaffiltext{3}{Department of Electrical and Electronic Engineering,
 University of Melbourne, Parkville, VIC 3010, Australia}

\begin{abstract}
\noindent 
Pulsar timing experiments typically generate a phase-connected
timing solution from a sequence of times-of-arrival (TOAs)
by absolute pulse numbering,
i.e.\ by fitting an integer number of pulses between TOAs
in order to minimize the residuals with respect to a 
parametrized phase model.
In this observing mode, rotational glitches are discovered,
when the residuals of the no-glitch phase model diverge after some epoch,
and glitch parameters are refined by Bayesian follow-up.
Here an alternative, complementary approach is presented
which tracks the pulse frequency $f$ and its time derivative $\dot{f}$
with a hidden Markov model (HMM),
whose dynamics include stochastic spin wandering (timing noise) 
and impulsive jumps in $f$ and $\dot{f}$ (glitches).
The HMM tracks spin wandering explicitly,
as a specific realization of a discrete-time Markov chain.
It discovers glitches by comparing the Bayes factor for glitch
and no-glitch models.
It ingests standard TOAs for convenience and,
being fully automated,
allows performance bounds to be calculated quickly
via Monte Carlo simulations.
Practical, user-oriented plots are presented of the false alarm probability 
and detection threshold (e.g.\ minimum resolvable glitch size)
versus observational scheduling parameters
(e.g.\ TOA uncertainty, mean delay between TOAs)
and glitch parameters
(e.g.\ transient and permanent jump sizes, exponential recovery time-scale).
The HMM is also applied to $\sim 1\,{\rm yr}$
of real data bracketing the 2016 December 12 glitch
in PSR J0835$-$4510 as a proof of principle.
It detects the known glitch 
and confirms that no other glitch exists
in the same data with size $\gtrsim 10^{-7} f$.
\end{abstract}


\section{Introduction 
 \label{sec:hmm1}}
The exceptional rotational stability of pulsars allows a terrestrial observer
armed with an accurate clock to construct a phase-connected timing solution
by absolute pulse numbering,
even when the observations are irregularly spaced over decades
and separated by many pulse periods
\citep{lyn12}.
Traditionally the timing solution is constructed in three stages.
(i) 
The pulse train is folded by cross-correlating against a 
template profile in the frequency domain to generate a sequence
of times-of-arrival (TOAs) 
\citep{tay92}.
(ii)
A phase model is stipulated, which includes frame-of-reference terms
(e.g.\ Solar System barycenter),
the pulsar's intrinsic spin evolution
(e.g.\ spin frequency and its time derivatives,
packaged as the coefficients of a Taylor series),
astrometric terms
(e.g.\ sky position and proper motion),
dispersion in the interstellar plasma,
Keplerian orbital elements
(if the pulsar is in a binary),
and post-Keplerian corrections
\citep{edw06}.
(iii) 
The parameters of the phase model are inferred by fitting the TOAs
using a weighted least-squares algorithm,
such that the residuals are white and minimzed, if the model is perfect.
The approach (i)--(iii) has proved highly successful.
It forms the backbone of observational studies in pulsar astronomy
on a wide range of topics,
including tests of general relativity 
\citep{tay92,sta03},
magnetospheric electrodynamics and coherent emission
\citep{mic91,mel17b},
interstellar scintillation
\citep{ric90},
population synthesis and binary evolution
\citep{fau06},
and the search for nanohertz gravitational waves
\citep{len15c,sha15,arz16,hob17}.

One intriguing phenomenon revealed by phase-connected timing
is rotational glitches:
impulsive, erratically occurring, spin-up events
which interrupt the secular, electromagnetic spin down of a 
rotation-powered pulsar.
Traditionally a glitch is discovered, 
when the residuals with respect to a glitchless phase model
diverge after a certain epoch,
e.g.\ a jump in spin frequency causes a linear phase ramp.
Once the glitch is discovered,
two separate, glitchless phase models are fitted
to the TOAs before and after the relevant epoch.
The differences between the models define
the parameters of the glitch,
e.g.\ the jump in spin frequency and its derivatives
\citep{lyn00,esp11}.
It is hard to do this uniquely,
because the phase evolution includes stochastic spin wandering,
known as timing noise
\citep{cor85},
which is often covariant with glitch-related features like
post-glitch recoveries
\citep{lyn96}.
Moreover the gaps between observations can be long and irregular,
leading to degeneracies.
Work has been undertaken recently to address these issues
by applying Bayesian model selection to glitch detection,
e.g.\ using software like {\sc temponest}
\citep{len14,sha16,yu17,low18,low19}.
Bayesian methods are promising but relatively expensive;
they have not been applied to most pulsars to date.

The physical mechanism that triggers glitch activity
remains a mystery;
see \citet{has15} for a recent review.
Broadly speaking, however, it is thought to involve the sudden relaxation 
of spin-down-driven elastic stress and differential rotation
by local, stick-slip processes such as starquakes \citep{mid06,chu10b}
and superfluid vortex avalanches \citep{war11}.
In this picture, glitches and their recoveries probe the 
material properties of bulk matter at nuclear densities,
e.g.\ the shear modulus and superfluid energy gap,
under physical conditions which cannot be replicated on Earth
\citep{yak99,lat07,van10,wat15}.
In particular,
the statistics of glitch sizes and waiting times carry
important information
\citep{mel08,ful17,ash17,mel18,fue19,car19b}.
Expanding the glitch database
\footnote{
Electronic access to up-to-date glitch catalogues is available at the
following locations on the World Wide Web:
{\tt http://www.jb.man.ac.uk/pulsar/glitches/gTable.html} 
(Jodrell Bank Centre for Astrophysics)
and
{\tt http://www.atnf.csiro.au/people/pulsar/psrcat/glitchTbl.html}
(Australia Telescope National Facility).
\label{foot:hmm1}
}
is essential for achieving a better understanding of the
nuclear physics involved.

In this paper, we develop a fast approach to glitch detection
and estimation, which complements the existing approach
and contributes new insights into performance bounds
and spin wandering, as explained in detail in \S\ref{sec:hmm1a}.
Standard TOAs, generated by cross-correlating the pulse train against a
template profile, are still the starting point.
Existing software like {\sc tempo2} and {\sc psrchive} can be used unaltered.
The TOAs are analysed with a hidden Markov model (HMM),
which tracks the underlying evolution of the pulsar's rotation,
including the secular and stochastic components associated with
electromagnetic spin down and spin wandering respectively.
Glitches are detected by Bayesian model selection,
by comparing the evidence for HMMs with and without glitches
(cf.\ {\sc temponest}).
The paper is structured as follows.
In \S\ref{sec:hmm1a} we motivate the algorithm by explaining clearly 
how it fits with existing approaches and what open issues it addresses.
In \S\ref{sec:hmm2} we define the logical components of the HMM-based
phase tracker
and map each component to its corresponding measurement or model
variable in a pulsar timing experiment.
In \S\ref{sec:hmm3} we present and justify an algorithm for
converting the HMM output into Bayesian evidence 
in order to select rigorously between
phase models with and without glitches.
The performance of the glitch-finding algorithm is then tested.
Synthetic data are generated according to the procedure discussed
in \S\ref{sec:hmm4}.
An introductory worked example is presented in an appendix.
Performance metrics such as 
receiver operating characteristic (ROC) curves are evaluated systematically 
as functions of the astrophysical and measurement noises,
secular spin-down parameters,
and glitch parameters in \S\ref{sec:hmm6}.
The figures in \S\ref{sec:hmm6} are designed to be practical.
Together they can be used to plan a glitch discovery campaign 
as a function of experimental variables such as TOA uncertainties and
the desired measurement resolution,
e.g.\ of glitch sizes and recovery time-scales.

The paper is framed as a method paper. 
Most of the tests are done on synthetic data deliberately, 
to study the behavior of the algorithm under controlled conditions.
The next step is to apply the HMM to real data,
a larger project which is under way.
A quick foretaste of what is possible in presented in \S\ref{sec:hmm7}
using public data from PSR J0835$-$4510
as a worked example.
Theoretical aspects of the algorithm are explored further
by \citet{suv18} in a general signal processing context.

\section{Motivation
 \label{sec:hmm1a}}
Before introducing the HMM in \S\ref{sec:hmm2} we explain firstly
what issues in glitch detection the new algorithm seeks to address, 
under what circumstances it proves useful
(and when it does not),
and how it complements traditional glitch detection methods.
Existing approaches enjoy a long record of success,
so it is important to articulate what specific contributions
the new algorithm makes.
The main contributions are
(i) a fast recipe for generating systematic performance bounds,
and (ii) a sophisticated way to distinguish spin wandering and glitches.

Some interesting questions remain unanswered about the
performance bounds of traditional glitch searches
based on software packages such as {\sc tempo2} 
\citep{hob06,edw06},
{\sc psrchive}
\citep{van12},
{\sc temponest}
\citep{len14},
and their relatives.
Given the spin wandering amplitude and TOA measurement uncertainty
in a particular pulsar, 
as well as a glitch size detection threshold,
what is the false alarm probability,
when a traditional glitch search is performed?
Are all catalogued glitches real (see footnote \ref{foot:hmm1}),
or are some of the smaller events actually spin wandering
\citep{jon90,dal95,jan06,yu17}?
What is the smallest event that a traditional glitch search can detect,
as a function of the false alarm and false dismissal probabilities?
How does the detection limit vary between objects with different 
spin wandering amplitudes?
Some work has been done to develop quantitative answers to these questions.
\citet{jan06} conducted Monte Carlo simulations to estimate the 
minimum glitch size resolvable in PSR J1740$-$3015;
see also \citet{wat15} with reference to the Square Kilometer Array.
\citet{sha16} and \citet{low18} reanalysed TOAs from PSR J0835$-$4510
and PSR J1709$-$4429 respectively within a Bayesian framework 
to look for false alarms and false dismissals,
and a similar, multi-object project is under way using data collected by the 
Molonglo Observatory Synthesis Telescope
\citep{jan19,low19}.
\footnote{M. E. Lower, private communication.}
\citet{yu17} performed the largest study of glitch detection probabilities so far,
again within a Bayesian framework,
involving 165 pulsars timed by the Parkes Observatory between 1990 and 2011
\citep{yu13,yu17}.
The latter authors argued persuasively, that the study should be extended
to more pulsars.
However, the task is not easy.
One rigorous approach in signal processing
is to construct a ROC curve for the search algorithm in question,
by plotting the detection probability against the false alarm probability.
This entails many Monte Carlo simulations,
which are prohibitive to analyse,
when traditional algorithms still rely on human supervision
(e.g.\ by-eye inspection of post-fit residuals)
even when aided by software like {\sc temponest}.
Crowdsourcing offers one possible solution,
perhaps by leveraging the infrastructure of the PULSE@Parkes project
\citep{hob09},
but it brings its own logistical challenges.
Consequently few if any ROC curves have been published for
traditional glitch finding schemes.

How does the new algorithm relate to traditional methods of glitch detecton?
The HMM formulation shares some common features with
recent work developing a new, Bayesian, pulsar timing infrastructure
based on pulse domain analysis and/or model selection
\citep{len14,len15a,len15b,len17a,len17b,len18,ash19}.
The main similarity is that a glitch is discovered,
when the Bayes factor comparing glitch and no-glitch phase models
surmounts a user-selected threshold, as with {\sc temponest}.
However there are differences.
(i) The HMM plugs into the traditional infrastructure for generating TOAs.
\footnote{
This is also true for many {\sc temponest} analyses to date.
}
It does not operate in the pulse domain,
in order to maximize the use of existing software.
It can be extended to the pulse domain in the future,
if there is enough demand.
(ii) The HMM does not treat spin wandering as ``noise'';
it tracks it explicitly.
In other words,
it evaluates the likelihood of the specific spin wandering pattern observed
(i.e.\ a specific realization of a discrete-time Markov chain,
in the language of stochastic processes),
whereas {\sc temponest} and related algorithms analyse
the ensemble statistics of the spin wandering
[e.g.\ the timing noise power spectral density \citep{col11}].
(iii) The HMM is fast.
It requires $\sim 10^{12}$ floating point operations 
[$\sim 0.1$ central processing unit (CPU) hours]
per target per year of observations,
starting from an approximate, glitchless timing solution generated
by traditional methods.

We emphasize that the approach developed here does not supplant
traditional timing methods nor the newer pulse domain approach.
{\em All three approaches complement each other
and are more powerful when deployed in tandem.}
For example, when the goal is to measure 
a slow, secular phase evolution described faithfully by a Taylor expansion
(e.g.\ in binary pulsar tests of general relativity),
the HMM formulation is unnecessary, because there is no covariance
between stochastic spin wandering and the secular dynamics
(e.g.\ binary orbital decay).
On the other hand, when spin wandering is covariant with
other short-time-scale phenomena like glitches and their recoveries,
the HMM offers an alternative perspective on whether a glitch occurs,
by tracking the spin wandering directly within systematic performance bounds,
while ingesting standard TOAs for the sake of convenience. 

\section{Phase tracking
 \label{sec:hmm2}}
A HMM is a scheme for inferring the trajectory of a system
through a sequence of unobservable (hidden) states by measuring
observables related probabilistically to the hidden states.
In the pulsar context,
the observables are the TOAs,
and the hidden state is the underlying rotational state of the pulsar 
(e.g.\ its spin frequency and instantaneous derivatives
with respect to time), 
which cannot be measured uniquely from a single TOA
or the interval between a TOA pair.
In \S\ref{sec:hmm2a},
we describe how to formulate the pulsar timing problem in terms of a HMM,
which converts TOAs into a phase-connected timing solution.
The state structure of the HMM is defined precisely in \S\ref{sec:hmm2b}.
We then relate the TOAs probabilistically to the pulsar's rotational state
in \S\ref{sec:hmm2c}
and describe how the rotational state evolves stochastically
under the action of electromagnetic spin down, timing noise, and glitches
in \S\ref{sec:hmm2d}.
Resolution and gridding issues are discussed in \ref{sec:hmm2e}.
An efficient algorithm for solving the HMM numerically
is set out in Appendix \ref{sec:hmmappa}.
The presentation follows closely the formal derivation by \citet{suv18}.

\subsection{HMM formulation
 \label{sec:hmm2a}}
A HMM is a probabilistic finite-state automaton
\footnote{
HMMs with infinite state spaces exist but are not relevant here.
} 
specified by a hidden state variable $q(t)$,
which can take on $N_Q$ discrete values;
an observation variable $o(t)$,
which is not necessarily discrete;
and a sequence of times $t_1 \leq \dots \leq t_{N_T}$
when snapshots of the system are taken.
In general, $q(t_n)$ and $o(t_n)$ are multi-dimensional vectors,
and the times $t_n$ are unequally spaced.

The probability for the system to jump from hidden state $q_i$
at time $t_n$ to hidden state $q_j$ at time $t_{n+1}$
is called the transition probability.
It is given by
\begin{equation}
 A_{q_j q_i}
 = 
 {\rm Pr}[ q(t_{n+1})=q_j | q(t_n)=q_i ]~.
\label{eq:hmm1}
\end{equation}
The probability of measuring the datum $o(t_n)$ at time $t_n$,
if the system is in state $q(t_n)=q_i$, 
is called the emission probability.
It is given by
\begin{equation}
 L_{o(t_n)q_i}
 =
 {\rm Pr}[ o(t_{n}) | q(t_n)=q_i ]~.
\label{eq:hmm2}
\end{equation}
Writing $Q_{1:N_T} = \{ q(t_1),\dots,q(t_{N_T}) \}$ and
$O_{1:N_T} = \{ o(t_1),\dots,o(t_{N_T}) \}$,
we can express the total probability that the observed sequence $O_{1:N_T}$
arises from the hidden sequence $Q_{1:N_T}$ as
\begin{equation}
 \Pr(Q_{1:N_T}|O_{1:N_T})
 =
 \Pi_{q(t_1)} L_{o(t_1) q(t_1)}
 \prod_{n=2}^{N_T}
 A_{q(t_{n}) q(t_{n-1})} L_{o(t_n) q(t_n)}~,
\label{eq:hmm3}
\end{equation}
where
\begin{equation}
 \Pi_{q_i} = {\rm Pr} [ q(t_1) = q_i ]
\label{eq:hmm4}
\end{equation}
denotes the prior probability.

Three essential questions of practical value can be asked
about a HMM of the above form
\citep{rab89,qui01}.
First, given the observed sequence $O_{1:N_T}$ and a model 
$M=\{ A_{q_j q_i}, L_{o(t_n)q_i}, \Pi_{q_i} \}$,
what is $\Pr(O_{1:N_T}|M)$,
i.e.\ what is the Bayesian evidence for $M$?
Knowing $\Pr(O_{1:N_T}|M)$, one can select between different models.
Second, given $O_{1:N_T}$ and $M$, 
what is the optimal hidden sequence $Q_{1:N_T}$ which best explains
the data according to some meaningful metric?
Third, given $O_{1:N_T}$, what model $M$ maximizes $\Pr(O_{1:N_T}|M)$?

The first and second questions in the previous paragraph are fundamental to the 
glitch-finding problem studied in this paper.
Efficient algorithms to solve them 
are presented in Appendix \ref{sec:hmmappa}
and assembled into a systematic glitch-finding scheme in \S\ref{sec:hmm3}.
There is no unique answer to the second question.
One possible solution is
$Q_{1:N_T}^\ast
 = {\rm arg\,max\,} \Pr(Q_{1:N_T}|O_{1:N_T},M)$,
which maximizes $\Pr(Q_{1:N_T}|O_{1:N_T},M)$ sequence-wise
\citep{qui01}.
Another possible solution is
$\hat{q}(t_n) = {\rm arg\, max \,} \Pr[q(t_n)|O_{1:N_T},M]$
for $1\leq  n \leq N_T$,
which maximizes $\Pr[q(t_n)|O_{1:N_T},M]$ point-wise
\citep{rab89}.
The third question, which corresponds here to learning 
a dynamical model of glitches statistically from the data,
can be solved by iterative methods like the Baum-Welch algorithm
\citep{rab89}
but lies outside the scope of this work.

\subsection{Summary of HMM components
 \label{sec:hmm2b}}
In the pulsar timing context, 
the components of the HMM are the following.
\begin{enumerate}
\item
{\em Hidden state.} 
In this paper, we track the instantaneous
frequency $f(t)$ and its first time derivative $\dot{f}(t)$.
Future work can easily include higher-order derivatives,
e.g.\ the secular component of the second derivative 
$\langle \ddot{f} \rangle =n\dot{f}^2/f$
describing electromagnetic braking,
where $1\lesssim n \leq 3$ is the electromagnetic braking index
\citep{mel97,arc16}.
The stochastic component of $\ddot{f}$,
whose magnitude usually exceeds $n\dot{f}^2/f$
\citep{arz94,joh99},
is absorbed in the wandering of $f$.
We also define 
(but do not track; see \S\ref{sec:hmm2d} and \S\ref{sec:hmm3})
a Boolean variable, $g(t)$,
which equals unity
if a glitch occurs at time $t$ and zero otherwise.
In summary, therefore, the hidden state is
$q(t)=[f(t),\dot{f}(t),g(t)]$.
\item
{\em Observable.}
In this paper, the HMM time sequence $\{ t_1, \dots , t_{N_T} \}$ is defined 
to map one-to-one onto the measured, unequally spaced TOAs, 
starting from the second TOA.
The measurement variable at time $t_n$ is defined to equal
the displacement between consecutive TOAs,
viz.\ $o(t_n) = t_n - t_{n-1}$,
where $t_0$ corresponds to the first TOA;
henceforth we write $x_n=t_n - t_{n-1}$ for brevity.
Future refinements include
augmenting $o(t_n)$ with auxiliary information,
e.g.\ tagging it with the pulse period measured locally at each TOA.
\item
{\em Emission probability.}
Given $x_n$ and an associated measurement error,
whose variance equals $2\sigma_{\rm TOA}^2$,
there exists a limited but degenerate set of $(f,\dot{f})$ pairs,
which produce an integer number of pulses in the interval $x_n$.
An explicit formula for the emission probability for arbitrary $f$ and $\dot{f}$
and Gaussian measurement errors is given in \S\ref{sec:hmm2c}
in terms of the von Mises distribution.
By way of illustration,
in the artificial special case 
with $\dot{f}=0$ and $\sigma_{\rm TOA}=0$,
the emission probability is proportional to a sum of delta functions,
$\delta(f-1/x_n)+ \delta(f-2/x_n) +\dots$.
\item
{\em Transition probability.}
In this paper, we track the rotational phase on three time-scales:
(i) secular, electromagnetic braking on the longest time-scale,
$f/\dot{f} \gtrsim 10^3\,{\rm yr}$,
which greatly exceeds the total observation span,
$T_{\rm obs}\lesssim 10^2\,{\rm yr}$;
(ii) 
spin wandering (timing noise) on an intermediate time-scale,
stretching from days to years \citep{cor85,pri12,nam19,par19,gon19,low20};
and (iii)
glitches, i.e.\ unresolved jumps in $f$ and $\dot{f}$,
whose rise times are much shorter than $\min_n x_n$.
The stochastic dynamics of $q(t)=[f(t),\dot{f}(t),g(t)]$,
which determine $A_{q_j q_i}$,
are modeled as biased Brownian motion with process variance
per unit time $\sigma^2$
via a Langevin equation in \S\ref{sec:hmm2d} and Appendix \ref{sec:hmmappb}.
Note that glitches are often followed by quasiexponential recoveries,
which last days to years
\citep{van10}.
The recoveries can be incorporated into the phase model in future work.
Here we absorb them into the timing noise,
which occurs on a similar time-scale,
and show a posteriori that this is an effective approach in practice,
with the algorithm successfully detecting glitches in synthetic data
containing recoveries
(see \S\ref{sec:hmm4}).
\item
{\em Prior.}
A uniform prior is adopted on $f$ and $\dot{f}$ within a restricted domain,
known as the domain of interest (DOI; see \S\ref{sec:hmm2e}).
Practically the DOI for any pulsar
is defined by traditional phase-connected timing methods, 
e.g.\ a standard {\sc tempo2} fit,
as well as prior astrophysical knowledge,
e.g.\ population-based constraints on glitch sizes
\citep{mel08,esp11,how18}.
In general $\Pr(Q_{1:N_T} | O_{1:N_T})$ is insensitive to the choice of
a uniform prior,
because $\Pi_{q_i}$ is just one factor out of $2N_T \gg 1$
in what is usually a large product in (\ref{eq:hmm3})
\citep{suv16,suv17,abb17}.
\end{enumerate}

\subsection{Emission probability
 \label{sec:hmm2c}}
Given a displacement $x_n$,
what can we say probabilistically about
the rotational state of the pulsar at $t_{n}$?
For $\dot{f}(t_n)=0$, without measurement noise,
we can infer the instantaneous frequency,
$f(t_{n})$,
to be an integer multiple of $x_{n}^{-1}$.
For $\dot{f}(t_n) \neq 0$,
a particular combination of $x_n$, $f(t_n)$, and $\dot{f}(t_n)$
is inferred to be an integer.
The combination is unique, as long as $x_n$ is short enough (see below).
When measurement noise is switched on,
these statements continue to hold true, but the estimates are ``fuzzy''.

In the absence of measurement noise and discontinuous glitches, 
and with $\ddot{f}=0$ over a short enough time-scale,
we can approximate the frequency evolution 
in the interval $t_{n-1}\leq t \leq t_n$
as a {\em backward} Taylor series,
$f(t)= f(t_n) + (t-t_n) \dot{f}(t_n)$,
and then integrate $d\phi/dt = 2\pi f(t)$ to get the phase,
\begin{equation}
 \phi(t_{n})
 =
 \phi(t_{n-1}) 
 + 2\pi x_{n} f(t_n)
 - \pi x_n^2 \dot{f}(t_n)~.
\label{eq:hmm5}
\end{equation}
The minus sign in the last term arises,
because we use a backward difference scheme.
Let $N_n$ be the number of pulses between $t_{n-1}$ and $t_n$.
By the definition of the TOAs, $N_n$ is an integer,
and we have $N_n=\Phi(x_n)$
with $\Phi(x_n) = x_n f(t_n) - x_n^2 \dot{f}(t_n)/2 $.
This equation corresponds to a line in the 
$f(t_n)$-$\dot{f}(t_n)$ plane given $x_n$ and $N_n$.

If each TOA has a Gaussian measurement error with zero mean 
and variance $\sigma_{\rm TOA}^2$, 
then $x_n$ also has a Gaussian measurement error,
denoted by $w_n$, with twice the variance.
We write the measurement equation as
\begin{equation}
 x_n = \Phi^{-1}(N_n) + w_n~,
\label{eq:hmm6}
\end{equation}
where $\Phi^{-1}$ is the inverse function of $\Phi$,
not its reciprocal.
In practice,
$x_n$ is always short enough,
i.e.\ $x_n \ll 2f(t_n)/ | \dot{f}(t_n) |$,
so that $\Phi$ is uniquely invertible up to an integer multiple.
The inversion is unique,
even when the timing noise is strong ($|\ddot{f}| \gg n \dot{f}^2/f$),
unlike higher-order Taylor expansions,
where the inversion is multi-valued (modulo the integer multiples)
for $x_n^2 \gtrsim 6 f(t_n) / | \ddot{f}(t_n) |$.
In this paper, timing noise is tracked explicitly
via the HMM transition probability, as described in \S\ref{sec:hmm2d}.

The emission probability is proportional to the probability density
function (PDF) of the observed variable $x_n$.
\citet{suv18} showed that the PDF of $\Phi(x_n)$ is approximately 
a wrapped Gaussian, because the phase is $2\pi$-periodic;
see Appendix A of the latter reference.
\citet{suv18} also showed that the wrapped Gaussian can be approximated
accurately by a von Mises distribution
\citep{mar09}, 
which is more convenient to evaluate numerically. 
Hence one can write
\begin{equation}
 L_{x_n q(t_n)}
 =
 [2\pi I_0(\kappa)]^{-1}
 \exp\{ \kappa \cos[2\pi \Phi(x_n) ] \}
\label{eq:hmm7}
\end{equation}
with
\begin{equation}
 \kappa = [ 2 \sigma_{\rm TOA}^2 f(t_n)^2 ]^{-1}~.
\label{eq:hmm8}
\end{equation}
In (\ref{eq:hmm7}), $I_0(\kappa)$ symbolizes a modified Bessel function
of the first kind.
It is approximated by 
$I_0(\kappa) \approx (2\pi\kappa)^{-1/2} \exp(\kappa)$
in the regime $\kappa \gg 1$
to avoid underflow errors in the computation.
Intuitively $\kappa^{-1/2}$ is the number of pulses squeezed into
a time interval lasting as long as the uncertainty in $x_n$.
Note that the hidden state $q(t_n)$ enters (\ref{eq:hmm7}) through $\Phi(x_n)$,
which depends on $f(t_n)$ and $\dot{f}(t_n)$.
By contrast, $N_n$ does not enter (\ref{eq:hmm7}) explicitly;
the HMM does not count the number of pulses in the interval 
$t_{n-1} \leq t \leq t_n$ explicitly,
although this information can always be extracted {\em post factum}
using (\ref{eq:hmm6}),
once the HMM is solved to obtain $Q_{1:N_T}$.

\begin{figure}
\begin{center}
\includegraphics[width=14cm,angle=0]{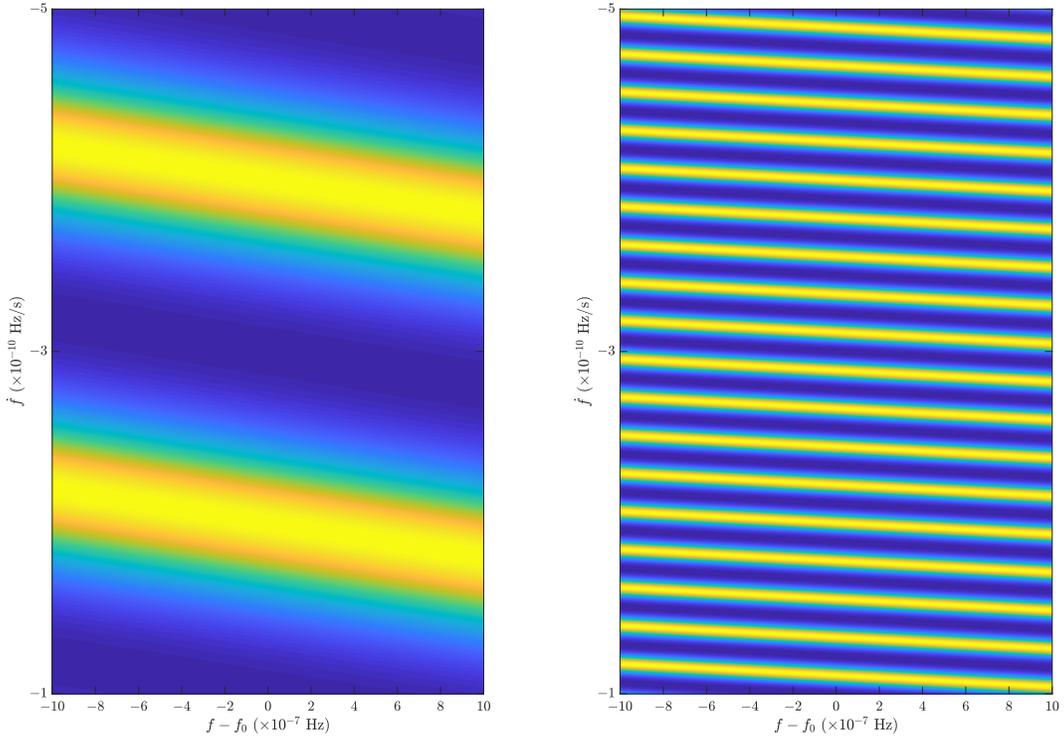}
\end{center}
\caption{
Contour map of the emission probability $L_{x_n q(t_n)}$
[equation (\ref{eq:hmm7})]
(arbitrary color scale; yellow high, blue low)
as a function of the hidden state components 
$f(t_n)$ (units: Hz) 
and $\dot{f}(t_n)$ (units: ${\rm Hz\,s^{-1}}$)
for measurements $x_n = 1\times 10^5\,{\rm s}$ ({\em left panel})
and $x_n = 3\times 10^5\,{\rm s}$ ({\em right panel}).
Both panels are centered on 
$(f_0,\dot{f}_0)=
 (5\,{\rm Hz},-3\times 10^{-10}\,{\rm Hz\,s^{-1}})$.
Measurement uncertainty:
$\sigma_{\rm TOA}^2 = 2\times 10^{-2}\,{\rm s^2}$,
i.e.\ $\kappa=1$.
}
\label{fig:hmm1}
\end{figure}

Figure \ref{fig:hmm1} displays a sample of $L_{x_n q(t_n)}$ contours
in the $f(t_n)$-$\dot{f}(t_n)$ plane for two $x_n$ values.
Each stripe corresponds to a peak of $L_{x_n q(t_n)}$ along the line
$N_n=\Phi(x_n) = (2\pi)^{-1} [ x_n f(t_n) - x_n^2 \dot{f}(t_n)/2 ]$.
Its slope, $2/x_n$, decreases as $x_n$ increases.
Formally speaking,
equation (\ref{eq:hmm7}) has an infinite number of equal-height peaks,
each corresponding to an integer value of $N_n$.
In practice,
the number of peaks within the DOI 
(drawn arbitrarily here as the figure frame)
is finite.
Without extra information,
e.g.\ a phase-connected solution constructed by traditional means,
all the peaks are equally likely.
As $x_n$ increases three-fold from the left panel to the right panel, 
two things happen:
the minimum $f(t_n)$ (corresponding to $N_n=1$) decreases,
and the separation of the peaks decreases.
On the one hand, therefore, there is greater ambiguity,
because there are more peaks to interrogate in the DOI.
On the other hand, the estimate of $q(t_n)$ is more accurate,
once the HMM finds the optimal peak,
because the peaks are narrower.
\footnote{
The peaks are narrower because they are more closely separated,
not because their width decreases relative to their separation.
The argument of the cosine in (\ref{eq:hmm7}) depends on $x_n$,
but the factor $\kappa$ multiplying the cosine does not.
}
In Figure \ref{fig:hmm1},
the number of yellow stripes in the frame increases from two to 18, 
as $x_n$ increases from $1\times 10^5\,{\rm s}$ in the left panel
to $3\times 10^5\,{\rm s}$ in the right panel.
The full-width half-maximum (FWHM) per stripe 
projected on the $\dot{f}(t_n)$ axis
decreases from $7.2\times 10^{-11}\,{\rm Hz\,s^{-1}}$ in the left panel 
to $7.9\times 10^{-12}\,{\rm Hz\,s^{-1}}$ in the right panel.

Equations (\ref{eq:hmm7}) and (\ref{eq:hmm8}) assume that the 
state space is continuous.
In practice, the $f(t_n)$-$\dot{f}(t_n)$ plane is divided into a grid.
A generalized version of (\ref{eq:hmm8}) 
that accounts for gridding
and lets $\sigma_{\rm TOA}$ vary with $t_n$
is discussed in \S\ref{sec:hmm2e}.

\subsection{Transition probability
 \label{sec:hmm2d}}
The equation of motion obeyed by $q(t)=[f(t),\dot{f}(t),g(t)]$ 
in a real pulsar is unknown.
Instead we construct an idealized model for how $q(t)$
evolves during the HMM step $t_{n-1} \leq t \leq t_n$.
Away from a glitch,
we assume that the system obeys a continuous Wiener process described by
the Langevin equation
\begin{equation}
 \frac{d^2 f}{dt^2} = \xi(t)~,
\label{eq:hmm9}
\end{equation}
where $\xi(t)$ is a fluctuating torque derivative 
with white noise statistics satisfying
$\langle \xi(t) \rangle = 0$
and 
$\langle \xi(t) \xi(t') \rangle = \sigma^2 \delta(t-t')$,
and $\sigma$ is a tunable parameter
(units: ${\rm Hz\,s^{-3/2}}$).
At the instant when a glitch occurs,
the continuous evolution is interrupted,
and $f(t)$ and $\dot{f}(t)$ undergo impulsive permanent changes 
$\Delta f_{\rm p}$ and $\Delta\dot{f}_{\rm p}$ respectively.

We emphasize that (\ref{eq:hmm9}) is designed mainly with the practical needs
of the HMM in mind; it should not be viewed as a physical model of a pulsar.
Nevertheless it does embody the three physical time-scales discussed in point 4
in \S\ref{sec:hmm2c}:
long (electromagnetic braking),
intermediate (timing noise),
and short (glitches).
Electromagnetic braking enters through the initial conditions;
the secular spin-down torque sets $\dot{f}(t_{n-1})$.
We neglect $\langle \ddot{f} \rangle = n \dot{f}^2/f$
in (\ref{eq:hmm9}) as discussed in \S\ref{sec:hmm2b}.
Timing noise enters through the right-hand side of (\ref{eq:hmm9}).
Its amplitude is set by $\sigma$,
which satisfies 
$\langle [ \dot{f}(t_n) - \dot{f}(t_{n-1}) ]^2 \rangle = \sigma^2 x_n$
as for any Wiener process.
Glitch-driven jumps in the frequency and frequency derivative enter 
through the initial conditions 
$f(t_{n-1})$ and $\dot{f}(t_{n-1})$ respectively.
Glitches can occur anywhere within a TOA gap,
because a discrete-time HMM only registers state changes at $t_1,\dots,t_{N_T}$
by definition. 
The TOA gap defines the uncertainty on the estimated glitch epoch,
once the HMM detects a glitch.
Note that $\xi(t)$ corresponds to a fluctuating torque derivative,
whereas theories of timing noise often invoke a fluctuating torque
\citep{cor80,cor85,mel10,mel14}
as well as frequency and phase fluctuations
\citep{cor80}.
The distinction is unimportant in many HMM tracking problems.
\footnote{
For example, a simple transition probability matrix of the form
$A_{q_{i+1}q_i} = A_{q_i q_i} = A_{q_{i-1} q_i} = 1/3$
successfully tracks various complicated random walks
in gravitational wave applications
\citep{suv16};
cf.\ \citet{bay19}.
}
The degree to which it matters when doing model selection,
as in this paper,
is tested empirically in \S\ref{sec:hmm6}.

The forward Fokker-Planck equation corresponding to (\ref{eq:hmm9})
can be solved to find the PDF of $q(t_n)$ given $q(t_{n-1})$
\citep{gar94},
as required by (\ref{eq:hmm3}).
The result, derived in Appendix \ref{sec:hmmappb}, is
\begin{equation}
 A_{q(t_{n})q(t_{n-1})}
 =
 (2\pi)^{-1/2} | {\rm det}\Sigma|^{-1/2}
 N_G^{-1}
 \sum_{(\Delta f_{\rm p},\Delta\dot{f}_{\rm p}) \in G}
 \exp\{ - [q(t_{n})-\mu]^{\rm T} \Sigma^{-1} [q(t_{n})-\mu] \}~,
\label{eq:hmm10}
\end{equation}
where the superscript T denotes the matrix transpose,
the secular evolution is described by the mean vector 
$\mu=(\mu_f,\mu_{\dot{f}})$,
with
\begin{eqnarray}
 \mu_f
 & = &
 f(t_{n-1}) + x_n \dot{f}(t_{n-1})
 + g(t_{n-1}) ( \Delta f_{\rm p} + x_n \Delta\dot{f}_{\rm p} )~,
\label{eq:hmm11}
 \\
 \mu_{\dot{f}}
 & = &
 \dot{f}(t_{n-1}) + g(t_{n-1}) \Delta\dot{f}_{\rm p}~,
\label{eq:hmm12}
\end{eqnarray}
the dispersion is described by the covariance matrix,
\begin{equation}
 \Sigma
 = 
 \sigma^2 \left(
 \begin{tabular}{cc}
  $x_n^3/3$ & $x_n^2/2$ \\
  $x_n^2/2$ & $x_n$
 \end{tabular}
 \right)~,
\label{eq:hmm13}
\end{equation}
and $\Sigma^{-1}$ is the matrix inverse of $\Sigma$.
In (\ref{eq:hmm10}),
$G=G[g(t_{n-1})]$ denotes the set of jump pairs 
$(\Delta f_{\rm p},\Delta\dot{f}_{\rm p})$
searched by the HMM at step $t_{n-1}$,
and $N_G$ is the cardinality of $G$.
If a glitch does not occur,
we have $g(t_{n-1})=0$, $\Delta f_{\rm p}=0$,
$\Delta \dot{f}_{\rm p}=0$, and $N_G=1$.
If a glitch does occur,
$\Delta f_{\rm p}$ and $\Delta \dot{f}_{\rm p}$ are constrained
to lie within the DOI defined by astrophysical priors,
e.g.\ historical glitch observations,
and $N_G$ is determined by the grid resolution within the DOI
(see \S\ref{sec:hmm2e}).

\begin{figure}
\begin{center}
\includegraphics[width=14cm,angle=0]{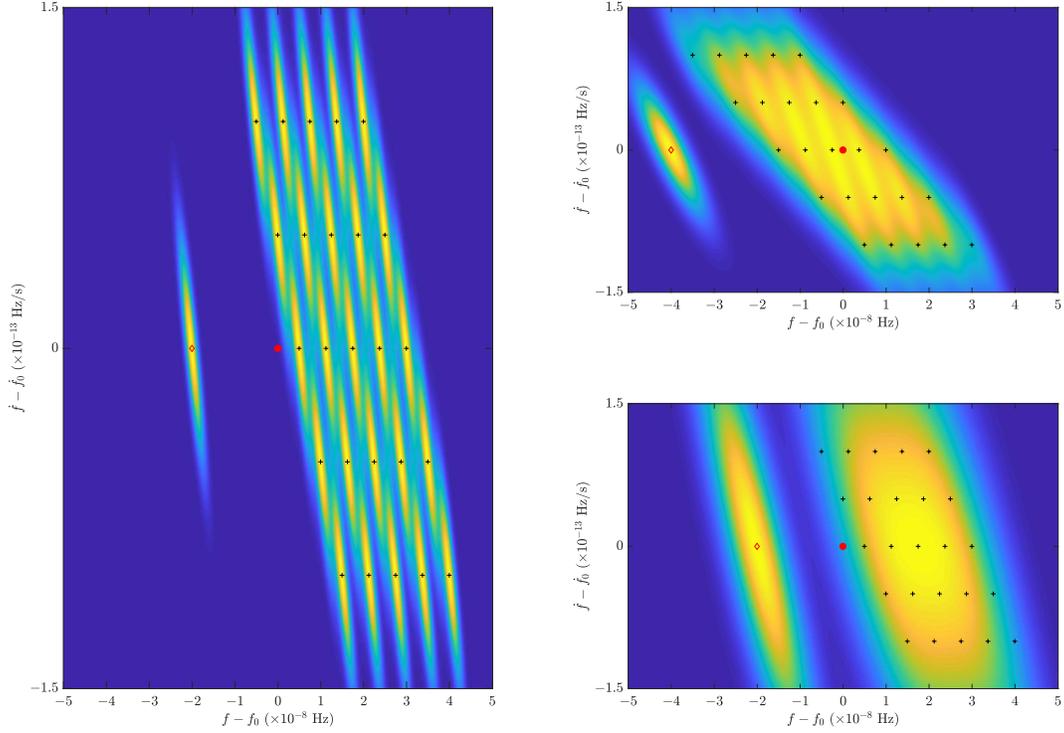}
\end{center}
\caption{
Contours of the transition probability $A_{q(t_n)q(t_{n-1})}$ 
(arbitrary color scale; yellow high, blue low)
plotted versus the endpoint $f(t_n)$ and $\dot{f}(t_n)$
of a HMM step.
The axes are centered on the red dot at
$q(t_{n-1})=(f_0,\dot{f}_0)
 =(5\,{\rm Hz},-2\times 10^{-13}\,{\rm Hz\,s^{-1}})$.
Every cross-centered ellipse corresponds to one term in (\ref{eq:hmm10}),
i.e.\ one combination of $\Delta f_{\rm p}$ and $\Delta \dot{f}_{\rm p}$
if a glitch occurs at $t_{n-1}$.
The allowed jumps in the set $G$ are restricted to the ranges
$2.5 \leq \Delta f_{\rm p} / (10^{-8}\,{\rm Hz}) \leq 5.0$
and
$-1.0 \leq \Delta \dot{f}_{\rm p} / (10^{-13}\,{\rm Hz\,s^{-1}}) \leq 1.0$
for the sake of readability.
The red diamond indicates the probability peak,
when no glitch occurs at $t_{n-1}$.
({\em Left panel.}) 
$x_n = 1\times 10^5\,{\rm s}$, $\sigma = 1\times 10^{-16} \, {\rm Hz\,s^{-3/2}}$.
({\em Right top panel.})
$x_n = 2\times 10^5\,{\rm s}$, $\sigma = 1\times 10^{-16} \, {\rm Hz\,s^{-3/2}}$.
({\em Right bottom panel.}) 
$x_n = 1\times 10^5\,{\rm s}$, $\sigma = 5\times 10^{-16} \, {\rm Hz\,s^{-3/2}}$.
}
\label{fig:hmm2}
\end{figure}

Figure \ref{fig:hmm2} displays a sample of $A_{q(t_n) q(t_{n-1})}$ contours
in the $f(t_n)$-$\dot{f}(t_n)$ plane for a
representative choice of $q(t_{n-1})$ 
(centered on the red dot in each panel) and two values each of $x_n$ and $\sigma$.
Every cross-centered ellipse in the figure 
corresponds to one term in (\ref{eq:hmm10}),
i.e.\ one choice of $\Delta f_{\rm p}$ and $\Delta \dot{f}_{\rm p}$
in $G$ and the DOI.
The principal axes of the ellipses are determined by $x_n$ through
the covariance matrix $\Sigma$,
as can be seen by comparing the left ($x_n=1\times 10^5\,{\rm s}$)
and right top ($x_n = 2\times 10^5\,{\rm s}$) panels.
The ellipse marked with a red diamond corresponds to no glitch,
i.e.\ $g(t_{n-1})=0$;
the other ellipses have $g(t_{n-1})=1$.
The DOI is drawn artificially small,
so that the reader can see its boundaries within the figure 
while still making out the ellipses individually;
in practice one would expect typically $\gtrsim 10^3$ ellipses
within the frame of Figure \ref{fig:hmm2}.
For $\sigma$ relatively low, as in the left panel,
the ellipses are narrow and nearly disjoint.
For $\sigma$ relatively high, as in the right bottom panel,
the ellipses broaden and overlap;
$\sigma$ increases five-fold in passing from the left to the right bottom panel.
In the high-$\sigma$ regime, $A_{q(t_n) q(t_{n-1})}$ can be approximated
as uniform across the DOI for $g(t_{n-1})=1$,
with all glitch terms contributing equally to the sum in (\ref{eq:hmm10}),
while the $g(t_{n-1})=0$ term stands apart.

The Boolean component $g(t)$ of $q(t)$ does not appear explicitly
in (\ref{eq:hmm9});
in other words, we do not track it.
Partly this is because its true evolution is unknown.
Are glitches a Poisson process, for instance, and what is the rate?
Empirically some pulsars show Poisson-like glitch activity,
but others do not 
\citep{mel08,esp11,how18,fue19}.
Moreover, it is inefficient computationally to track $g(t)$.
Historical glitch data imply $g(t)=0$ most of the time,
with one glitch being discovered among every $\gtrsim 10^3$ TOAs,
at least for glitches of a size that traditional timing methods can resolve
\citep{jan06}.
\footnote{
It is possible that the glitches observed to date,
with $\Delta f_{\rm p} \gtrsim 10^{-10} f$, 
represent the ``tip of the iceberg'',
and there exists a (e.g.\ power-law) population of microglitches 
below the resolution limit of current experiments
\citep{mel08,onu16}.
Indeed it has been argued that microglitches collectively
add up to produce timing noise 
\citep{dal95}.
On the other hand, there is evidence that the lower cut-off of the
glitch size PDF is resolved observationally in PSR J0534$+$2200
\citep{esp14}.
The existence of microglitches remains an open question
at the time of writing.
}
In this paper, therefore, we do not assume anything about the
distribution of glitch waiting times.
Instead we incorporate $g(t)$ into the Bayesian
model selection procedure described in \S\ref{sec:hmm3}.
We run the HMM for a glitchless model with $g(t_n)=0$
and a single-glitch model with $g(t_n)=\delta_{nk}$ 
for fixed $k$
($1\leq n \leq N_T$)
and compute the odds ratio to test for the existence
of a glitch at $t_k$.
We then repeat the exercise $N_T$ times for $1\leq k \leq N_T$.
A full discussion of the procedure is given in
\S\ref{sec:hmm3} and Appendix \ref{sec:hmmappa}.

Quasiexponential post-glitch recoveries do not feature in the
hidden state evolution described by (\ref{eq:hmm9})--(\ref{eq:hmm13}),
even though they are observed in reality.
In this paper, we include post-glitch recoveries in the synthetic data 
generated according to \S\ref{sec:hmm4} and show that the HMM
performs well at finding synthetic glitches with recoveries,
even though the recoveries are not built into $A_{q_j q_i}$.
Of course $A_{q_j q_i}$ can be extended to include recoveries, 
at the expense of introducing at least two extra parameters into the HMM model,
viz.\ recovery fraction and recovery time-scale,
and weakening the Markov approximation.
Neither parameter is known a priori and would need to be searched over, 
increasing the complexity of the HMM.
We postpone developing this capability,
until the data sets grow to the point, when it is genuinely needed.

\subsection{Grid resolution and DOI
 \label{sec:hmm2e}}
The DOI relevant to a particular pulsar 
is the region in the $f$-$\dot{f}$ plane that contains
all possible hidden state sequences $Q_{1:N_T}$ consistent with
the observed TOAs.
The set of hidden states is constructed
by dividing the DOI into a grid,
whose spacing is chosen to resolve essential features 
like electromagnetic spin down, timing noise, and glitches
(see \S\ref{sec:hmm2b}).
In Appendix \ref{sec:hmmappc} we offer one practical recipe 
for gridding the DOI.
It is not unique;
the reader is encouraged to modify it, as the experiment demands.
We distinguish carefully between the DOI and the set $G$ in (\ref{eq:hmm10}).
The DOI encompasses the trajectories of the HMM,
starting from the subset of the $f$-$\dot{f}$ plane covered
by the uniform prior (see \S\ref{sec:hmm2b}).
It is chosen at the outset so that 
it does not exclude any admissible HMM trajectory consistent 
with the observed TOAs and the phase model in \S\ref{sec:hmm2d}.
In contrast,
$G$ defines the set of glitch-related jumps in $q(t)$
consistent with \S\ref{sec:hmm2d} and the requirement
that $q(t)$ stays within the DOI at all times.
It is updated at each $t_n$ and depends on
$q(t_{n-1})$ via (\ref{eq:hmm10}).
Appendix \ref{sec:hmmappc} describes how discretization 
affects $G$ and modifies the formulas (\ref{eq:hmm7}) and (\ref{eq:hmm8})
for the emission probability.

\section{Glitch detection by Bayesian model selection
 \label{sec:hmm3}}
Once the phase tracker in \S\ref{sec:hmm2} is implemented,
the task of discovering a glitch reduces to comparing,
given the data,
the probability of a phase model with one or more glitches
against the probability of a glitchless phase model.
From a Bayesian perspective,
the comparison reduces to calculating the evidence ratio 
(or marginal likelihood ratio) of the competing models.
In \S\ref{sec:hmm3a} and Appendix \ref{sec:hmmappa},
we describe how to calculate the evidence ratio using the
HMM forward algorithm.
In \S\ref{sec:hmm3aa} we generalize the evidence ratio calculation
to multiple glitches.
In \S\ref{sec:hmm3b} and Appendix \ref{sec:hmmappa},
we describe how to infer the optimal ephemeris using the
HMM forward-backward algorithm,
once the preferred model (the one with the highest evidence ratio) 
is identified.
The preferred model may or may not contain a glitch.
In \S\ref{sec:hmm3d} we present a preliminary survey of
the computational cost.
Finally, for the sake of completeness, we outline briefly 
in Appendix \ref{sec:hmmappd}
a related approach to discovering glitches,
known as a jump Markov model,
and explain why it is not used here.

The HMM does not prefer a particular {\em physical} model of glitches.
Any physical mechanism which conforms to the idealized transition
probability (\ref{eq:hmm10})--(\ref{eq:hmm13}) falls within the ambit
of the HMM.
Equations (\ref{eq:hmm10})--(\ref{eq:hmm13}) take a generic form
and are motivated observationally,
so they automatically embrace many microphysical mechanisms,
which have been developed to explain observed glitch activity,
including superfluid vortex avalanches
\citep{and75,war11},
starquakes
\citep{mid06,chu10b},
and hydrodynamic instabilities
\citep{gla09};
see \citet{has15} for a recent review.
Equations (\ref{eq:hmm10})--(\ref{eq:hmm13}) also embrace
many microphysics-agnostic meta-models,
which have been developed to make falsifiable predictions
about long-term glitch statistics
\citep{ful17,mel18,car19b}.
In this sense, the HMM is robust towards physical mechanisms
in the literature; it accommodates all the main classes.
By the same token, it cannot discriminate between the classes;
that is not its function;
it is a glitch detector, not a sieve for physical mechanisms.
In what follows, the term ``model''
refers to a sequence $g(t_0), \dots, g(t_{N_T-1})$
admissible by the Markov process
(\ref{eq:hmm10})--(\ref{eq:hmm13}),
not a codification of a physical mechanism.
The sequence preferred by the data is the one with
the highest evidence ratio, as noted above and in \S\ref{sec:hmm3a}.
The reader is encouraged to experiment with alternatives to 
(\ref{eq:hmm10})--(\ref{eq:hmm13}) 
and explore their effect on glitch detection.

\subsection{Model evidence
 \label{sec:hmm3a}}
Let $M_0$ denote the model, where no glitch occurs in the interval
$t_0 \leq t \leq t_{N_T}$,
i.e.\ we have $g(t_{n-1})=0$ for all $1\leq n \leq N_T$.
Let $M_1(k)$ denote the model, where one glitch occurs
in the interval $t_{k-1}\leq t \leq t_k$,
i.e.\ we have $g(t_{n-1})=\delta_{nk}$ for all $1\leq n \leq N_T$
($\delta_{nk}$ is the Kronecker delta symbol).
Let $M_2(k,l)$ denote the model, where one glitch occurs
in the interval $t_{k-1} \leq t \leq t_k$,
and another glitch occurs in the nonoverlapping interval $t_{l-1} \leq t \leq t_l$
i.e.\ we have $g(t_{n-1})=\delta_{nk}+\delta_{nl}$ with $k\neq l$.
In the tests in this paper we consider a maximum of one glitch
in the interval $t_0 \leq t \leq N_T$,
except in the worked example involving PSR J0835$-$4510 in \S\ref{sec:hmm7},
where we briefly consider a maximum of two glitches.
In practice, when analysing real data,
one can generalize the model family to an arbitrary number of glitches
using a greedy hierarchical algorithm \citep{suv18},
discussed in \S\ref{sec:hmm3aa}.
Alternatively one can subdivide the data into multiple segments,
each of which is likely to contain one glitch at most,
based on history or the outcome of a preliminary {\sc tempo2} fit.
The exact subdivision is left to the analyst's discretion;
e.g.\ for PSR J0534$+$2200 and PSR J0537$-$6910, 
one might choose segments of
$\approx 1\,{\rm yr}$ and $\approx 0.3\,{\rm yr}$ respectively.

We can compare the relative plausibility of two models
by calculating their evidence ratio or Bayes factor.
The evidence for a model $M$ is defined as the probability
$\Pr(O_{1:N_T} | M)$ of measuring the data $O_{1:N_T}$ given $M$.
\footnote{
The definition of the evidence depends on the form of 
Bayes's Theorem under consideration.
If we consider 
$\Pr(Q_{1:N_T} | O_{1,N_T}, M)
 = \Pr(O_{1:N_T} | Q_{1,N_T},M) \Pr(Q_{1:N_T} | M) / \Pr(O_{1:N_T} |M)$
for fixed $M$, then
$\Pr(O_{1:N_T} | Q_{1,N_T},M)$ is the likelihood,
and
$\Pr(O_{1:N_T} |M)$ in (\ref{eq:hmm16}) is the evidence,
as in this paper.
If we consider
$\Pr(M | O_{1,N_T})
 = \Pr(O_{1:N_T} | M) \Pr(M) / \Pr(O_{1:N_T})$
after marginalizing over $Q_{1:N_T}$,
then
$\Pr(O_{1:N_T} |M)$ is the likelihood,
and 
$\Pr(O_{1:N_T}) = \sum_M \Pr(O_{1:N_T} | M) \Pr(M)$
is the evidence.
}
In the HMM context,
$\Pr(O_{1:N_T} | M)$ equals the probability of measuring $O_{1:N_T}$
given a hidden state sequence $Q_{1:N_T}$,
multiplied by the probability of $Q_{1:N_T}$, 
marginalized over all admissible sequences:
\begin{equation}
 \Pr(O_{1:N_T} | M)
 =
 \sum_{Q_{1:N_T}} 
 \Pr(O_{1:N_T} | Q_{1:N_T}, M) 
 \Pr(Q_{1:N_T}, M)~.
\label{eq:hmm16}
\end{equation}
There exist $N_Q^{N_T}$ possible sequences $Q_{1:N_T}$ in general
but they all pass through the same set of $N_Q$ states at each HMM step.
Therefore the sum in (\ref{eq:hmm16}) can be computed efficiently
from partial sums accounting 
for the $N_Q^2$ possible transitions at each step.
Appendix \ref{sec:hmmappa} explains how to do this 
using the HMM forward algorithm 
\citep{rab89},
which calculates
$\Pr [q(t_{n+1})=q_i, O_{1:n+1} | M]$ from
$\Pr [q(t_n)=q_j, O_{1:n} | M]$ by induction for $1\leq i,j \leq N_Q$,
with
\begin{equation}
 \Pr [q(t_{n+1})=q_i, O_{1:n+1} | M]
 =
 L_{o(t_{n+1}) q_i}
 \sum_{j=1}^{N_Q} A_{q_i q_j}
 \Pr [q(t_n)=q_j, O_{1:n} | M]~.
\label{eq:hmm17}
\end{equation}
Pseudocode for the HMM forward algorithm 
is presented in Appendix \ref{sec:hmmappa}.
\footnote{
To increase accuracy and avoid arithmetic underflow
when computing products with many factors, such as (\ref{eq:hmm3}),
we take advantage of the log-sum-exp approximation
\citep{cal14}.
}

If the Bayes factor exceeds a threshold, 
the model in the numerator is preferred.
There is no unique way to set the threshold.
On the popular Jeffreys scale
\citep{jef98}, 
a Bayes factor above $10$ counts as ``strong'' evidence,
and a Bayes factor between $10^{1/2}$ and 10 counts as ``substantial''.
In this paper, we arbitrarily regard a glitch as having occurred
in the interval $t_{k-1} \leq t \leq t_k$,
if we obtain
$\Pr[O_{1:N_T} | M_1(k) ] > 10^{1/2} \Pr(O_{1:N_T} | M_0)$.
Tests with arbitrarily higher thresholds
(up to $10^2$, which counts as ``decisive'' on the Jeffreys scale)
yield qualitatively similar results.

\subsection{Multiple glitches
 \label{sec:hmm3aa}}
To search for multiple glitches in data which are not subdivided,
\citet{suv18} proposed a greedy hierarchical algorithm,
which works as follows.
For $m=1,2,\dots$ in increasing order, construct the sequence of
Bayes factors
\begin{equation}
 K_m(k)
 =
 \frac{\Pr[O_{1:N_T} | M_m(k_1^\ast, \dots, k_{m-1}^\ast, k)]}
  {\Pr[O_{1:N_T} | M_{m-1}(k_1^\ast, \dots, k_{m-1}^\ast)]}~,
\label{eq:hmm17a}
\end{equation}
where $k_{m'}^\ast$ indexes the TOA corresponding to the $m'$-th
detected glitch,
and the no-glitch model $M_0$ has no arguments.
In other words, $K_m(k)$ evaluates, as a function of $k$,
the evidence for a model with $m$ glitches
at $\{ k_1^\ast,\dots, k_{m-1}^\ast, k \}$
compared to the evidence for a model with $m-1$ glitches
at $\{ k_1^\ast,\dots,  k_{m-1}^\ast \}$.
Starting from $m=1$, if $K_m(k)$ exceeds the user-selected threshold for some $k$
[e.g.\ $K_m(k) > 10^{1/2}$ for some $k$], 
we set $k_m^\ast = {\rm arg\,max\,}_k K_m(k)$ and increment $m$.
The iteration halts, when we obtain $K_m(k) < 10^{1/2}$ for all $k$.

\subsection{Optimal ephemeris
 \label{sec:hmm3b}}
Once the preferred model is identified out of the set
$\{ M_0, M_1(k), M_2(k,l), \dots \}$,
the next step is to compute the ephemeris which fits the data best,
given the preferred model.
There is no unique definition of ``best'',
as discussed in \S\ref{sec:hmm2a} and Appendix \ref{sec:hmmappa}.
In this paper, we stipulate that the optimal ephemeris is the one
constructed from the most probable state at each HMM step,
given by
\begin{eqnarray}
 \hat{q}(t_n)
 & = &
 \underset{q(t_n)} {\rm arg\,\,max\,\,}
 \sum_{Q_{1:n-1}}
 \Pi_{q(t_1)} L_{o(t_1) q(t_1)}
 \prod_{m=2}^{n-1}
 A_{q(t_{m-1}) q(t_m)} L_{o(t_m) q(t_m)}
 \nonumber \\
 & &
 \times
 A_{q(t_{n-1}) q(t_n)} L_{o(t_n) q(t_n)}
 \sum_{Q_{n+1:N_T}}
 \prod_{m=n+1}^{N_T}
 A_{q(t_{m-1}) q(t_m)} L_{o(t_m) q(t_m)}~.
\label{eq:hmm18}
\end{eqnarray}
Equation (\ref{eq:hmm18}) takes sequences of the form
$\{ Q_{1:n-1}, q(t_n) = q_i, Q_{n+1,N_T} \}$,
calculates their probabilities according to (\ref{eq:hmm3}) for $q_i$ fixed,
sums the probabilities over $Q_{1:n-1}$ and $Q_{n+1:N_T}$,
then maximizes over $1\leq i \leq N_Q$.
It can be evaluated efficiently by the
HMM forward-backward algorithm, whose pseudocode is presented
in Appendix \ref{sec:hmmappa}.
The approach maximizes the number of
most probable states in the ephemeris.
It also generates the PDF of $q(t_n)$ automatically as a by-product,
allowing one to examine the states in the neighborhood of the peak,
to see how much $\hat{q}(t_n)$ stands out.
The results can be checked for broad consistency against 
$Q^\ast_{1:N_T}$ (see \S\ref{sec:hmm2a}).
The subtle difference between
$Q^\ast_{1:N_T}$ and $\{ \hat{q}(t_1),\dots,\hat{q}(t_{N_T}) \}$,
along with the Viterbi algorithm which computes the former sequence efficiently,
are described in Appendix \ref{sec:hmmappa}.

\subsection{Computational cost
 \label{sec:hmm3d}}
From a practical standpoint, the computational cost of the glitch detector
depends on what astrophysical experiment is being attempted.
For example, a search for three glitches in a stretch of data
with the greedy hierarchical algorithm in \S\ref{sec:hmm3aa}
involves passing the data through the HMM $3N_T$ times:
$N_T$ times to calculate $K_1(k)$ for model $M_1(k)$ and $1\leq k \leq N_T$,
$N_T$ times to calculate $K_2(k)$ 
for model $M_2(k_1^\ast,k)$ and $1\leq k \leq N_T$,
and $N_T$ times to calculate $K_3(k)$ 
for model $M_3(k_1^\ast,k_2^\ast,k)$ and $1\leq k \leq N_T$.
In order to embrace a variety of experiments,
we present below a rough cost estimate for the key computational step
which is common to all of them: 
a single pass of the HMM forward algorithm through the full data
to calculate one Bayes factor, e.g.\ $K_1(k)$.
The HMM backward algorithm, 
which calculates the associated optimal ephemeris,
costs roughly the same.

The cost of the HMM forward algorithm is of order $N_Q^2 N_T$,
as described in Appendix \ref{sec:hmmappa}.
Importantly, it does not depend on the data or the model parameters,
with one exception which we discuss below.
The HMM addresses each of the $N_Q^2 N_T$ links in the HMM trellis once
without discretion and without reference to any tolerances;
iterative convergence does not play a role.
\footnote{
In other algorithms like Markov chain Monte Carlo samplers,
convergence is an issue, 
and the run time depends on the shape of the posterior distribution,
the form of the proposal function, and the tolerance.
}
Preliminary benchmarking tests,
characteristic of the computations in \S\ref{sec:hmm6}
and done on a consumer-grade, quad-core 
Intel CPU with $2.7\,{\rm GHz}$ clock speed, 
indicate that the run time for one pass of the HMM forward algorithm
scales approximately as
\begin{equation}
 T_{\rm CPU} 
 =
 9 
 \left( \frac{N_f}{10^3} \right)^2
 \left( \frac{N_{\dot{f}}}{10} \right)^2
 \left( \frac{N_T}{10^2} \right)
 \, {\rm s}~,
\label{eq:hmm18a}
\end{equation}
where $N_f$ and $N_{\dot{f}}$ are the number of $f$ and $\dot{f}$ bins
in the DOI respectively
(see Appendix \ref{sec:hmmappc}).
Hence, from (\ref{eq:hmm18a}),
a typical experiment searching for a single glitch among $N_T$ TOAs takes 
$N_T T_{\rm CPU} \approx 9\times 10^2
 ({N_f} / {10^3})^2
 ({N_{\dot{f}}} / {10})^2
 ({N_T} / {10^2})^2
 \, {\rm s}$,
independent of $\sigma$ and $x_n$.

In the transition probability $A_{q(t_n)q(t_{n-1})}$ in (\ref{eq:hmm10}),
each Gaussian term in the sum over $G$ extends formally across the whole DOI. 
To accelerate the computation, we truncate $A_{q(t_n)q(t_{n-1})}$
at three standard deviations along the $f$ axis.
If the truncated $A_{q(t_n)q(t_{n-1})}$ spans multiple frequency bins,
the computational cost scales according to (\ref{eq:hmm18a}).
If the truncated $A_{q(t_n)q(t_{n-1})}$ fits wholly within
one frequency bin,
the scaling with $N_f$ is linear instead,
and one finds
$T_{\rm CPU} \approx 0.8
 ({N_f} / {10^3})
 ({N_{\dot{f}}} / {10})^2
 ({N_T} / {10^2})
 \, {\rm s}$.
The latter scaling prevails over (\ref{eq:hmm18a}), when
$\approx 3 \sigma \langle x_n \rangle^{3/2}$ drops below
the frequency bin width.
The latter dependence on $\sigma$ and $x_n$ is the exception 
foreshadowed in the previous paragraph.
It stems from an implementation trick
and is not fundamental to the HMM forward algorithm.

Further study of the computational cost is postponed to future work
as it raises the role of graphics processing units (GPUs),
a topic outside the scope of this paper.
GPUs have proved effective in accelerating HMM-based searches 
for continuous gravitational wave signals with
the Laser Interferometer Gravitational Wave Observatory
\citep{abb19,dun20}.
Acceleration by a factor of $\sim 40$ is achieved
in the latter references.

\section{Synthetic data
 \label{sec:hmm4}}
We now quantify the performance of the HMM systematically
through a suite of Monte Carlo tests based on synthetic data.
Many valid recipes exist to generate the synthetic data;
the physical origin and hence the statistics
of the fluctuating torque are unknown from first principles and
cannot be inferred uniquely from pulsar timing noise studies
\citep{cor85,hob04}.
In this paper, we take an empirical approach and generate data
consistent with glitch templates
derived from traditional pulsar timing studies
\citep{mcc87,won01},
without seeking to relate the output to an underlying physical model.
The TOAs are sampled according to a Poisson observing process
for simplicity, as described in Appendix \ref{sec:hmmappe},
to ensure that they do not coincide artificially with a glitch,
but any reasonable sampling algorithm (e.g.\ uniform spacing)
does just as well.
When analysing real data, the TOAs are referred first
to the Solar System barycenter using standard methods 
\citep{tay92}.
We do not consider the orbital motion of binary pulsars in this paper.

Let $T>0$ be an epoch, when a glitch occurs.
Consider a time interval containing $t=T$,
which is short enough that spin wandering 
and the secular component of $\ddot{f}$ can be ignored temporarily,
i.e.\ $\ddot{f}=0$.
Traditional pulsar timing studies based on empirical fits to the data
propose that the system evolves according to
\citep{mcc87,won01}
\begin{equation}
 f(t) =
 f(0) + \dot{f}(0) t
 + \{ \Delta f_{\rm p} + \Delta \dot{f}_{\rm p} (t-T)
 + \Delta f_1 \exp[-(t-T)/\tau] \} H(t-T)~,
\label{eq:hmm19}
\end{equation}
where $H(\dots)$ symbolizes the Heaviside step function,
and $\Delta f_1$ and $\tau$ are the amplitude and $e^{-1}$ 
recovery time-scale respectively
of the transient component of the frequency jump following the glitch.
Now suppose that the time interval is long enough,
that spin wandering cannot be neglected.
Then (\ref{eq:hmm19}) still describes the deterministic evolution
before and after the glitch 
(neglecting $\langle \ddot{f} \rangle$; see \S\ref{sec:hmm2a})
but with a random walk added.
The random walk can be generated in many valid ways.
In Appendix \ref{sec:hmmappe} we present and justify a systematic recipe,
which involves solving a system of two stochastic differential equations,
one of which [see equation (\ref{eq:hmm20})] takes the form
\begin{equation}
 \frac{df}{dt} = {\rm deterministic \, \, terms} + \zeta(t)~,
\label{eq:hmm20a}
\end{equation}
with
\begin{equation}
 \langle \zeta(t) \zeta(t') \rangle
 = 
 \sigma_{\rm TN}^2 \delta(t-t')~.
\label{eq:hmm22a}
\end{equation}
In (\ref{eq:hmm20a}) and (\ref{eq:hmm22a}),
the deterministic terms model secular spin down
and glitch-related jumps and recoveries,
$\zeta(t)$ is a zero-mean, white-noise torque,
$\delta(\dots)$ is the Dirac delta function,
and
$\sigma_{\rm TN}$ is the timing noise amplitude 
(units: ${\rm Hz\,s^{-1/2}}$).
The white torque noise is filtered by the deterministic terms
in (\ref{eq:hmm20a}) to produce red frequency noise in $f(t)$.
Multiple exponential recoveries can be added to (\ref{eq:hmm19})
and are discussed in Appendix \ref{sec:hmmappe}.

As a prelude to the systematic performance tests in \S\ref{sec:hmm6},
we walk the reader through a practical, representative worked example,
where the HMM detects a glitch injected into synthetic data.
The worked example is laid out in Appendix \ref{sec:hmmappf}.
It presents graphically the output of the key intermediate steps
in \S\ref{sec:hmm2} and \S\ref{sec:hmm3},
including setting the DOI and grid spacing,
calculating the Bayes factor $K_1(k)$ as a basis for model selection,
and calculating the point-wise and sequence-wise optimal ephemerides
to estimate the injected jumps in $f$ and $\dot{f}$.

\section{ROC curves
 \label{sec:hmm6}}
Glitch detection by model selection involves asking
if the Bayes factor relating two models exceeds a threshold.
The threshold determines the false alarm probability, $P_{\rm fa}$.
Given $P_{\rm fa}$,
the detection probability, $P_{\rm d}$,
can be expressed as a function of the signal parameters,
e.g.\ glitch size $\Delta f_{\rm p}$.
One can set the threshold by fiat, as in \S\ref{sec:hmm3a},
and infer $P_{\rm fa}$ or vice versa.
In this section, we evaluate the HMM's performance 
by constructing ROC curves
($P_{\rm d}$ versus $P_{\rm fa}$,
with signal parameters fixed)
and detection probability curves
($P_{\rm d}$ versus one or more signal parameters,
with $P_{\rm fa}$ fixed)
for a range of representative values
of the intrinsic astrophysical and measurement noises in the system
(\S\ref{sec:hmm6a}),
secular spin-down parameters, e.g. $f_{\rm LS}$ and $\dot{f}_{\rm LS}$
(\S\ref{sec:hmm6c}),
and glitch parameters, e.g.\ size and recovery time-scale
(\S\ref{sec:hmm6d}).
A short, preliminary analysis of the impact on performance of
the observational scheduling strategy, e.g.\ mean inter-TOA interval,
is presented in Appendix \ref{sec:hmmappg}.
Optimizing the observational schedule is a subtle exercise,
which we will take up more fully in future work.

\subsection{Intrinsic and measurement noises
 \label{sec:hmm6a}}
The detectability of a glitch is connected to its size
relative to the noise,
which comes in two flavors.
A glitch may be drowned out by TOA measurement errors;
if $\kappa$ is too small,
the peaks in $L_{x_n q(t_n)}$ in (\ref{eq:hmm7}) blur together.
A glitch may also be obscured by astrophysical timing noise,
if $\Delta f_{\rm p}$ is relatively small,
$\sigma_{\rm TN}$ is relatively large,
and there are long delays between TOAs.
Conversely, a random walk with large $\sigma_{\rm TN}$ 
may masquerade as a step $\Delta f_{\rm p}\neq 0$ during a subset of TOAs,
triggering a false alarm.

\begin{figure}
\begin{center}
\includegraphics[width=8cm,angle=0]{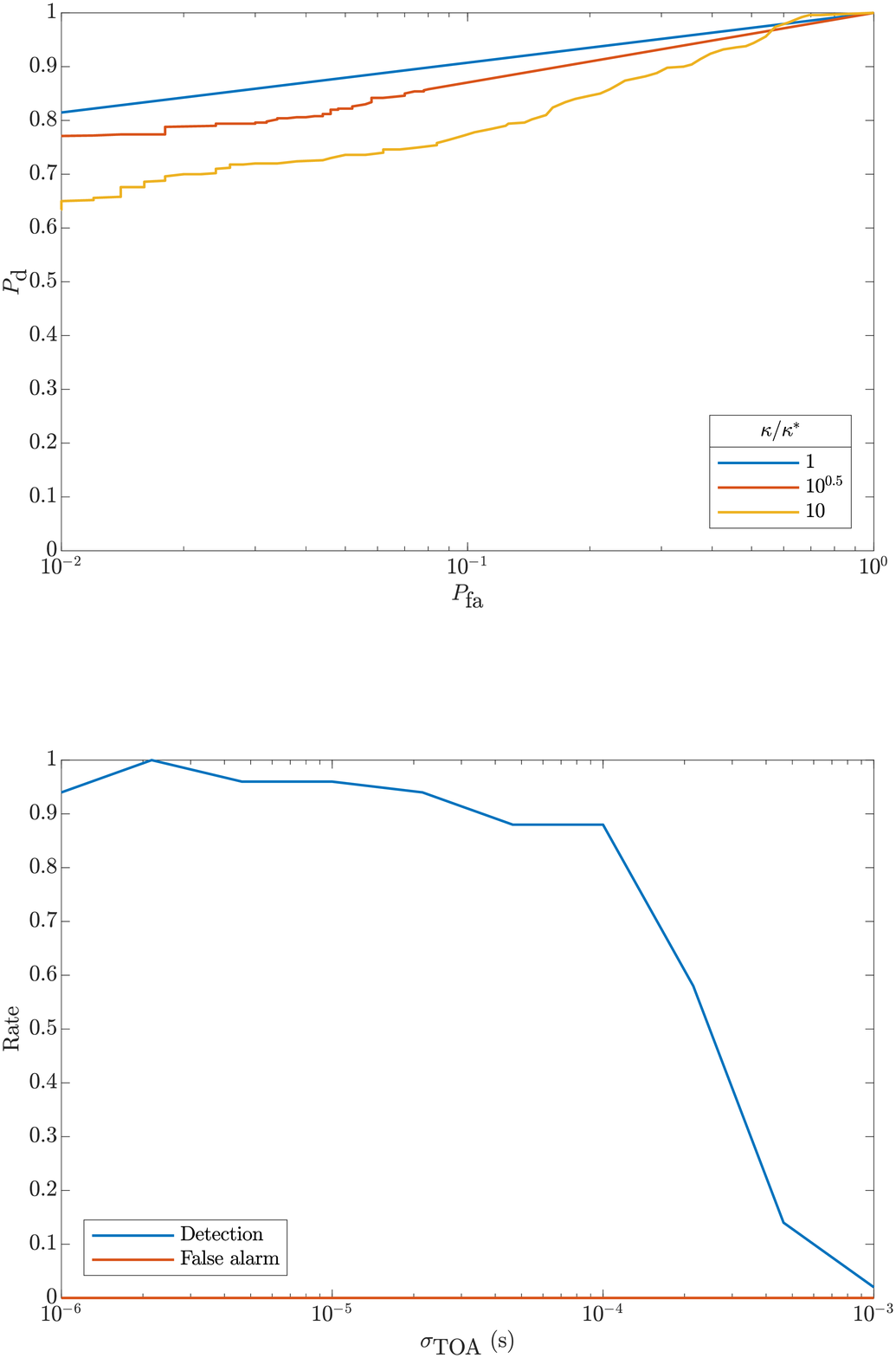}
\end{center}
\caption{
HMM glitch detector performance as a function of measurement uncertainty
$\sigma_{\rm TOA}$.
({\em Top panel.}) 
ROC curve ($P_{\rm d}$ versus $P_{\rm fa}$) 
for three values of $\kappa/\kappa^\ast$
in the range $1 \leq \kappa/\kappa^\ast \leq 10$,
with $\kappa^\ast$ adjusted as a function of $\sigma_{\rm TOA}$
according to the recipe in Appendix \ref{sec:hmmappc}.
({\em Bottom panel.})
Detection probability $P_{\rm d}$ (blue curve) and 
false alarm probability $P_{\rm fa}$ (red curve, obscured by horizontal axis)
versus $\sigma_{\rm TOA}$ (units: s),
with the Bayes factor threshold kept at $10^{1/2}$
to give $P_{\rm fa} \approx 10^{-2}$ on average
across the plotted range.
Parameters:
as in penultimate (typical) column in Table \ref{tab:hmm1},
except with
$10^{-6} \leq \sigma_{\rm TOA} / (1\,{\rm s}) \leq 10^{-3}$.
Realizations: $1.5\times 10^3$ per ROC curve.
}
\label{fig:hmm7}
\end{figure}

Figure \ref{fig:hmm7} illustrates how the task of detection 
is affected by $\sigma_{\rm TOA}$.
Measurement errors enter through $L_{x_n q(t_n)}$,
which depends on $\sigma_{\rm TOA}$ through $\kappa$ 
as defined by (\ref{eq:hmm7}) and (\ref{eq:hmm8})
or (\ref{eq:hmm8a}) after gridding.
The top panel in Figure \ref{fig:hmm7} displays ROC curves
for three values of $\kappa$ ranging from $\kappa^\ast$
to $10\kappa^\ast$,
where $\kappa^\ast$ is the fiducial value calculated according to
the recipe in \S\ref{sec:hmm2c} and Appendix \ref{sec:hmmappc}.
The results are encouraging.
For $\kappa = \kappa^\ast$,
we obtain $P_{\rm d} \geq 0.8$ for $P_{\rm fa}\geq 10^{-2}$
and $P_{\rm d} \geq 0.9$ for $P_{\rm fa}\geq 10^{-1}$.
The HMM's performance varies mildly with $\kappa$,
e.g.\ $P_{\rm d}$ drops by $\lesssim 0.15$ 
across the ROC curve for $\kappa = 10 \kappa^\ast$;
it does not depend sensitively,
on how one estimates $\kappa^\ast$ from $\sigma_{\rm TOA}$.
The bottom panel summarizes the behavior in a practical fashion
by graphing $P_{\rm d}$ and $P_{\rm fa}$ versus $\sigma_{\rm TOA}$
for parameters matching the penultimate column in Table \ref{tab:hmm1},
with $\kappa = \kappa^\ast$ updated according to
\S\ref{sec:hmm2c} and Appendix \ref{sec:hmmappc}.
The Bayes factor threshold is kept at $10^{1/2}$
and maintains $P_{\rm fa} \approx 10^{-2}$
across the plotted $\sigma_{\rm TOA}$ range;
false alarms are not sensitive to $\sigma_{\rm TOA}$,
when $\kappa$ is updated.
The detection probability drops off,
as $\sigma_{\rm TOA}$ increases,
with $P_{\rm d} \leq 0.9$ for $\sigma_{\rm TOA} \gtrsim 3 \times 10^{-5}\,{\rm s}$.
Roughly speaking, one requires $\sigma_{\rm TOA} \propto \Delta f_{\rm p}$
to maintain a desired $P_{\rm d}$ value.

\begin{table}
\begin{center}
\begin{tabular}{lccccc}
\hline
 Quantity & Symbol & Units & Min & Typical & Max \\
\hline
 {\em Noise} & & & & & \\
 Timing noise amplitude & $\sigma_{\rm TN}$ & ${\rm Hz \, s^{-1/2}}$ &
  $10^{-15}$ & $10^{-13}$ & $10^{-11}$ \\
 TOA measurement uncertainty & $\sigma_{\rm TOA}$ & ${\rm s}$ &
  $10^{-6}$ & $10^{-5}$ & $10^{-3}$ \\
\hline
 {\em Scheduling} & & & & & \\
 Mean waiting time & $\langle x_n \rangle$ & {\rm d} & 
  $10^{-2}$ & $13$ & 116 \\
 Number of sessions & --- & --- &
  5 & 51 & $10^2$ \\
\hline
 {\em Secular spin down} & & & & & \\
 Frequency & $f_{\rm LS}$ & ${\rm Hz}$ & 
  $10^{0}$ & 5.435 & $10^{2}$ \\
 Frequency derivative & $-\dot{f}_{\rm LS}$ & ${\rm Hz\,s^{-1}}$ & 
  $10^{-15}$ & $10^{-15}$ & $10^{-11}$ \\
\hline
 {\em Glitch} & & & & & \\
 Permanent frequency jump & $\Delta f_{\rm p}$ & ${\rm Hz}$ &
  $10^{-10}$ & $10^{-8}$ & $10^{-7}$ \\
 Transient frequency jump & $\Delta f_1$ & ${\rm Hz}$ &
  0 & 0 & $10^{-8}$ \\
 Recovery time-scale & $\tau$ & ${\rm s}$ &
  $10^3$ & $10^{10}$ & $10^{10}$ \\
 Permanent frequency derivative jump & $\Delta \dot{f}_{\rm p}$ & ${\rm Hz\,s^{-1}}$ &
  $-10^{-12}$ & $10^{-15}$ & $10^{-15}$ \\
\hline
\end{tabular}
\end{center}
\caption{
Parameters used to generate synthetic data to test the HMM,
classified as astrophysical and measurement noise,
scheduling of observations, secular spin down, and glitch parameters.
}
\label{tab:hmm1}
\end{table}

The HMM also contends with astrophysical timing noise.
An important practical issue is how to select the HMM parameter $\sigma$
for a particular astrophysical target.
Again, there is no unique prescription,
and the final conclusions concerning glitch detection are conditional
on the choice made. 
\footnote{
The same applies to traditional timing methods or pulse domain analysis, 
where the conclusions concerning glitch detection are conditional
on the phase model, e.g.\ Taylor expansion.
}
A useful rule of thumb is to match the root mean square
phase residual $\langle \delta \phi(t_n)^2 \rangle^{1/2}$
accumulated by the random walk in the HMM,
derived by integrating (\ref{eq:hmm9}),
with the phase residual accumulated by the timing noise in the pulsar,
derived by integrating (\ref{eq:hmm20a}) and (\ref{eq:hmm22a}).
The latter quantities are of order
$\sigma \langle x_n \rangle^{5/2}$
and
$\sigma_{\rm TN} \langle x_n \rangle^{3/2}$
respectively when integrated over the mean TOA gap,
$\langle x_n \rangle$,
which implies
$\sigma \approx \langle x_n \rangle^{-1} \sigma_{\rm TN}
 = \sigma^\ast$.
In practice $\langle x_n \rangle$ is dominated by the TOA intervals
between rather than within observation sessions.

\begin{figure}
\begin{center}
\includegraphics[width=8cm,angle=0]{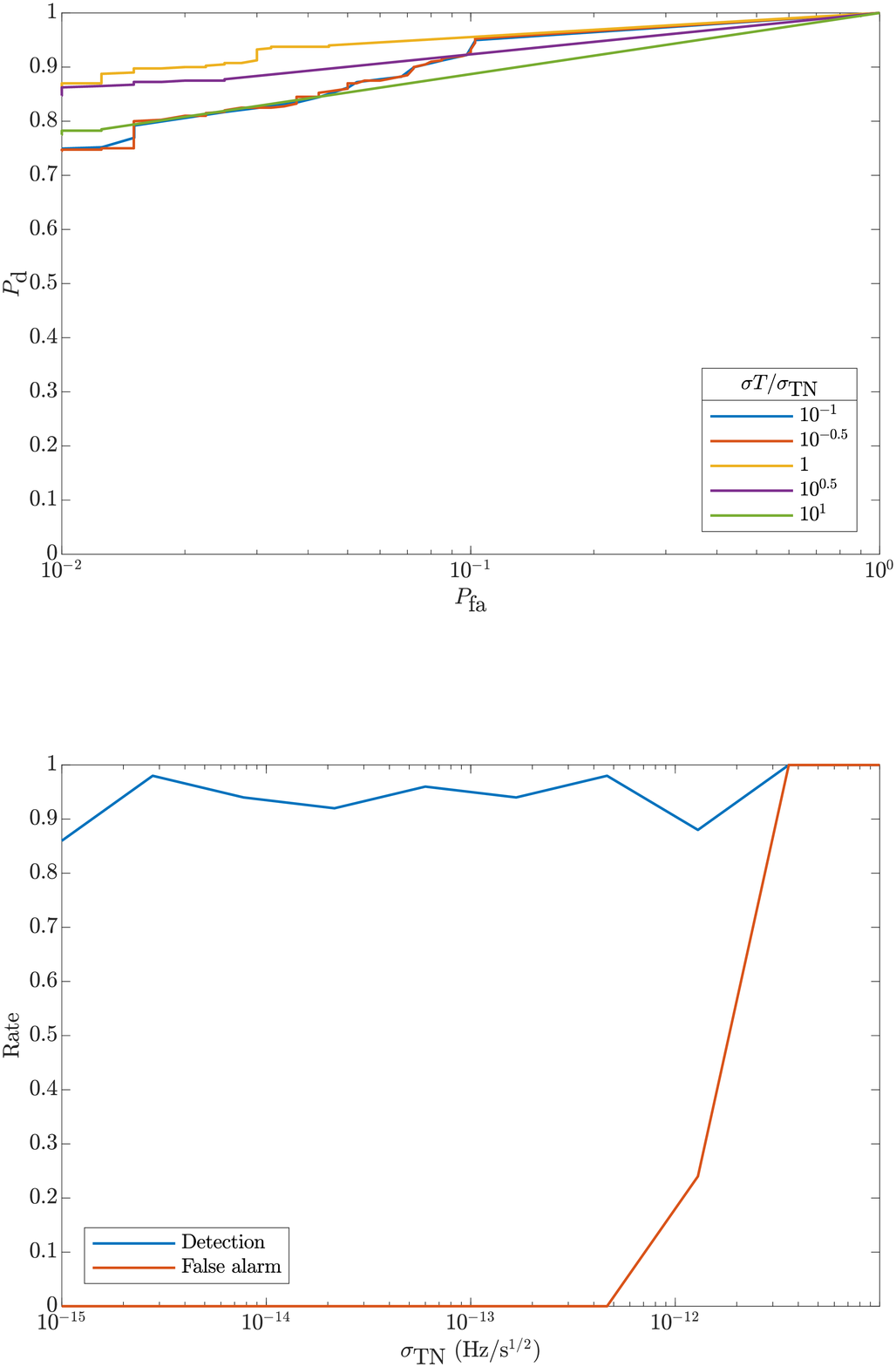}
\end{center}
\caption{
HMM glitch detector performance as a function of HMM noise parameter $\sigma$
and timing noise amplitude $\sigma_{\rm TN}$.
({\em Top panel.}) 
ROC curve ($P_{\rm d}$ versus $P_{\rm fa}$) 
for five values of $\sigma/\sigma^\ast$
in the range $10^{-1} \leq \sigma/\sigma^\ast \leq 10^1$,
with $\sigma^\ast$ set according to the recipe in \S\ref{sec:hmm6a}
as a function of $\sigma_{\rm TN}$ and $\langle x_n \rangle$.
({\em Bottom panel.})
Detection probability $P_{\rm d}$ (blue curve) and 
false alarm probability $P_{\rm fa}$ (red curve)
versus $\sigma_{\rm TN}$ (units: ${\rm Hz\,s^{-1/2}}$),
with the Bayes factor threshold held at $10^{1/2}$.
The curves are restricted to 
$\sigma_{\rm TN} \leq 3 \times 10^{-12}\, {\rm Hz\,s^{-1/2}}$
by the phase wandering mismatch described in \S\ref{sec:hmm6a}.
Parameters:
as in Figure \ref{fig:hmm7},
except with
$10^{-15} \leq \sigma_{\rm TN} / (1 \, {\rm Hz\,s^{-1/2}}) \leq 10^{-11}$.
Realizations: $1.5\times 10^3$ per ROC curve.
}
\label{fig:hmm8}
\end{figure}

Figure \ref{fig:hmm8} illustrates how the HMM's performance varies,
as $\sigma$ moves away from $\sigma^\ast$.
The top panel displays five ROC curves for 
$10^{-1} \leq \sigma/\sigma^\ast \leq 10^1$.
For $\sigma=\sigma^\ast$,
we obtain $P_{\rm d} \geq 0.87$ for $P_{\rm fa}\geq 10^{-2}$
and $P_{\rm d} \geq 0.95$ for $P_{\rm fa}\geq 10^{-1}$.
The results do not change much near the optimum,
with $P_{\rm d}$ changing by $\lesssim 0.1$ over the range
$10^{-1} \leq \sigma/\sigma^\ast \leq 10^{1}$
for $P_{\rm fa} \geq  10^{-2}$.
The bottom panel in Figure \ref{fig:hmm8}
summarizes the results in practical, observation-ready terms.
Adjusting $\sigma=\sigma^\ast$ as a function of $\sigma_{\rm TN}$
and $\langle x_n \rangle$ as in the previous paragraph,
we find that the detection probability stays roughly constant,
as $\sigma_{\rm TN}$ increases,
with $P_{\rm d} \gtrsim 0.9$ (and $P_{\rm fa} \approx 10^{-2}$) for 
$\sigma_{\rm TN} \leq 10^{-12} (\Delta f_{\rm p} / 10^{-8}\,{\rm Hz}) 
 \, {\rm Hz\,s^{-1/2}}$.
We also find that $P_{\rm fa}$ rises steeply for 
$\sigma_{\rm TN} \gtrsim 10^{-12} (\Delta f_{\rm p} / 10^{-8}\,{\rm Hz}) 
 \, {\rm Hz\,s^{-1/2}}$,
as the HMM misinterprets strong timing noise as glitches.
For reference, observations yield
$10^{-15} \lesssim \sigma_{\rm TN} / (1\,{\rm Hz \, s^{-1/2}}) 
 \lesssim 10^{-11}$
typically for nonrecycled pulsars;
for example, PSR J0534$+$2200 has
$\sigma_{\rm TN}^2 = 6\times 10^{-23} \, {\rm Hz^2 \, s^{-1}}$
\citep{cor80b}.
Therefore, even at the upper end of the measured $\sigma_{\rm TN}$ range,
the HMM can detect glitches with 
$\Delta f_{\rm p} \gtrsim 10^{-7}\,{\rm Hz}$.
Note that timing noise is red over long time-scales
but approximately white over $\langle x_n \rangle$ in most observations.
Care must be exercised when estimating
$\sigma_{\rm TOA}$ and $\sigma_{\rm TN}$
indirectly from {\sc tempo2} residuals,
because the Taylor expansion phase model correlates $\sigma_{\rm TOA}$
and $\sigma_{\rm TN}$ in a complicated way.

In practice the rule of thumb
$\sigma \approx \langle x_n \rangle^{-1} \sigma_{\rm TN}$
is modified by two factors: 
binning, and the functional form of the spin wandering.
Binning errors in $\dot{f}$ impel $f(t)$ to drift.
By setting $\sigma$ high enough to accommodate the drift
in the transition probability,
we ensure that the HMM self-corrects via the timing noise channel.
The frequency residuals generated by $\dot{f}$ binning and timing noise
are of order $\eta_{\dot{f}} \langle x_n \rangle$ and
$\sigma \langle x_n \rangle^{3/2}$
respectively when integrated over the mean TOA gap,
where $\eta_{\dot{f}}$ is the grid spacing in $\dot{f}$,
implying 
$\sigma \geq \sigma_{\rm min} = \langle x_n \rangle^{-1/2} \eta_{\dot{f}}$
and hence
$\sigma^\ast =
 \max(\sigma_{\rm min},\langle x_n \rangle^{-1} \sigma_{\rm TN})$.
Additionally,
the HMM is sensitive to the mismatch in phase wandering between the data 
(e.g.\ white noise in $\dot{f}$; see \S\ref{sec:hmm4})
and the transition probabilities 
(white noise in $\ddot{f}$; see \S\ref{sec:hmm2d}).
The mean square phase residuals arising from the two processes
are given by
$\sigma_{\rm TN}^2 x_n^3$ and
$\sigma^2  x_n^5$
respectively when integrated over a specific TOA gap $x_n$.
Substituting the rule of thumb
$\sigma \approx \langle x_n \rangle^{-1} \sigma_{\rm TN}$,
we calculate the frequency mismatch to be
$\approx (\langle x_n \rangle^{-1} x_n  - 1 )
 x_n^{1/2} \sigma_{\rm TN}$,
which exceeds the frequency bin size $\eta_f$ for certain combinations
of $x_n$, $\langle x_n \rangle$, and $\sigma_{\rm TN}$.
For the parameters in the penultimate column of Table \ref{tab:hmm1},
with $\eta_f = 6\times 10^{-10}\,{\rm Hz}$,
the effect becomes significant for 
$\sigma_{\rm TN} \gtrsim 3\times 10^{-12} \, {\rm Hz\,s^{-1/2}}$,
which corresponds to the cut-off in the curves 
in the bottom panel of Figure \ref{fig:hmm8}.

\subsection{Secular spin down
 \label{sec:hmm6c}}
\begin{figure}
\begin{center}
\includegraphics[width=14cm,angle=0]{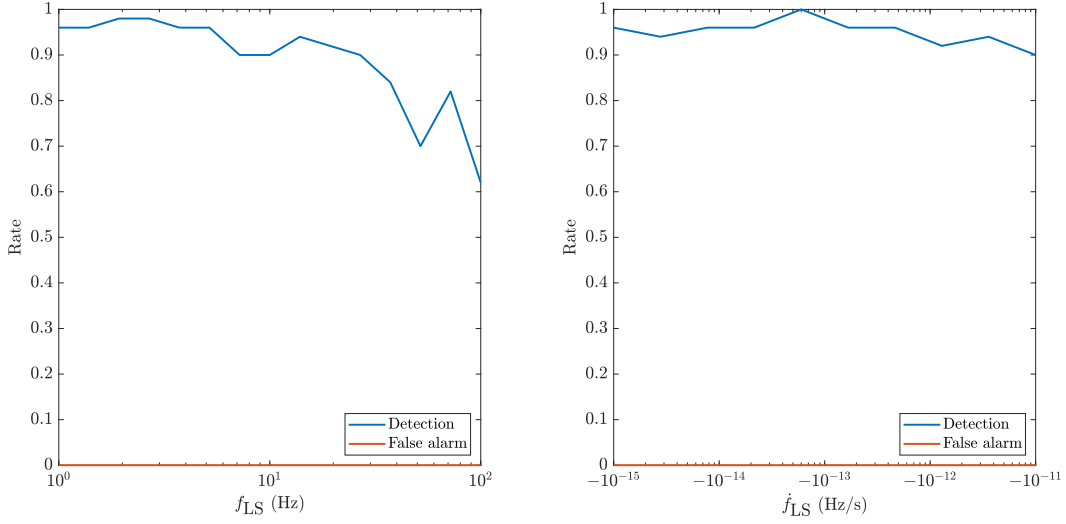}
\end{center}
\caption{
Detection probability (blue curve) and false alarm probability (red curve)
versus secular spin-down parameters:
$f_{\rm LS}$ (in ${\rm Hz}$) (left panel)
and $\dot{f}_{\rm LS}$ (in ${\rm Hz\,s^{-1}}$) (right panel).
Other parameters:
see penultimate (typical) column in Table \ref{tab:hmm1}.
}
\label{fig:hmm11}
\end{figure}

Glitch detection is fundamentally an exercise in tracking fluctuations
around the secular spin-down trend and distinguishing statistically
between a continuous random walk (timing noise) and discontinuous jumps (glitches).
One therefore expects the HMM's performance to be approximately
independent of the secular trend itself, 
i.e.\ $f_{\rm LS}$ and $\dot{f}_{\rm LS}$,
as long as
$\kappa \propto (\sigma_{\rm TOA} f_{\rm LS})^{-2}$
is held fixed, 
\footnote{
The rough proportionality 
$\kappa \propto (\sigma_{\rm TOA} f_{\rm LS})^{-2}$
governs how accurately $N_n$ can be inferred
through (\ref{eq:hmm7}) and (\ref{eq:hmm8})
before the modifications introduced by gridding 
(see Appendix \ref{sec:hmmappc}).
}
while the spin-down parameters vary.
Figure \ref{fig:hmm11} confirms that $P_{\rm d}$ stays approximately
constant for
$1 \leq f_{\rm LS} / (1\,{\rm Hz}) \lesssim 20$
and drops away for
$f_{\rm LS} \gtrsim 20\,{\rm Hz}$
for the parameters in Table \ref{tab:hmm1},
because $\kappa$ decreases with $f_{\rm LS}$,
when $\sigma_{\rm TOA}$ is held fixed.
The roll-over shifts right, as $\sigma_{\rm TOA}$ decreases,
and depends on $x_n$, $\eta_f$, and $\eta_{\dot{f}}$;
there is nothing unique about $f_{\rm LS} \gtrsim 20\,{\rm Hz}$.
Figure \ref{fig:hmm11} also confirms that $P_{\rm d} \geq 0.9$
stays approximately constant across the plotted range
$10^{-15} \leq \dot{f}_{\rm LS} / (1\,{\rm Hz\,s^{-1}}) \leq 10^{-11}$,
with $P_{\rm fa}\approx 10^{-2}$.
Trials indicate that, for certain parameter combinations,
$\sigma_{\rm TOA}$ is effectively underestimated when
interpreted according to (\ref{eq:hmm8}),
leading to high $K_1(k)$ values and false alarms.
As a precaution, we correct this behavior by taking $\sigma_{\rm TOA}$
to be $\approx 5$ times the fiducial {\sc tempo2} value.
The correction factor is set empirically;
it cannot be predicted analytically at present.
It is conservative, 
as it reduces $P_{\rm d}$ marginally (by $\lesssim 10\%$)
while nullifying the spike in $P_{\rm fa}$.

\subsection{Glitch parameters
 \label{sec:hmm6d}}
\begin{figure}
\begin{center}
\includegraphics[width=14cm,angle=0]{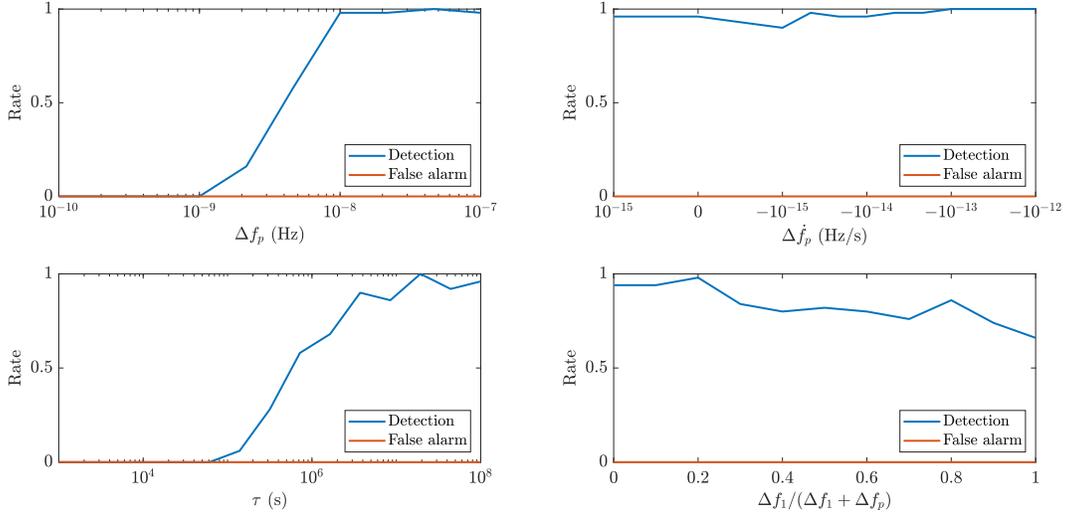}
\end{center}
\caption{
Detection probability (blue curve) and false alarm probability (red curve)
versus glitch parameters:
$\Delta f_{\rm p}$ (in ${\rm Hz}$) (top left panel),
$\Delta \dot{f}_{\rm p}$ (in ${\rm Hz\,s^{-1}}$) (top right panel),
$\tau$ (in ${\rm s}$) (bottom left panel),
and recovery fraction $\Delta f_1 / (\Delta f_{\rm p} + \Delta f_1)$
(bottom right panel).
Glitch parameters:
$\Delta f_1 = \Delta f_{\rm p} = 5\times 10^{-9}\,{\rm Hz}$
(bottom left panel);
$\Delta f_1 + \Delta f_{\rm p} = 1\times 10^{-8}\,{\rm Hz}$,
$\tau=1\times 10^5\,{\rm s}$
(bottom right panel).
Other parameters:
see penultimate (typical) column in Table \ref{tab:hmm1}.
}
\label{fig:hmm12}
\end{figure}

The size of the smallest glitch detectable by the HMM 
is governed chiefly by the user-selected probabilities
$P_{\rm fa}$ and $P_{\rm d}$.
In general, the permanent jump $\Delta f_{\rm p}$ is partially covariant 
with other glitch parameters (e.g.\ $\Delta\dot{f}_{\rm p}$,
$\Delta f_1$, and $\tau$)
as well as the non-glitch parameters 
discussed in \S\ref{sec:hmm6a}--\S\ref{sec:hmm6c}.
However, we find that $\Delta f_{\rm p}$ affects $P_{\rm d}$
more strongly than the other parameters.
Figure \ref{fig:hmm12} illustrates this behavior.
The top left panel shows that $P_{\rm d}$ rises steeply 
to $P_{\rm d} \geq 0.9$
for $\Delta f_{\rm p} \geq 8\times 10^{-9}\,{\rm Hz}$
and the parameters in the penultimate column of Table \ref{tab:hmm1}.
The top right panel shows that $P_{\rm d}$ is roughly independent
of $\Delta \dot{f}_{\rm p}$ in the range
$-10^{-12} \leq \Delta \dot{f}_{\rm p} / (1\,{\rm Hz\,s^{-1}}) \leq 10^{-15}$,
where $\Delta \dot{f}_{\rm p}$ values of both signs are tested.

The bottom panels in Figure \ref{fig:hmm12} illustrate how the HMM's
performance depends on the form and duration of the glitch recovery.
In the bottom left panel,
we observe that $P_{\rm d}$ rises to $P_{\rm d}\geq 0.5$
for $\tau \geq 2\times 10^6 \, {\rm s}$,
i.e.\ a glitch with a slower recovery is easier to detect.
The plotted example involves a substantial transient component
$\Delta f_1 = \Delta f_{\rm p}$,
which explains why $P_{\rm d}$ depends on $\tau$.
The phase deviation produced by $\Delta f_1$ relative to the glitchless model 
builds up during the recovery and asymptotes to a
constant value $\approx \tau \Delta f_1$,
unlike the permanent component, 
whose phase deviation grows indefinitely as $\approx t \Delta f_{\rm p}$.
The bottom right panel graphs $P_{\rm d}$ as a function of
the transient fraction, 
$\Delta f_1/(\Delta f_1 + \Delta f_{\rm p})$,
holding $\Delta f_1 + \Delta f_{\rm p}$ and $\tau$ fixed.
We find $P_{\rm d} \leq 0.5$ for 
$\Delta f_1 \gtrsim 0.6 (\Delta f_1 + \Delta f_{\rm p})$.
That is, when the permanent fraction drops below some value,
which depends on $\tau$,
the glitch ceases to be detectable,
if the transient component cannot be detected in its own right,
i.e.\ if $\tau \Delta f_1$ is too low.
Conversely, if $\Delta f_{\rm p}$ is large enough,
the phase deviation crosses the detection threshold eventually,
irrespective of $\Delta f_1$ and $\tau$.

\section{Representative worked example: PSR J0835$-$4510
 \label{sec:hmm7}}
The tests in this method paper are restricted deliberately to synthetic data,
in order to quantify the performance of the HMM under controlled conditions.
We look forward to applying the HMM to real, astrophysical data 
in the near future.
As a foretaste, we analyse a publicly available subset of 
490 TOAs from the regularly timed object PSR J0835$-$4510
from MJD 57427 to MJD 57810
\citep{sar17a,sar17b}.
The data are preprocessed to cull the closest spaced TOAs
(with $x_n \leq 8.9\times 10^4\,{\rm s}$ for definiteness),
as these TOA clusters exhibit excess white noise in {\sc tempo2}.
Results are presented below for the preprocessed data,
comprising $N_T = 212$ TOAs,
after checking for consistency against the 490 original TOAs.

\subsection{2016 December 12 glitch
 \label{sec:hmm7a}}
\begin{figure}
\begin{center}
\includegraphics[width=14cm,angle=0]{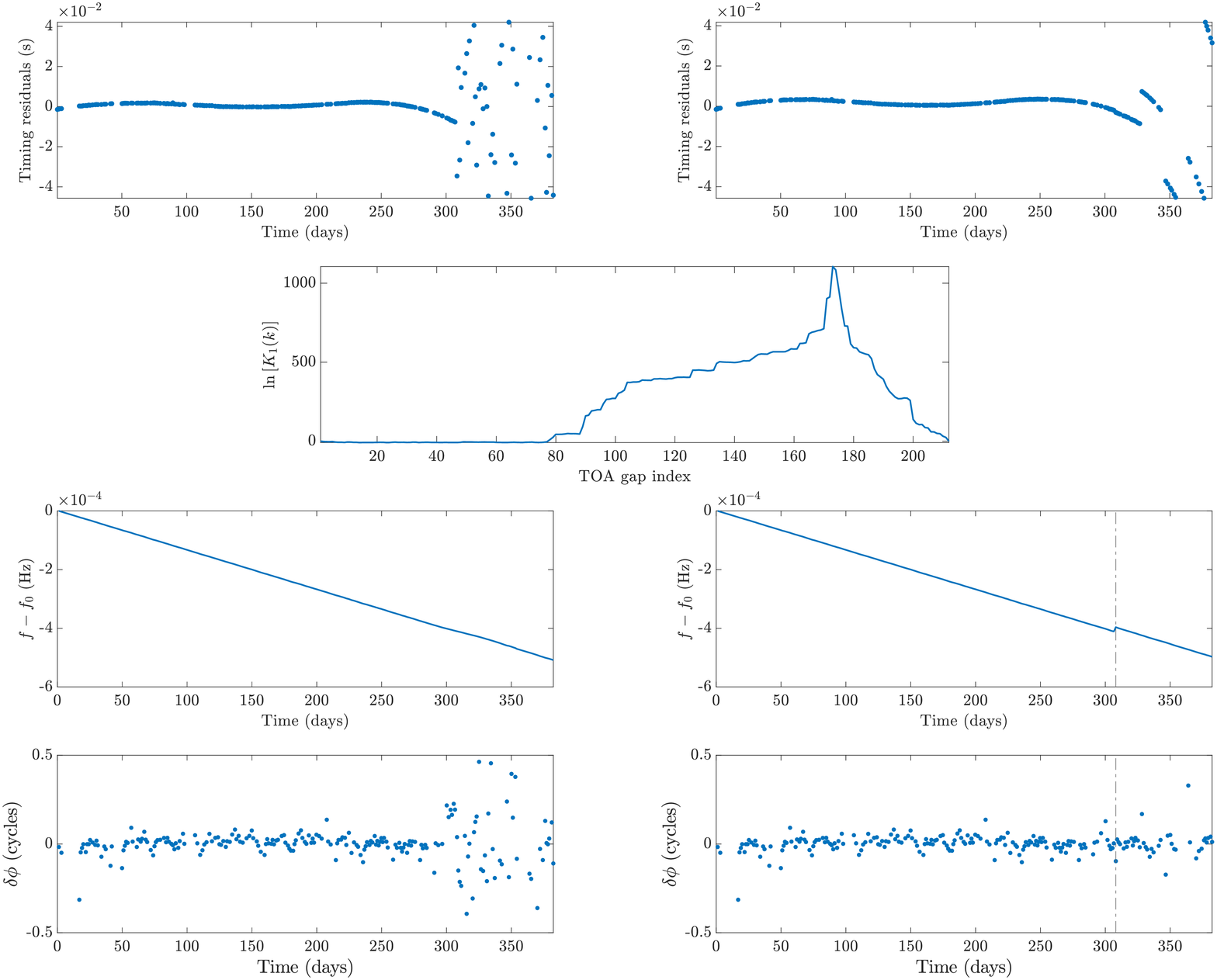}
\end{center}
\caption{
HMM analysis of $212$ TOAs measured for PSR J0835$-$4510
from MJD 57427 to MJD 57810
\citep{sar17a,sar17b}.
({\em First row.})
Phase residuals $\delta\phi(t_n)$ versus TOA $t_n$ computed with 
{\sc tempo2} for no-glitch (left panel)
and one-glitch (right panel) models,
with
$T = 57734.54\,{\rm MJD}$,
$\Delta f_{\rm p} = 1.596\times 10^{-5}\,{\rm Hz}$,
$\Delta \dot{f}_{\rm p} = -4.4\times 10^{-13}\,{\rm Hz\,s^{-1}}$
in the one-glitch model.
({\em Second row.})
Logarithm of the Bayes factor, $K_1(k)$, versus TOA index, $k$,
computed with the HMM using
$\sigma=5\times 10^{-16}\,{\rm Hz\,s^{-3/2}}$
and the DOI
$f_{\rm LS} = 11.1868550196 \,{\rm Hz}$,
$\dot{f}_{\rm LS} = 1.55886\times 10^{-11} \,{\rm Hz\,s^{-1}}$,
$-5.5\times 10^{-4} \leq (f-f_{\rm LS})/(1\,{\rm Hz})
 \leq 1\times 10^{-5}$,
$-2\times 10^{-12} \leq (\dot{f}-\dot{f}_{\rm LS})/(1\,{\rm Hz\,s^{-1}})
 \leq 2\times 10^{-12}$,
$\eta_f = 5.606\times 10^{-7}\,{\rm Hz}$ ($10^3$ bins)
and
$\eta_{\dot{f}} = 4\times 10^{-14}\,{\rm Hz\,s^{-1}}$ (101 bins).
({\em Third row.})
Recovered frequency $f(t_n)$ versus TOA $t_n$ for the no-glitch
(left panel) and one-glitch (right panel) HMM models $M_0$ and $M_1$.
The vertical dashed line marks the glitch.
({\em Fourth row.})
Unsummed per-gap phase residuals $\delta\phi(t_n)$
versus TOA $t_n$ for the HMM forward-backward sequences
in the third row.
}
\label{fig:hmm20}
\end{figure}

The results of applying the HMM to PSR J0835$-$4510 are presented 
in Figure \ref{fig:hmm20}.
The first row displays the phase residuals arising from 
traditional {\sc tempo2} fits to the data.
In the left panel, where the ephemeris does not incorporate a glitch,
the phase wraps violently beyond the glitch epoch.
In the right panel, where the ephemeris does incorporate a glitch,
the phase wraps more slowly,
because we do not correct for the quasiexponential post-glitch recovery
in this panel.
[The correction is performed by \citet{sar17a} and \citet{sar17b}.]
The second row of the figure displays the logarithm of the Bayes factor,
$K_1(k) = \Pr[O_{1:N_T} | M_1(k)] / \Pr(O_{1:N_T} | M_0)$,
for $1\leq k \leq N_T$
and $\sigma = 5\times 10^{-16}\,{\rm Hz\,s^{-3/2}}$.
The value of $\sigma$ is
estimated from the {\sc tempo2} residuals and the gridding bound
$\sigma \geq 1.3\times 10^{-16}$ 
[dominated by $\eta_{\dot{f}}$ in (\ref{eq:hmm8a})] 
plus a conservative safety factor.
The one-glitch model $M_1(173)$ is preferred strongly over $M_0$
and all $M_1(k)$ with $k\neq 173$ 
($k \neq 383$ before preprocessing).
The HMM glitch epoch, $T=57734.54\,{\rm MJD}$, 
approaches that obtained by traditional methods,
which yield $T=57734.4855(4)\,{\rm MJD}$
\citep{pal16b,sar17b,ash19}.
The maximum Bayes factor is huge, with 
$\ln K_1(173) = 1.1\times 10^3$,
a testament to the discriminating power of the HMM.
The third row displays $\hat{f}(t)$ versus $t$,
inferred using the HMM forward-backward algorithm,
for $M_0$ (left panel) and $M_1(173)$ (right panel).
The frequency step in the right panel is clearly visible.
The fourth row displays the associated phase residuals,
which wrap violently for $M_0$ at $t>T$
while remaining roughly constant for $M_1(173)$.
The results confirm,
that $M_1(173)$ offers a good description of the 2016 December 12 event
despite modeling it as a step for simplicity,
without a post-glitch recovery.

\begin{figure}
\begin{center}
\includegraphics[width=14cm,angle=0]{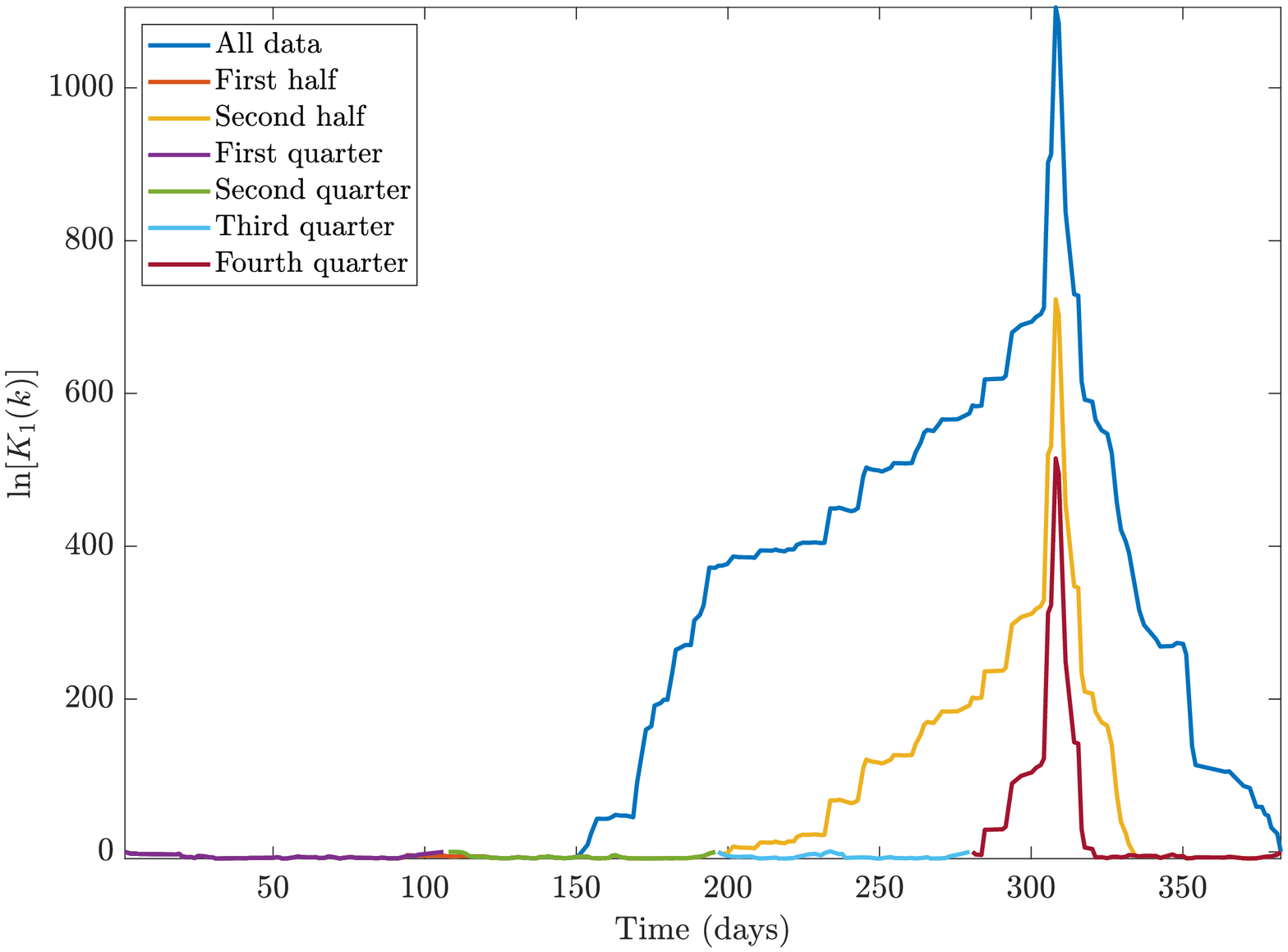}
\end{center}
\caption{
Logarithm of the Bayes factor, $K_1(k)$, versus TOA index, $k$,
for PSR J0835$-$4510 from MJD 57427 to MJD 57810,
computed with the same HMM parameters as in Figure \ref{fig:hmm20},
but with the data segmented into halves
(red and yellow curves)
and quarters
(purple, green, light blue, and brown curves).
The blue curve, incorporating all the data,
is copied from Figure \ref{fig:hmm20} for comparison.
}
\label{fig:hmm21}
\end{figure}

In order to check the robustness of the conclusion, 
that $M_1(173)$ is the preferred model,
we subdivide the data set into halves and quarters
and plot $K_1(k)$ versus $k$ in Figure \ref{fig:hmm21}.
The results for each subdivision are color-coded according to the legend.
No matter how the data are subdivided, the conclusion is the same:
$M_1(173)$ is strongly preferred over $M_0$ and $M_1(k)$ with $k\neq 173$ 
in the data segments that include $t_{173}$,
and $M_1(173)$ is not rivaled by a better alternative
in the data segments that do not include $t_{173}$.

\begin{figure}
\begin{center}
\includegraphics[width=10cm,angle=0]{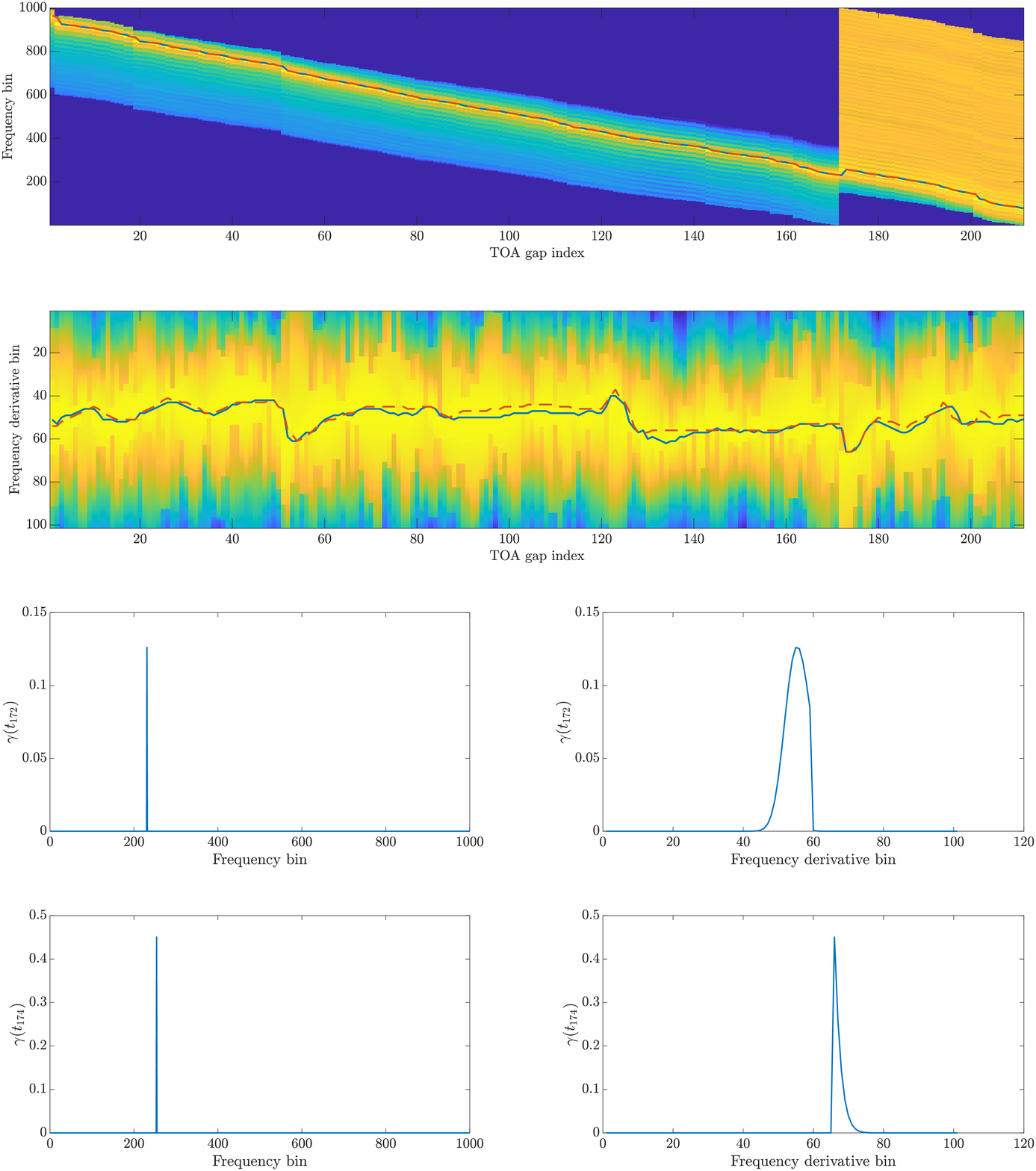}
\end{center}
\caption{
Evolution of the posterior PDF $\gamma_{q_i}(t_n)$
defined by (\ref{eq:hmmappa12}),
for PSR J0835$-$4510 from MJD 57427 to MJD 57810,
in the vicinity of the optimal values 
$\hat{f}(t_n)$ and $\hat{\dot{f}}(t_n)$.
({\em First row.})
Contours of $\gamma_{q_i}(t_n)$ marginalized over $\dot{f}$
(arbitrary color scale; yellow high, blue low)
versus TOA index $n$, with the point-wise optimal
(forward-backward) state sequence overplotted (red curve).
({\em Second row.})
Contours of $\gamma_{q_i}(t_n)$ marginalized over $f$
versus TOA index $n$.
({\em Third row.})
Cross-section of $\gamma_{q_i}(t_n)$ marginalized 
over $\dot{f}$ (left column) and $f$ (right column)
at $n=172$, i.e.\ at the TOA preceding the recovered glitch.
The horizontal axes display numbers of bins.
({\em Fourth row.})
Cross-section of $\gamma_{q_i}(t_n)$ marginalized 
over $\dot{f}$ (left column) and $f$ (right column)
at $n=174$, i.e.\ at the TOA following the recovered glitch.
Parameters: see Figure \ref{fig:hmm20}.
}
\label{fig:hmm22}
\end{figure}

How does the recovered ephemeris compare with the traditional timing solution,
now that the glitch is detected?
Figure \ref{fig:hmm22} presents the evolution of the posterior PDF
$\gamma_{q_i}(t_n)$ before and after the glitch.
The first and second rows display contours of $\gamma_{q_i}(t_n)$
marginalized over $\dot{f}$ and $f$ respectively,
graphed as functions of $t_n$,
together with the point-wise optimal sequences
$\hat{f}(t_n)$ and $\hat{\dot{f}}(t_n)$ respectively.
Both marginalized posteriors are strongly and singly peaked 
around the optimal sequences.
The jump in $\hat{f}(t_n)$ is visible in the top row.
The third and fourth rows display orthogonal cross-sections 
taken through the posterior PDF just before 
($t_{172}$; third row)
and after
($t_{174}$; fourth row)
the glitch.
In the left column, 
where $\gamma_{q_i}(t_n)$ is marginalized over $\dot{f}$,
there is an upward shift in frequency,
with
$\Delta f_{\rm p} = 
 \hat{f}(t_{174}) - \hat{f}(t_{172}) = 1.596\times 10^{-5}\,{\rm Hz}$.
The cross-sections are narrow,
spanning $\lesssim 4$ bins before and after the glitch.
In the right column, 
where $\gamma_{q_i}(t_n)$ is marginalized over $f$,
there is a downward shift in frequency derivative,
with
$\Delta \dot{f}_{\rm p} = 
 \hat{\dot{f}}(t_{174}) - \hat{\dot{f}}(t_{172}) 
 = -4.4\times 10^{-13}\,{\rm Hz\,s^{-1}}$.
The shift is significant in the sense that it exceeds the dispersion,
which actually decreases during the event
(FWHM $\lesssim 8$ bins before, cf.\ $\lesssim 2$ bins after).
The inferred jumps agree with traditional
pulsar timing methods, which give
$\Delta f_{\rm p} = 1.6044(2) \times 10^{-5}\,{\rm Hz}$
and
$\Delta \dot{f}_{\rm p} = -1.21(3)\times 10^{-13} \,{\rm Hz\,s^{-1}}$
\citep{pal16b,sar17b},
after allowing for the fact that the HMM transition probabilities 
do not include the post-glitch relaxation with $\tau = 0.96(17)\,{\rm d}$.
(Including the relaxation is straightforward but lies outside the scope of
this paper.)
All in all, the optimal sequence stands out clearly 
above its nearest competitors.

\subsection{Additional glitches
 \label{sec:hmm7b}}
A systematic search for multiple glitches lies outside the scope
of this paper. Nonetheless, again as a foretaste of what is feasible, 
we search for a second glitch in PSR J0835$-$4510 from MJD 57427 to MJD 57810
by applying the greedy hierarchical algorithm introduced in \S\ref{sec:hmm3aa}
\citep{suv18}.
The analysis is presented in Appendix \ref{sec:hmmapph}.
We conclude that no statistically significant second event exists,
in accord with previous analyses
\citep{sar17a,sar17b}.

We look forward to applying the HMM to more real data sets.
In particular, a fuller search for glitches in PSR J0835$-$4510
over several decades of continuous monitoring
using the greedy hierarchical algorithm in \S\ref{sec:hmm3aa}
will be undertaken in future work;
the relevant data are not currently at our disposal.
If additional events are found,
they can be cross-checked in many ways.
One can apply the HMM to data taken with a different telecope,
e.g.\ the higher cadence Mount Pleasant Radio Observatory
for PSR J0835$-$4510
\citep{pal18},
just as when checking the output of traditional timing methods.
The Mount Pleasant data were analysed recently by Bayesian methods
to study the pulse-to-pulse dynamics of the 2016 December 12 glitch
\citep{ash19}.
One can also test, how the Bayes factor changes, 
as one tunes HMM parameters like $\kappa$, $\sigma$, and $P_{\rm fa}$;
see \S\ref{sec:hmm6a} for details.
It is faster to do such tests systematically with the HMM
than with traditional timing methods.

\section{Conclusion
 \label{sec:hmm8}}
In this method paper, a new, systematic scheme is presented for detecting 
pulsar glitches given a sequence of standard TOAs.
The scheme is structured around a HMM,
which tracks the evolution of the pulse frequency and its first time derivative
on long (electromagnetic spin down), intermediate (timing noise),
and short (glitches) time-scales.
The emission probability of the HMM obeys a von Mises distribution.
The transition probability obeys a Gaussian distribution derived from
the Fokker-Planck equation for an unbiased Wiener process.
The HMM forward algorithm is used to compute and compare the Bayesian evidence 
for models with and without glitches.
Once the preferred model is selected,
the HMM forward-backward algorithm is used to compute the associated,
point-wise optimal ephemeris, 
composed of the most probable hidden state $\hat{q}(t_n)$ at each $t_n$
given all the observations $O_{1:N_T}$.
The algorithm and testing procedure
are documented in Appendices \ref{sec:hmmappa}--\ref{sec:hmmapph}
for the sake of reproducibility.

Monte Carlo simulations demonstrate that the HMM detects glitches accurately
in synthetic data for a range of realistic intrinsic and measurement noises
($\sigma_{\rm TN}$, $\sigma_{\rm TOA}$;
see \S\ref{sec:hmm6a}),
secular spin-down parameters
($f_{\rm LS}$, $\dot{f}_{\rm LS}$;
see \S\ref{sec:hmm6c}),
glitch parameters
($\Delta f_{\rm p}$, $\Delta \dot{f}_{\rm p}$, $\Delta f_1$, $\tau$;
see \S\ref{sec:hmm6d}),
and observational schedules
($\langle x_n \rangle$, $N_T$;
see Appendix \ref{sec:hmmappg}).
The performance of the HMM,
in particular the trade off between $P_{\rm fa}$ and $P_{\rm d}$,
is quantified systematically in terms of ROC curves constructed
as functions of the above parameters.
Success is achieved, even though 
(i) the HMM approximates glitches as instantaneous steps in $f$ and $\dot{f}$ 
without any post-glitch recovery,
and (ii) the HMM models timing noise as white noise in the torque derivative
(and hence red noise in the filtered torque),
an approximation which applies to some but not all pulsars
\citep{cor80,cor85}
and is violated deliberately when generating the synthetic data in this paper
in order to challenge the robustness of the HMM.
Several trends of practical utility are identified.
(i) The HMM performs stably,
neither overestimating nor underestimating the number of glitches,
for $\sigma \approx \langle x_n \rangle^{-1} \sigma_{\rm TN}$,
with $\sigma_{\rm TN}$ computed from the {\sc tempo2} phase residuals,
as described in \S\ref{sec:hmm6a}.
(ii) In order to detect a glitch of size $\Delta f_{\rm p}$,
it is recommended to schedule observations with 
$\langle x_n \rangle^{1/2} \lesssim \sigma_{\rm TN}^{-1} \Delta f_{\rm p}$,
independent of the number of TOAs per continuous observing session.
Roughly equal spacing is preferable,
as false alarms occur more commonly adjacent to longer TOA gaps.
(iii) Performance is essentially unaffected by $\dot{f}_{\rm LS}$
and depends roughly on the product $\sigma_{\rm TOA} f_{\rm LS}$.
(iv) The size of the smallest detectable glitch is governed 
mainly by $\Delta f_{\rm p}$ and depends weakly on $\tau$,
when the phase deviation produced by $\Delta f_{\rm p}$ exceeds
that produced by the transient ($\approx \tau \Delta f_1$).
(v) Recipes for setting the DOI and grid resolution
are set out in Appendix \ref{sec:hmmappc}.

The performance tests in this paper are restricted deliberately 
to synthetic data in order to establish performance bounds 
systematically under controlled conditions.
Nevertheless, as a foretaste of what can be achieved with astronomical data,
we also apply the HMM to 490 publicly available TOAs 
from PSR J0835$-$4510,
covering the interval from MJD 57427 to MJD 57810
\citep{sar17a,sar17b}.
We confirm the existence of the large glitch on 2016 December 12,
with log Bayes factor $\approx 1.1\times 10^3$,
and rule out with high statistical confidence the existence 
of a second glitch during the same interval.
The inferred ephemeris,
including $\Delta f_{\rm p}$ and $\Delta \dot{f}_{\rm p}$,
agrees with that yielded by traditional timing methods,
after allowing for the fact that the introductory HMM 
in this paper does not include post-glitch recoveries.
We look forward to applying the HMM to other pulsars,
both to detect glitches and to improve the
sensitivity of nanohertz gravitational wave searches
with pulsar timing arrays
\citep{len15c,sha15,arz16,hob17}.

In closing, we reaffirm that the HMM scheme developed in this paper 
complements --- but does not replace --- traditional glitch finding approaches 
based on least-squares fitting 
of a Taylor-expanded phase model plus glitch template.
Indeed, the HMM ingests standard TOAs
and leverages the outputs of existing software
[e.g.\ $f_{\rm LS}$, $\dot{f}_{\rm LS}$, and phase residuals
$\delta\phi(t_n)$ from {\sc tempo2}]
to demarcate its state space (DOI).
It complements existing Bayesian approaches,
e.g.\ {\sc temponest} \citep{len14,sha16,low18},
by tracking the observed spin wandering explicitly,
as a specific realization of a discrete-time Markov chain,
instead of estimating its ensemble statistics 
(e.g.\ power spectral density).
Every approach has advantages and disadvantages.
The HMM is unsupervised,
so its performance bounds (e.g.\ $P_{\rm fa}$, $P_{\rm d}$)
can be computed efficiently.
It is fast,
requiring $\sim 10^{12}$ floating point operations ($\sim 0.1$ CPU hours)
per pulsar per year of observations.
It discriminates accurately between spin wandering and glitches 
by tracking both phenomena explicitly with a Markov chain.
On the other hand,
when spin wandering and glitches are negligible, 
the HMM is superfluous.
Pulse domain methods ultimately promise the best sensitivity
but they expend a lot of computational effort 
correcting for random pulse-to-pulse profile variations
and do not ingest standard TOAs.
They may be strongest when combined with an HMM
similar to the one described here.
If the problem allows, it is wise to apply several methods simultaneously.
There is no purely objective answer to the question of whether or not
a data set contains a glitch.
The question is fundamentally statistical and can only be answered
in the context of a user-selected false alarm probability.
The results in this paper show concretely and systematically
how to define, compute, and set $P_{\rm fa}$ for the HMM.

\acknowledgments
The authors thank Stefan Oslowski and Marcus Lower
for pointing out important references
and for providing access to data from the
Molonglo Synthesis Radio Telescope 
for experimention while developing the  HMM algorithm. 
The PSR J0835$-$4510 data analysed in \S\ref{sec:hmm7}
are described by \citet{sar17a} and \citet{sar17b}.
This research was supported by the Australian Research Council
Centre of Excellence for Gravitational Wave Discovery (OzGrav),
grant number CE170100004.

\bibliographystyle{mn2e}
\bibliography{glitchstat}

\begin{thebibliography}{}

\bibitem[\protect\citeauthoryear{{Abbott}, {Abbott}, {Abbott}, {Abraham},
  {Acernese}, {Ackley}, {Adams}, {Adhikari}, {Adya}, {Affeldt} \& et
  al.}{{Abbott} et~al.}{2019}]{abb19}
{Abbott} B.~P.,  {Abbott} R.,  {Abbott} T.~D.,  {Abraham} S.,  {Acernese} F.,
  {Ackley} K.,  {Adams} C.,  {Adhikari} R.~X.,  {Adya} V.~B.,  {Affeldt} C.,
  et al. 2019, \prd, 100, 122002

\bibitem[\protect\citeauthoryear{{Abbott}, {Abbott}, {Abbott}, {Acernese},
  {Ackley}, {Adams}, {Adams}, {Addesso}, {Adhikari}, {Adya} \& et al.}{{Abbott}
  et~al.}{2017}]{abb17}
{Abbott} B.~P.,  {Abbott} R.,  {Abbott} T.~D.,  {Acernese} F.,  {Ackley} K.,
  {Adams} C.,  {Adams} T.,  {Addesso} P.,  {Adhikari} R.~X.,  {Adya} V.~B.,
  et al. 2017, \prd, 95, 122003

\bibitem[\protect\citeauthoryear{{Anderson} \& {Itoh}}{{Anderson} \&
  {Itoh}}{1975}]{and75}
{Anderson} P.~W.,  {Itoh} N.,  1975, \nat, 256, 25

\bibitem[\protect\citeauthoryear{{Archibald}, {Gotthelf}, {Ferdman}, {Kaspi},
  {Guillot}, {Harrison}, {Keane}, {Pivovaroff}, {Stern}, {Tendulkar} \&
  {Tomsick}}{{Archibald} et~al.}{2016}]{arc16}
{Archibald} R.~F.,  {Gotthelf} E.~V.,  {Ferdman} R.~D.,  {Kaspi} V.~M.,
  {Guillot} S.,  {Harrison} F.~A.,  {Keane} E.~F.,  {Pivovaroff} M.~J.,
  {Stern} D.,  {Tendulkar} S.~P.,    {Tomsick} J.~A.,  2016, \apjl, 819, L16

\bibitem[\protect\citeauthoryear{{Arzoumanian}, , {Brazier}, {Burke-Spolaor},
  {Chamberlin}, {Chatterjee}, {Christy}, {Cordes}, {Cornish}, {Crowter},
  {Demorest} \& {et al.}}{{Arzoumanian} et~al.}{2016}]{arz16}
{Arzoumanian} Z.,   {Brazier} A.,  {Burke-Spolaor} S.,  {Chamberlin} S.~J.,
  {Chatterjee} S.,  {Christy} B.,  {Cordes} J.~M.,  {Cornish} N.~J.,  {Crowter}
  K.,  {Demorest} P.~B.,    {et al.} 2016, \apj, 821, 13

\bibitem[\protect\citeauthoryear{{Arzoumanian}, {Nice}, {Taylor} \&
  {Thorsett}}{{Arzoumanian} et~al.}{1994}]{arz94}
{Arzoumanian} Z.,  {Nice} D.~J.,  {Taylor} J.~H.,    {Thorsett} S.~E.,  1994,
  \apj, 422, 671

\bibitem[\protect\citeauthoryear{{Ashton}, {Lasky}, {Graber} \&
  {Palfreyman}}{{Ashton} et~al.}{2019}]{ash19}
{Ashton} G.,  {Lasky} P.~D.,  {Graber} V.,    {Palfreyman} J.,  2019, Nature
  Astronomy, 3, 1143

\bibitem[\protect\citeauthoryear{{Ashton}, {Prix} \& {Jones}}{{Ashton}
  et~al.}{2017}]{ash17}
{Ashton} G.,  {Prix} R.,    {Jones} D.~I.,  2017, \prd, 96, 063004

\bibitem[\protect\citeauthoryear{{Bayley}, {Messenger} \& {Woan}}{{Bayley}
  et~al.}{2019}]{bay19}
{Bayley} J.,  {Messenger} C.,    {Woan} G.,  2019, \prd, 100, 023006

\bibitem[\protect\citeauthoryear{{Baym}, {Pethick}, {Pines} \&
  {Ruderman}}{{Baym} et~al.}{1969}]{bay69}
{Baym} G.,  {Pethick} C.,  {Pines} D.,    {Ruderman} M.,  1969, \nat, 224, 872

\bibitem[\protect\citeauthoryear{{Calafiore} \& {El Ghaoui}}{{Calafiore} \& {El
  Ghaoui}}{2014}]{cal14}
{Calafiore} G.~C.,  {El Ghaoui} L.,  2014, {Optimization Models}.
Cambridge: Cambridge University Press

\bibitem[\protect\citeauthoryear{{Carlin} \& {Melatos}}{{Carlin} \&
  {Melatos}}{2019}]{car19b}
{Carlin} J.~B.,  {Melatos} A.,  2019, \mnras, 483, 4742

\bibitem[\protect\citeauthoryear{{Carlin}, {Melatos} \& {Vukcevic}}{{Carlin}
  et~al.}{2019}]{car19}
{Carlin} J.~B.,  {Melatos} A.,    {Vukcevic} D.,  2019, \mnras, 482, 3736

\bibitem[\protect\citeauthoryear{{Chugunov} \& {Horowitz}}{{Chugunov} \&
  {Horowitz}}{2010}]{chu10b}
{Chugunov} A.~I.,  {Horowitz} C.~J.,  2010, \mnras, 407, L54

\bibitem[\protect\citeauthoryear{{Coles}, {Hobbs}, {Champion}, {Manchester} \&
  {Verbiest}}{{Coles} et~al.}{2011}]{col11}
{Coles} W.,  {Hobbs} G.,  {Champion} D.~J.,  {Manchester} R.~N.,    {Verbiest}
  J.~P.~W.,  2011, \mnras, 418, 561

\bibitem[\protect\citeauthoryear{{Cordes}}{{Cordes}}{1980}]{cor80}
{Cordes} J.~M.,  1980, \apj, 237, 216

\bibitem[\protect\citeauthoryear{{Cordes} \& {Downs}}{{Cordes} \&
  {Downs}}{1985}]{cor85}
{Cordes} J.~M.,  {Downs} G.~S.,  1985, \apjs, 59, 343

\bibitem[\protect\citeauthoryear{{Cordes} \& {Helfand}}{{Cordes} \&
  {Helfand}}{1980}]{cor80b}
{Cordes} J.~M.,  {Helfand} D.~J.,  1980, \apj, 239, 640

\bibitem[\protect\citeauthoryear{{D'Alessandro}, {McCulloch}, {Hamilton} \&
  {Deshpande}}{{D'Alessandro} et~al.}{1995}]{dal95}
{D'Alessandro} F.,  {McCulloch} P.~M.,  {Hamilton} P.~A.,    {Deshpande} A.~A.,
   1995, \mnras, 277, 1033

\bibitem[\protect\citeauthoryear{{Dunn}, {Clearwater}, {Melatos} \&
  {Wette}}{{Dunn} et~al.}{2020}]{dun20}
{Dunn} L.,  {Clearwater} P.,  {Melatos} A.,    {Wette} K.,  2020, Classical and
  Quantum Gravity, p. submitted

\bibitem[\protect\citeauthoryear{{Edwards}, {Hobbs} \& {Manchester}}{{Edwards}
  et~al.}{2006}]{edw06}
{Edwards} R.~T.,  {Hobbs} G.~B.,    {Manchester} R.~N.,  2006, \mnras, 372,
  1549

\bibitem[\protect\citeauthoryear{{Espinoza}, {Antonopoulou}, {Stappers},
  {Watts} \& {Lyne}}{{Espinoza} et~al.}{2014}]{esp14}
{Espinoza} C.~M.,  {Antonopoulou} D.,  {Stappers} B.~W.,  {Watts} A.,    {Lyne}
  A.~G.,  2014, \mnras, 440, 2755

\bibitem[\protect\citeauthoryear{{Espinoza}, {Lyne}, {Stappers} \&
  {Kramer}}{{Espinoza} et~al.}{2011}]{esp11}
{Espinoza} C.~M.,  {Lyne} A.~G.,  {Stappers} B.~W.,    {Kramer} M.,  2011,
  \mnras, 414, 1679

\bibitem[\protect\citeauthoryear{{Faucher-Gigu{\`e}re} \&
  {Kaspi}}{{Faucher-Gigu{\`e}re} \& {Kaspi}}{2006}]{fau06}
{Faucher-Gigu{\`e}re} C.-A.,  {Kaspi} V.~M.,  2006, \apj, 643, 332

\bibitem[\protect\citeauthoryear{{Fuentes}, {Espinoza} \&
  {Reisenegger}}{{Fuentes} et~al.}{2019}]{fue19}
{Fuentes} J.~R.,  {Espinoza} C.~M.,    {Reisenegger} A.,  2019, \aap, 630, A115

\bibitem[\protect\citeauthoryear{{Fulgenzi}, {Melatos} \& {Hughes}}{{Fulgenzi}
  et~al.}{2017}]{ful17}
{Fulgenzi} W.,  {Melatos} A.,    {Hughes} B.~D.,  2017, \mnras, 470, 4307

\bibitem[\protect\citeauthoryear{{Gardiner}}{{Gardiner}}{1994}]{gar94}
{Gardiner} C.~W.,  1994, {Handbook of stochastic methods for physics, chemistry
  and the natural sciences}

\bibitem[\protect\citeauthoryear{{Glampedakis} \& {Andersson}}{{Glampedakis} \&
  {Andersson}}{2009}]{gla09}
{Glampedakis} K.,  {Andersson} N.,  2009, \prl, 102, 141101

\bibitem[\protect\citeauthoryear{{Goncharov}, {Zhu} \& {Thrane}}{{Goncharov}
  et~al.}{2019}]{gon19}
{Goncharov} B.,  {Zhu} X.-J.,    {Thrane} E.,  2019, arXiv e-prints, p.
  arXiv:1910.05961

\bibitem[\protect\citeauthoryear{{Haskell} \& {Melatos}}{{Haskell} \&
  {Melatos}}{2015}]{has15}
{Haskell} B.,  {Melatos} A.,  2015, International Journal of Modern Physics D,
  24, 1530008

\bibitem[\protect\citeauthoryear{{Helfand}, {Manchester} \& {Taylor}}{{Helfand}
  et~al.}{1975}]{hel75}
{Helfand} D.~J.,  {Manchester} R.~N.,    {Taylor} J.~H.,  1975, \apj, 198, 661

\bibitem[\protect\citeauthoryear{{Hobbs} \& {Dai}}{{Hobbs} \&
  {Dai}}{2017}]{hob17}
{Hobbs} G.,  {Dai} S.,  2017, ArXiv e-prints

\bibitem[\protect\citeauthoryear{{Hobbs}, {Edwards} \& {Manchester}}{{Hobbs}
  et~al.}{2006}]{hob06}
{Hobbs} G.,  {Edwards} R.,    {Manchester} R.,  2006, Chinese Journal of
  Astronomy and Astrophysics Supplement, 6, 189

\bibitem[\protect\citeauthoryear{{Hobbs}, {Hollow}, {Champion}, {Khoo} \& {et
  al.}}{{Hobbs} et~al.}{2009}]{hob09}
{Hobbs} G.,  {Hollow} R.,  {Champion} D.,  {Khoo} J.,    {et al.} 2009, \pasa,
  26, 468

\bibitem[\protect\citeauthoryear{{Hobbs}, {Lyne}, {Kramer}, {Martin} \&
  {Jordan}}{{Hobbs} et~al.}{2004}]{hob04}
{Hobbs} G.,  {Lyne} A.~G.,  {Kramer} M.,  {Martin} C.~E.,    {Jordan} C.,
  2004, \mnras, 353, 1311

\bibitem[\protect\citeauthoryear{{Howitt}, {Melatos} \& {Delaigle}}{{Howitt}
  et~al.}{2018}]{how18}
{Howitt} G.,  {Melatos} A.,    {Delaigle} A.,  2018, \apj, 867, 60

\bibitem[\protect\citeauthoryear{{Jankowski}, {Bailes}, {van Straten}, {Keane}
  \& {et al.}}{{Jankowski} et~al.}{2019}]{jan19}
{Jankowski} F.,  {Bailes} M.,  {van Straten} W.,  {Keane} E.~F.,    {et al.}
  2019, \mnras, 484, 3691

\bibitem[\protect\citeauthoryear{{Janssen} \& {Stappers}}{{Janssen} \&
  {Stappers}}{2006}]{jan06}
{Janssen} G.~H.,  {Stappers} B.~W.,  2006, \aap, 457, 611

\bibitem[\protect\citeauthoryear{{Jeffreys}}{{Jeffreys}}{1998}]{jef98}
{Jeffreys} H.,  1998, {The Theory of Probability}.
Oxford: Oxford University Press (3rd ed.)

\bibitem[\protect\citeauthoryear{{Johnston} \& {Galloway}}{{Johnston} \&
  {Galloway}}{1999}]{joh99}
{Johnston} S.,  {Galloway} D.,  1999, \mnras, 306, L50

\bibitem[\protect\citeauthoryear{{Jones}}{{Jones}}{1990}]{jon90}
{Jones} P.~B.,  1990, \mnras, 246, 364

\bibitem[\protect\citeauthoryear{{Lattimer} \& {Prakash}}{{Lattimer} \&
  {Prakash}}{2007}]{lat07}
{Lattimer} J.~M.,  {Prakash} M.,  2007, \physrep, 442, 109

\bibitem[\protect\citeauthoryear{{Leaci} \& {Prix}}{{Leaci} \&
  {Prix}}{2015}]{lea15}
{Leaci} P.,  {Prix} R.,  2015, \prd, 91, 102003

\bibitem[\protect\citeauthoryear{{Lentati}, {Alexander} \& {Hobson}}{{Lentati}
  et~al.}{2015}]{len15a}
{Lentati} L.,  {Alexander} P.,    {Hobson} M.~P.,  2015, \mnras, 447, 2159

\bibitem[\protect\citeauthoryear{{Lentati}, {Alexander}, {Hobson}, {Feroz},
  {van Haasteren}, {Lee} \& {Shannon}}{{Lentati} et~al.}{2014}]{len14}
{Lentati} L.,  {Alexander} P.,  {Hobson} M.~P.,  {Feroz} F.,  {van Haasteren}
  R.,  {Lee} K.~J.,    {Shannon} R.~M.,  2014, \mnras, 437, 3004

\bibitem[\protect\citeauthoryear{{Lentati}, {Champion}, {Kramer}, {Barr} \&
  {Torne}}{{Lentati} et~al.}{2018}]{len18}
{Lentati} L.,  {Champion} D.~J.,  {Kramer} M.,  {Barr} E.,    {Torne} P.,
  2018, \mnras, 473, 5026

\bibitem[\protect\citeauthoryear{{Lentati} \& {et al.}}{{Lentati} \& {et
  al.}}{2017}]{len17a}
{Lentati} L.,  {et al.} 2017, \mnras, 466, 3706

\bibitem[\protect\citeauthoryear{{Lentati}, {Kerr}, {Dai}, {Shannon}, {Hobbs}
  \& {Os{\l}owski}}{{Lentati} et~al.}{2017}]{len17b}
{Lentati} L.,  {Kerr} M.,  {Dai} S.,  {Shannon} R.~M.,  {Hobbs} G.,
  {Os{\l}owski} S.,  2017, \mnras, 468, 1474

\bibitem[\protect\citeauthoryear{{Lentati} \& {Shannon}}{{Lentati} \&
  {Shannon}}{2015}]{len15b}
{Lentati} L.,  {Shannon} R.~M.,  2015, \mnras, 454, 1058

\bibitem[\protect\citeauthoryear{{Lentati}, {Taylor}, {Mingarelli}, {Sesana},
  {Sanidas}, {Vecchio}, {Caballero}, {Lee}, {van Haasteren}, {Babak} \& {et
  al.}}{{Lentati} et~al.}{2015}]{len15c}
{Lentati} L.,  {Taylor} S.~R.,  {Mingarelli} C.~M.~F.,  {Sesana} A.,  {Sanidas}
  S.~A.,  {Vecchio} A.,  {Caballero} R.~N.,  {Lee} K.~J.,  {van Haasteren} R.,
  {Babak} S.,    {et al.} 2015, \mnras, 453, 2576

\bibitem[\protect\citeauthoryear{{Lower}, {Bailes}, {Shannon}, {Johnston},
  {Flynn}, {Bateman}, {Campbell-Wilson}, {Day}, {Deller}, {Farah} \& {et
  al.}}{{Lower} et~al.}{2019}]{low19}
{Lower} M.~E.,  {Bailes} M.,  {Shannon} R.~M.,  {Johnston} S.,  {Flynn} C.,
  {Bateman} T.,  {Campbell-Wilson} D.,  {Day} C.~K.,  {Deller} A.,  {Farah} W.,
     {et al.} 2019, Research Notes of the American Astronomical Society, 3, 192

\bibitem[\protect\citeauthoryear{{Lower}, {Bailes}, {Shannon}, {Johnston},
  {Flynn}, {Os{\l}owski}, {Gupta}, {Farah}, {Bateman}, {Green}, {Hunstead},
  {Jameson}, {Jankowski}, {Parthasarathy}, {Price}, {Sutherland }, {Temby} \&
  {Krishnan}}{{Lower} et~al.}{2020}]{low20}
{Lower} M.~E.,  {Bailes} M.,  {Shannon} R.~M.,  {Johnston} S.,  {Flynn} C.,
  {Os{\l}owski} S.,  {Gupta} V.,  {Farah} W.,  {Bateman} T.,  {Green} A.~J.,
  {Hunstead} R.,  {Jameson} A.,  {Jankowski} F.,  {Parthasarathy} A.,  {Price}
  D.~C.,  {Sutherland } A.,  {Temby} D.,    {Krishnan} V.~V.,  2020, \mnras

\bibitem[\protect\citeauthoryear{{Lower}, {Flynn}, {Bailes}, {Barr}, {Bateman},
  {Bhandari}, {Caleb}, {Campbell-Wilson}, {Day}, {Deller} \& {et al.}}{{Lower}
  et~al.}{2018}]{low18}
{Lower} M.~E.,  {Flynn} C.,  {Bailes} M.,  {Barr} E.~D.,  {Bateman} T.,
  {Bhandari} S.,  {Caleb} M.,  {Campbell-Wilson} D.,  {Day} C.,  {Deller} A.,
   {et al.} 2018, Research Notes of the American Astronomical Society, 2, 139

\bibitem[\protect\citeauthoryear{{Lyne} \& {Graham-Smith}}{{Lyne} \&
  {Graham-Smith}}{2012}]{lyn12}
{Lyne} A.,  {Graham-Smith} F.,  2012, {Pulsar Astronomy}

\bibitem[\protect\citeauthoryear{{Lyne}, {Pritchard}, {Graham-Smith} \&
  {Camilo}}{{Lyne} et~al.}{1996}]{lyn96}
{Lyne} A.~G.,  {Pritchard} R.~S.,  {Graham-Smith} F.,    {Camilo} F.,  1996,
  \nat, 381, 497

\bibitem[\protect\citeauthoryear{{Lyne}, {Shemar} \& {Smith}}{{Lyne}
  et~al.}{2000}]{lyn00}
{Lyne} A.~G.,  {Shemar} S.~L.,    {Smith} F.~G.,  2000, \mnras, 315, 534

\bibitem[\protect\citeauthoryear{{Mardia} \& {Jupp}}{{Mardia} \&
  {Jupp}}{2009}]{mar09}
{Mardia} K.~C.,  {Jupp} P.~E.,  2009, {Directional Statistics}

\bibitem[\protect\citeauthoryear{{McCulloch}, {Klekociuk}, {Hamilton} \&
  {Royle}}{{McCulloch} et~al.}{1987}]{mcc87}
{McCulloch} P.~M.,  {Klekociuk} A.~R.,  {Hamilton} P.~A.,    {Royle} G.~W.~R.,
  1987, Australian Journal of Physics, 40, 725

\bibitem[\protect\citeauthoryear{{Melatos}}{{Melatos}}{1997}]{mel97}
{Melatos} A.,  1997, \mnras, 288, 1049

\bibitem[\protect\citeauthoryear{{Melatos}, {Howitt} \& {Fulgenzi}}{{Melatos}
  et~al.}{2018}]{mel18}
{Melatos} A.,  {Howitt} G.,    {Fulgenzi} W.,  2018, \apj, 863, 196

\bibitem[\protect\citeauthoryear{{Melatos} \& {Link}}{{Melatos} \&
  {Link}}{2014}]{mel14}
{Melatos} A.,  {Link} B.,  2014, \mnras, 437, 21

\bibitem[\protect\citeauthoryear{{Melatos} \& {Peralta}}{{Melatos} \&
  {Peralta}}{2010}]{mel10}
{Melatos} A.,  {Peralta} C.,  2010, \apj, 709, 77

\bibitem[\protect\citeauthoryear{{Melatos}, {Peralta} \& {Wyithe}}{{Melatos}
  et~al.}{2008}]{mel08}
{Melatos} A.,  {Peralta} C.,    {Wyithe} J.~S.~B.,  2008, \apj, 672, 1103

\bibitem[\protect\citeauthoryear{{Melrose}}{{Melrose}}{2017}]{mel17b}
{Melrose} D.~B.,  2017, Reviews of Modern Plasma Physics, 1, 5

\bibitem[\protect\citeauthoryear{{Michel}}{{Michel}}{1991}]{mic91}
{Michel} F.~C.,  1991, {Theory of neutron star magnetospheres}.
Chicago: University of Chicago Press

\bibitem[\protect\citeauthoryear{{Middleditch}, {Marshall}, {Wang}, {Gotthelf}
  \& {Zhang}}{{Middleditch} et~al.}{2006}]{mid06}
{Middleditch} J.,  {Marshall} F.~E.,  {Wang} Q.~D.,  {Gotthelf} E.~V.,
  {Zhang} W.,  2006, \apj, 652, 1531

\bibitem[\protect\citeauthoryear{{Namkham}, {Jaroenjittichai} \&
  {Johnston}}{{Namkham} et~al.}{2019}]{nam19}
{Namkham} N.,  {Jaroenjittichai} P.,    {Johnston} S.,  2019, \mnras, 487, 5854

\bibitem[\protect\citeauthoryear{{Onuchukwu} \& {Chukwude}}{{Onuchukwu} \&
  {Chukwude}}{2016}]{onu16}
{Onuchukwu} C.~C.,  {Chukwude} A.~E.,  2016, \apss, 361, 300

\bibitem[\protect\citeauthoryear{{Palfreyman}}{{Palfreyman}}{2016}]{pal16b}
{Palfreyman} J.,  2016, The Astronomer's Telegram, 9847

\bibitem[\protect\citeauthoryear{{Palfreyman}, {Dickey}, {Hotan}, {Ellingsen}
  \& {van Straten}}{{Palfreyman} et~al.}{2018}]{pal18}
{Palfreyman} J.,  {Dickey} J.~M.,  {Hotan} A.,  {Ellingsen} S.,    {van
  Straten} W.,  2018, \nat, 556, 219

\bibitem[\protect\citeauthoryear{{Palfreyman}, {Dickey}, {Ellingsen}, {Jones}
  \& {Hotan}}{{Palfreyman} et~al.}{2016}]{pal16}
{Palfreyman} J.~L.,  {Dickey} J.~M.,  {Ellingsen} S.~P.,  {Jones} I.~R.,
  {Hotan} A.~W.,  2016, \apj, 820, 64

\bibitem[\protect\citeauthoryear{{Parthasarathy}, {Shannon}, {Johnston},
  {Lentati}, {Bailes}, {Dai}, {Kerr}, {Manchester}, {Os{\l}owski}, {Sobey},
  {van Straten} \& {Weltevrede}}{{Parthasarathy} et~al.}{2019}]{par19}
{Parthasarathy} A.,  {Shannon} R.~M.,  {Johnston} S.,  {Lentati} L.,  {Bailes}
  M.,  {Dai} S.,  {Kerr} M.,  {Manchester} R.~N.,  {Os{\l}owski} S.,  {Sobey}
  C.,  {van Straten} W.,    {Weltevrede} P.,  2019, \mnras, 489, 3810

\bibitem[\protect\citeauthoryear{{Price}, {Link}, {Shore} \& {Nice}}{{Price}
  et~al.}{2012}]{pri12}
{Price} S.,  {Link} B.,  {Shore} S.~N.,    {Nice} D.~J.,  2012, \mnras, 426,
  2507

\bibitem[\protect\citeauthoryear{{Quinn} \& Hannan}{{Quinn} \&
  Hannan}{2001}]{qui01}
{Quinn} B.~G.,  Hannan E.~J.,  2001, {The estimation and tracking of
  frequency}.
Cambridge: Cambridge University Press

\bibitem[\protect\citeauthoryear{{Rabiner}}{{Rabiner}}{1989}]{rab89}
{Rabiner} L.~R.,  1989, Proceedings of the IEEE, 77, 257

\bibitem[\protect\citeauthoryear{{Rickett}}{{Rickett}}{1990}]{ric90}
{Rickett} B.~J.,  1990, \araa, 28, 561

\bibitem[\protect\citeauthoryear{{Sarkissian}, {Reynolds}, {Hobbs} \&
  {Harvey-Smith}}{{Sarkissian} et~al.}{2017a}]{sar17a}
{Sarkissian} J.,  {Reynolds} J.,  {Hobbs} G.,    {Harvey-Smith} L.,  2017a,
  CSIRO Data Collection; DOI 10.4225/08/59183e949e033

\bibitem[\protect\citeauthoryear{{Sarkissian}, {Reynolds}, {Hobbs} \&
  {Harvey-Smith}}{{Sarkissian} et~al.}{2017b}]{sar17b}
{Sarkissian} J.~M.,  {Reynolds} J.~E.,  {Hobbs} G.,    {Harvey-Smith} L.,
  2017b, \pasa, 34, e027

\bibitem[\protect\citeauthoryear{{Shannon}, {Lentati}, {Kerr}, {Johnston},
  {Hobbs} \& {Manchester}}{{Shannon} et~al.}{2016}]{sha16}
{Shannon} R.~M.,  {Lentati} L.~T.,  {Kerr} M.,  {Johnston} S.,  {Hobbs} G.,
  {Manchester} R.~N.,  2016, \mnras, 459, 3104

\bibitem[\protect\citeauthoryear{{Shannon}, {Ravi}, {Lentati}, {Lasky},
  {Hobbs}, {Kerr}, {Manchester}, {Coles}, {Levin}, {Bailes} \& {et
  al.}}{{Shannon} et~al.}{2015}]{sha15}
{Shannon} R.~M.,  {Ravi} V.,  {Lentati} L.~T.,  {Lasky} P.~D.,  {Hobbs} G.,
  {Kerr} M.,  {Manchester} R.~N.,  {Coles} W.~A.,  {Levin} Y.,  {Bailes} M.,
  {et al.} 2015, Science, 349, 1522

\bibitem[\protect\citeauthoryear{{Stairs}}{{Stairs}}{2003}]{sta03}
{Stairs} I.~H.,  2003, Living Reviews in Relativity, 6, 5

\bibitem[\protect\citeauthoryear{{Suvorova}, {Clearwater}, {Melatos}, {Sun},
  {Moran} \& {Evans}}{{Suvorova} et~al.}{2017}]{suv17}
{Suvorova} S.,  {Clearwater} P.,  {Melatos} A.,  {Sun} L.,  {Moran} W.,
  {Evans} R.~J.,  2017, \prd, 96, 102006

\bibitem[\protect\citeauthoryear{{Suvorova}, {Melatos}, {Evans} \&
  {Moran}}{{Suvorova} et~al.}{2018}]{suv18}
{Suvorova} S.,  {Melatos} A.,  {Evans} R.~J.,    {Moran} W.,  2018, IEEE
  Transactions on Signal Processing, p. submitted

\bibitem[\protect\citeauthoryear{{Suvorova}, {Sun}, {Melatos}, {Moran} \&
  {Evans}}{{Suvorova} et~al.}{2016}]{suv16}
{Suvorova} S.,  {Sun} L.,  {Melatos} A.,  {Moran} W.,    {Evans} R.~J.,  2016,
  \prd, 93, 123009

\bibitem[\protect\citeauthoryear{{Taylor}}{{Taylor}}{1992}]{tay92}
{Taylor} J.~H.,  1992, Philosophical Transactions of the Royal Society of
  London Series A, 341, 117

\bibitem[\protect\citeauthoryear{{van Eysden} \& {Melatos}}{{van Eysden} \&
  {Melatos}}{2010}]{van10}
{van Eysden} C.~A.,  {Melatos} A.,  2010, \mnras, 409, 1253

\bibitem[\protect\citeauthoryear{{van Straten}, {Demorest} \& {Oslowski}}{{van
  Straten} et~al.}{2012}]{van12}
{van Straten} W.,  {Demorest} P.,    {Oslowski} S.,  2012, Astronomical
  Research and Technology, 9, 237

\bibitem[\protect\citeauthoryear{{Warszawski} \& {Melatos}}{{Warszawski} \&
  {Melatos}}{2011}]{war11}
{Warszawski} L.,  {Melatos} A.,  2011, \mnras, 415, 1611

\bibitem[\protect\citeauthoryear{{Watts}, {Espinoza}, {Xu}, {Andersson},
  {Antoniadis}, {Antonopoulou}, {Buchner}, {Datta}, {Demorest}, {Freire},
  {Hessels}, {Margueron}, {Oertel}, {Patruno}, {Possenti}, {Ransom}, {Stairs}
  \& {Stappers}}{{Watts} et~al.}{2015}]{wat15}
{Watts} A.,  {Espinoza} C.~M.,  {Xu} R.,  {Andersson} N.,  {Antoniadis} J.,
  {Antonopoulou} D.,  {Buchner} S.,  {Datta} S.,  {Demorest} P.,  {Freire} P.,
  {Hessels} J.,  {Margueron} J.,  {Oertel} M.,  {Patruno} A.,  {Possenti} A.,
  {Ransom} S.,  {Stairs} I.,    {Stappers} B.,  2015, in Advancing Astrophysics
  with the Square Kilometre Array (AASKA14) {Probing the neutron star interior
  and the Equation of State of cold dense matter with the SKA}.
p.~43

\bibitem[\protect\citeauthoryear{{Wette}}{{Wette}}{2016}]{wet16}
{Wette} K.,  2016, \prd, 94, 122002

\bibitem[\protect\citeauthoryear{{Wong}, {Backer} \& {Lyne}}{{Wong}
  et~al.}{2001}]{won01}
{Wong} T.,  {Backer} D.~C.,    {Lyne} A.~G.,  2001, \apj, 548, 447

\bibitem[\protect\citeauthoryear{{Yakovlev}, {Levenfish} \&
  {Shibanov}}{{Yakovlev} et~al.}{1999}]{yak99}
{Yakovlev} D.~G.,  {Levenfish} K.~P.,    {Shibanov} Y.~A.,  1999, Physics
  Uspekhi, 42, 737

\bibitem[\protect\citeauthoryear{{Yu} \& {Liu}}{{Yu} \& {Liu}}{2017}]{yu17}
{Yu} M.,  {Liu} Q.-J.,  2017, \mnras, 468, 3031

\bibitem[\protect\citeauthoryear{{Yu}, {Manchester}, {Hobbs}, {Johnston},
  {Kaspi}, {Keith}, {Lyne}, {Qiao}, {Ravi}, {Sarkissian}, {Shannon} \&
  {Xu}}{{Yu} et~al.}{2013}]{yu13}
{Yu} M.,  {Manchester} R.~N.,  {Hobbs} G.,  {Johnston} S.,  {Kaspi} V.~M.,
  {Keith} M.,  {Lyne} A.~G.,  {Qiao} G.~J.,  {Ravi} V.,  {Sarkissian} J.~M.,
  {Shannon} R.,    {Xu} R.~X.,  2013, \mnras, 429, 688

\end{thebibliography}

\appendix
\section{Solving the HMM
 \label{sec:hmmappa}}
Let $M = \{ A_{q_j q_i}, L_{o(t_n) q_i}, \Pi_{q_i} \}$ be a HMM
with transition probability $A_{q_j q_i}$,
emission probability $L_{o(t_n) q_i}$,
and prior probability $\Pi_{q_i}$ defined according to 
(\ref{eq:hmm1}), (\ref{eq:hmm2}), and (\ref{eq:hmm4}) respectively.
Let $Q_{m:n}=\{ q(t_m),\dots, q(t_n) \}$ and
$O_{m:n}=\{ o(t_{m}),\dots, o(t_{n}) \}$
denote arbitrary, partial sequences of hidden and observed states respectively,
with $1\leq m \leq n \leq N_T$.
In this appendix, we present efficient numerical algorithms,
which exploit recursion to solve the two
fundamental HMM problems below.
\begin{enumerate}
\item
What is the Bayesian evidence $\Pr(O_{1:N_T}|M)$ for the model $M$,
given the full observed sequence, $O_{1:N_T}$?
This question reduces to calculating
\begin{eqnarray}
 \Pr(O_{1:N_T}|M)
 & = &
 \sum_{Q_{1:N_T}}
 \Pr(O_{1:N_T} | Q_{1:N_T},M) \Pr(Q_{1:N_T},M)
\label{eq:hmmappa1}
 \\
 & = &
 \sum_{Q_{1:N_T}}
 \Pi_{q(t_1)} L_{o(t_1) q(t_1)}
 \prod_{n=2}^{N_T}
 A_{q(t_{n-1}) q(t_n)} L_{o(t_n) q(t_n)}~.
\label{eq:hmmappa2}
\end{eqnarray}
\item
What is the optimal hidden sequence given $M$ and $O_{1:N_T}$?
This question reduces to calculating
\begin{eqnarray}
 \hat{q}(t_n)
 & = &
 \underset{q(t_n)} {\rm arg\,\,max\,\,}
 \sum_{Q_{1:n-1}}
 \Pi_{q(t_1)} L_{o(t_1) q(t_1)}
 \prod_{m=2}^{n-1}
 A_{q(t_{m-1}) q(t_m)} L_{o(t_m) q(t_m)}
 \nonumber \\
 & &
 \times
 A_{q(t_{n-1}) q(t_n)} L_{o(t_n) q(t_n)}
 \sum_{Q_{n+1:N_T}}
 \prod_{m=n+1}^{N_T}
 A_{q(t_{m-1}) q(t_m)} L_{o(t_m) q(t_m)}
\label{eq:hmmappa3a}
\end{eqnarray}
for $2\leq n \leq N_T$ and a uniform prior,
if one wishes to maximize 
$\Pr[q(t_n) | O_{1:N_T},M]$ point-wise, or
\begin{equation}
 Q_{1:N_T}^\ast
 =
 \underset{Q_{1:N_T}} {\rm arg\,\,max\,\,}
 \Pi_{q(t_1)} L_{o(t_1) q(t_1)}
 \prod_{n=2}^{N_T}
 A_{q(t_{n-1}) q(t_n)} L_{o(t_n) q(t_n)}~,
\label{eq:hmmappa3b}
\end{equation}
if one wishes to maximize
$\Pr(Q_{1:N_T} | O_{1:N_T})$ sequence-wise.
The difference between options 
(\ref{eq:hmmappa3a}) and (\ref{eq:hmmappa3b})
is explained below.
\end{enumerate}
The above problems are essential building blocks of the glitch-finding
algorithm in \S\ref{sec:hmm3}.
A third fundemantal problem --- given $O_{1:N_T}$, what model $M$ maximizes
the Bayesian evidence $\Pr(O_{1:N_T}|M)$? --- 
amounts to learning the optimal model (here, the glitch dynamics)
from the data.
It is of great interest but lies outside the scope of this paper.
The reader is referred to the excellent tutorial by \citet{rab89}
for a fuller treatment of the fundamental principles of HMMs.

\subsection{Forward algorithm
 \label{sec:hmmappaa}}
It may seem that evaluating the sum (\ref{eq:hmmappa2})
involves $\sim N_T N_Q^{N_T}$ floating point operations,
because each term is a product of $2N_T$ factors,
and there are $N_Q^{N_T}$ possible hidden sequences.
Fortunately recursive filtering offers a more efficient approach.

\begin{figure}
\begin{center}
\includegraphics[width=16cm,angle=0]{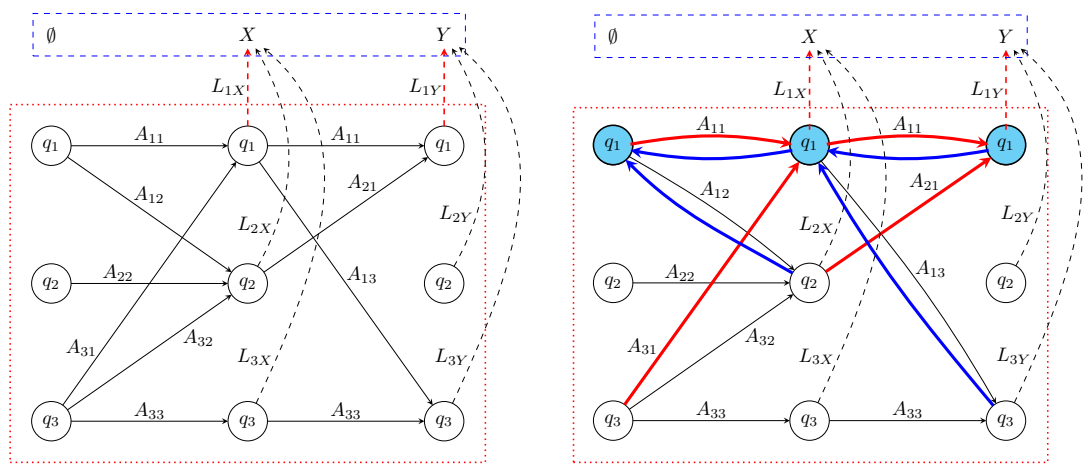}
\end{center}
\caption{
Schematic of the HMM trellis.
({\em Left panel.})
Subset of the links in the trellis.
Every circle denotes a hidden state ($q_1$, $q_2$, $q_3$)
at some time step (time increases to the right).
The top rectangle contains the data ($X$, $Y$).
Every unbroken arrow corresponds to a nonzero transition probability,
e.g.\ $A_{12}$ is the probability of transitioning from $q_1$ to $q_2$.
Every broken arrow corresponds to a nonzero emission probability,
e.g.\ $L_{2Y}$ is the probability of observing the data $Y$
at the third time-step while occupying hidden state $q_2$.
({\em Right panel.})
Sample of the links that go into evaluating the induction step
for $\gamma_{q_1}$ (circles shaded blue)
for the forward variable 
[red arrows; equation (\ref{eq:hmmappa6})]
and the backward variable
[blue arrows; equation (\ref{eq:hmmappa10})].
}
\label{fig:hmmappa1}
\end{figure}

Consider the forward variable
\begin{equation}
 \alpha_{q_i}(t_n)
 =
 \Pr[q(t_n)=q_i,O_{1:n} | M]~,
\label{eq:hmmappa4}
\end{equation}
i.e.\ $\alpha_{q_i}(t_n)$ equals the probability that one observes
the partial data $O_{1:n}$ during the interval $t_1 \leq t \leq t_n$,
and the system occupies the state $q_i$ at time $t=t_n$.
Notice that, at $t=t_n$, every one of the $N_Q$ hidden states 
is reached from the same $N_Q$ hidden states at $t=t_{n-1}$.
Hence one can calculate (\ref{eq:hmmappa2}) by addressing every link
in the trellis in Figure \ref{fig:hmmappa1} once,
instead of backtracking over every link multiple times while tracing 
all $N_Q^{N_T}$ hidden sequences separately.
The following algorithm achieves this economy by storing the partial results
at each forward step through the trellis
\citep{rab89,qui01}.
\begin{enumerate}
\item
{\em Initialization.} For $1\leq i \leq N_Q$, set
\begin{equation}
 \alpha_{q_i}(t_1)
 =
 \Pi_{q_i} L_{o(t_1) q_i}~.
\label{eq:hmmappa5}
\end{equation}
\item
{\em Induction.} For $1\leq n \leq N_T-1$ and $1\leq i \leq N_Q$, 
compute the forward variable
by summing over its values at the previous HMM step:
\begin{equation}
 \alpha_{q_i}(t_{n+1})
 =
 L_{o(t_{n+1}) q_i} 
 \sum_{j=1}^{N_Q} A_{q_i q_j} \alpha_{q_j}(t_n)~.
\label{eq:hmmappa6}
\end{equation}
\item
{\em Termination.} 
The Bayesian evidence is the sum of the forward variable
over the final states, viz.
\begin{equation}
 \Pr(O_{1:N_T} | M)
 =
 \sum_{i=1}^{N_Q} \alpha_{q_i}(t_{N_T})~.
\label{eq:hmmappa7}
\end{equation}
\end{enumerate}
The trellis contains $N_Q^2$ links per HMM transition,
and there are $N_T-1$ transitions,
so the computation involves $\sim N_T N_Q^2$ floating point operations in total,
a large saving.

\subsection{Forward-backward algorithm
 \label{sec:hmmappab}}
The maximization step in (\ref{eq:hmmappa3a}) can be
executed with the help of recursive smoothing,
without comparing the $N_Q^{N_T}$ hidden sequences severally.
This is achieved by introducing a backward variable,
analogous to the forward variable above,
and then maximizing the product of the forward and backward variables.

Consider the backward variable,
\begin{equation}
 \beta_{q_i}(t_n)
 =
 \Pr[O_{n+1:N_T} | q(t_n)=q_i, M]~,
\label{eq:hmmappa8}
\end{equation}
i.e.\ $\beta_{q_i}(t_n)$ equals the probability that one observes
the partial data $O_{n+1:N_T}$ during the interval $t_{n+1} \leq t \leq N_T$,
conditional on the system occupying the state $q_i$ at time $t=t_n$.
Every one of the hidden states at $t=t_n$ 
connects to the same set of hidden states at $t=t_{n+1}$,
so one can express $\beta_{q_i}(t_n)$ inductively in terms of
$\beta_{q_1}(t_{n+1}), \dots, \beta_{q_{N_Q}}(t_{n+1})$ 
by summing over the $N_Q$ possible transitions
from $q_i(t_n)$ to $q_1(t_{n+1}),\dots,q_{N_Q}(t_{n+1})$
\citep{rab89}.
\begin{enumerate}
\item
{\em Initialization.} For $1\leq i \leq N_Q$, set
\begin{equation}
 \beta_{q_i}(t_{N_T})
 =
 1~.
\label{eq:hmmappa9}
\end{equation}
\item
{\em Induction.} 
For $1\leq n \leq N_T-1$ and $1\leq i \leq N_Q$,
compute the backward variable
by summing over its values at the succeeding HMM step,
starting from $n=N_T-1$ and stepping back to $n=1$.
\begin{equation}
 \beta_{q_i}(t_n)
 =
 \sum_{j=1}^{N_Q}
 A_{q_i q_j} L_{o(t_{n+1}) q_j} \beta_{q_j}(t_{n+1})~.
\label{eq:hmmappa10}
\end{equation}
\end{enumerate}
The backward algorithm (\ref{eq:hmmappa8})--(\ref{eq:hmmappa10})
entails $\sim N_T N_Q^2$
floating point operations like the forward algorithm.

We now ask what hidden state is most likely to be occupied at $t=t_n$,
given the {\em entire} observed sequence $O_{1:N_T}$ and the model $M$.
Define
\begin{eqnarray}
 \gamma_{q_i}(t_n)
 & = &
 \Pr[q(t_n)=q_i | O_{1:N_T}, M]
\label{eq:hmmappa11}
 \\
 & = &
 \left[ 
  \sum_{j=1}^{N_Q} \alpha_{q_j}(t_n) \beta_{q_j}(t_n)
 \right]^{-1}
 \alpha_{q_i}(t_n) \beta_{q_i}(t_n)~,
\label{eq:hmmappa12}
\end{eqnarray}
where (\ref{eq:hmmappa12}) follows from (\ref{eq:hmmappa11}),
because the forward variable accounts for the hidden and observed
sequences $Q_{1:n}$ and $O_{1:n}$ terminating at $q(t_n)=q_i$,
and the backward variable accounts for the hidden and observed sequences
$Q_{n:N_T}$ and $O_{n:N_T}$ originating at $q(t_n)=q_i$.
Equation (\ref{eq:hmmappa12}) implies that the most likely
state at each HMM step is given by
\begin{equation}
 \hat{q}(t_n)
 =
 \underset{1\leq i \leq N_Q} {\rm arg\,\,max\,\,}
 \gamma_{q_i}(t_n)
\label{eq:hmmappa13}
\end{equation}
for $1\leq n \leq N_T$.
The denominator of (\ref{eq:hmmappa12}) is a normalization factor
which, for $t=N_T$, reduces to $\Pr(O_{1:N_T}|M)$ in (\ref{eq:hmmappa2})
via (\ref{eq:hmmappa7}) and (\ref{eq:hmmappa9}).
It can be ignored when maximizing over $1\leq i \leq N_Q$.
Equations (\ref{eq:hmmappa4})--(\ref{eq:hmmappa13}) together
constitute the HMM forward-backward algorithm.
The algorithm entails $\sim N_T N_Q^2$ floating point operations,
dominated by (\ref{eq:hmmappa10})--(\ref{eq:hmmappa12});
the final step (\ref{eq:hmmappa13}) reduces to 
$\sim N_T \ln N_Q$ operations with binary maximization.

The above solution of the HMM optimization problem is not unique.
It does maximize the number of most probable hidden states.
On the other hand,
there is no guarantee that the sequence generated thus is admissible, 
i.e.\ consistent with the transition probabilities.
For example, if we have $A_{q_j q_i}=0$ for some $q_j$ and $q_i$,
it may not be possible to connect the sequence 
$\{ \hat{q}(t_1), \dots, \hat{q}(t_{N_T}) \}$
generated by (\ref{eq:hmmappa13}).
In this sense,
$\{ \hat{q}(t_1), \dots, \hat{q}(t_{N_T}) \}$
differs subtly from 
$Q^\ast_{1:N_T}$ in (\ref{eq:hmmappa3b}).
The latter quantity is admissible by construction and
maximizes the probability of the whole sequence
rather than individual states along the sequence.
In general, both approaches (and indeed others not discussed here) are valid.
In this paper, we focus on $\{ \hat{q}(t_1), \dots, \hat{q}(t_{N_T}) \}$
for three reasons.
First, we wish to maximize the number of most probable hidden states
when generating an ephemeris.
Second,
we find by trial and error that inadmissibility
arises rarely in the glitch-finding application.
Third,
we wish to know the shape of the joint PDF of 
$f(t_n)$ and $\dot{f}(t_n)$ at each $t_n$,
in order to check how far the optimal sequence stands above its
nearest competitors.
This is done easily by plotting $\gamma_{q_i}(t_n)$ versus $q_i$,
whereas $Q^\ast_{1:N_T}$ gives the best sequence only.
Traditional, frequentist pulsar timing methods involve a mixture of
point-wise and sequence-wise optimization by minimizing
the squares of the point-by-point phase residuals 
summed over the entire sequence.

\subsection{Viterbi algorithm
 \label{sec:hmmappa1c}}
For the sake of completeness, we outline an algorithm
for calculating $Q^\ast_{1:N_T}$ in (\ref{eq:hmmappa3b}).
Known as the Viterbi algorithm,
and based on dynamic programming methods,
it exploits the property that any subsequence of the optimal sequence
is itself optimal in order to prune the trellis of admissible sequences
\citep{rab89,qui01}.
The pseudocode below matches closely the notation adopted by
\citet{rab89}
and in recent gravitational wave applications 
\citep{suv16};
cf.\ \citet{bay19}.

Consider the variable
\begin{eqnarray}
 \delta_{q_i}(t_n)
 & = & 
 \max_{Q_{1:n-1}} 
 \Pr[q(t_n)=q_i,Q_{1:n-1} | O_{1:n}, M]~,
\label{eq:hmmappa14}
 \\
 & =  &
 L_{o(t_n)q_i}
 \max_{q_j}
 A_{q_i q_j}
 \delta_{q_j}(t_{n-1})~,
\label{eq:hmmappa15}
\end{eqnarray}
which corresponds to the maximum probability,
that the HMM terminates in the hidden state $q_i$ at $t=t_n$
given the partial observation sequence $O_{1:n}$.
Let $\psi_{q_i}(t_n)$ denote the hidden state at $t=t_{n-1}$
from which $q_i$ is reached at $t=t_n$,
along the sequence that maximizes
$\Pr[q(t_n)=q_i,Q_{1:n-1} | O_{1:n}, M]$ 
in (\ref{eq:hmmappa14}), viz.\
\begin{equation}
 \psi_{q_i}(t_n)
 = 
 \underset{q_j} {\rm arg\,max\,}
 A_{q_i q_j}
 \delta_{q_j}(t_{n-1})~.
\label{eq:hmmappa16}
\end{equation}
The Viterbi algorithm evaluates $\delta_{q_i}(t_n)$ and $\psi_{q_i}(t_n)$
for all of the $N_Q N_T$ nodes in the trellis in Figure \ref{fig:hmmappa1}
and then backtracks to reconstruct $Q^\ast_{1:N_T}$.
It resembles the forward algorithm,
with the sum in (\ref{eq:hmmappa6}) replaced by the maximization steps
in (\ref{eq:hmmappa15}) and (\ref{eq:hmmappa16}).

\begin{enumerate}
\item
{\em Initialization.}
For $1\leq i \leq N_Q$, set
\begin{equation}
 \delta_{q_i}(t_1) = \Pi_{q_i} L_{o(t_1)q_i}~,
\label{eq:hmmappa17}
\end{equation}
Note that $\psi_{q_i}(t_1)$ is not initialized as it is never needed.
\item
{\em Forward recursion.} For $2\leq n \leq N_T$
and $1\leq i \leq N_Q$,
implement the induction step (\ref{eq:hmmappa15}) via
\begin{equation}
 \delta_{q_i}(t_n)
 = 
 L_{o(t_n)q_i} 
 \max_{1\leq j \leq N_Q} A_{q_i q_j} \delta_{q_j}(t_{n-1})
\label{eq:hmmappa18}
\end{equation}
and
\begin{equation}
 \psi_{q_i}(t_n)
 = 
 \underset{1\leq j\leq N_Q}{\rm arg\,max\,} 
  A_{q_i q_j} \delta_{q_j}(t_{n-1})~.
\label{eq:hmmappa19}
\end{equation}
\item
{\em Termination.} 
Identify the state $q^\ast(t_{N_T})$, where the optimal sequence ends.
\begin{equation}
 \Pr(Q^\ast_{1:N_T} | O_{1:N_T},M)
 =
 \max_{1\leq j \leq N_Q} \delta_{q_j}(t_{N_T})
\label{eq:hmmappa20}
\end{equation}
and
\begin{equation}
 q^\ast(t_{N_T})
 =
 \underset{1\leq j \leq N_Q}{\rm arg\,max\,} \delta_{q_j}(t_{N_T})~.
\label{eq:hmmappa21}
\end{equation}
\item
{\em Backward recursion.}
Backtrack through the trellis in Figure \ref{fig:hmmappa1}
to reconstruct the optimal sequence,
guided by $\psi_{q_j}(t_n)$.
For $1\leq n \leq N_T-1$, compute
\begin{equation}
 q^\ast(t_{n}) = \psi_{q^\ast(t_{n+1})}(t_{n+1})~,
\label{eq:hmmappa22}
\end{equation}
starting from $n=N_T-1$ and stepping back to $n=1$.
\end{enumerate}
The algorithm involves $\sim N_T N_Q \ln N_Q$ floating point operations
with binary maximization
\citep{qui01}.

\section{Hidden state evolution via a Langevin equation
 \label{sec:hmmappb}}
In this appendix we derive the transition probabilities
in \S\ref{sec:hmm2d} self-consistently by solving
a stochastic differential equation for $q(t)=[f(t),\dot{f}(t)]$
in the inter-step interval $t_{n-1} \leq t \leq t_n$.
Although the system is measured at discrete instants $t_n$,
its state evolves stochastically between the TOAs
due to timing noise
\citep{cor80}.
For now, we neglect the secular component of the torque derivative,
$\langle \ddot{f} \rangle$,
a good approximation provided that
$x_n^2 \lesssim 6 f(t_n) / | \ddot{f}(t_n) |$ 
is satisfied; see \S\ref{sec:hmm2a}.
This leaves an approximately constant secular torque,
which enters as an initial condition on $\dot{f}(t_{n-1})$,
and a fluctuating torque derivative $\xi(t)$,
which drives the Langevin equation,
\begin{equation}
 \frac{d^2f}{dt^2}
 =
 \xi(t)~,
\label{eq:hmmappb1}
\end{equation}
with white noise statistics,
\begin{equation}
 \langle \xi(t) \rangle = 0
\label{eq:hmmappb2}
\end{equation}
and
\begin{equation}
 \langle \xi(t) \xi(t') \rangle = 
 \sigma^2 \delta(t-t')~.
\label{eq:hmmappb3}
\end{equation}
Angular brackets denote an ensemble average over noise realizations.

Unlike $\sigma_{\rm TOA}^2$,
the variance $\sigma^2$ is not a measurement uncertainty.
It is a mean-square measure of the amplitude of the process noise 
driven by the fluctuating torque derivative,
which may arise physically from starquakes and superfluid vortex avalanches
for example
\citep{chu10b,war11,has15}.
Its units are ${\rm Hz^2 \,s^{-3}}$,
cf.\ $\sigma_{\rm TOA}^2$, which has units of ${\rm s}^2$.
Likewise, $\sigma^2$ is not the same as $\sigma_{\rm TN}^2$ 
in the synthetic data in \S\ref{sec:hmm4},
because
$\sigma_{\rm TN}^2$ equals the variance in the autocorrelation function
of the torque, not the torque derivative.
White noise fluctuations in the torque derivative,
as in (\ref{eq:hmmappb1}),
are not necessarily physical;
they are an artificial device to keep $A_{q_j q_i}$ finite.
By contrast, a fluctuating torque gives
$\langle \dot{f}(t) \dot{f}(t') \rangle
 \propto \delta(t-t')$,
which diverges in the limit $t\rightarrow t'$.
The tests in \S\ref{sec:hmm6} and Appendix \ref{sec:hmmappf}
confirm that (\ref{eq:hmmappb1}) works well empirically
when tracking synthetic data generated by a fluctuating torque.
This reflects a well-known property of HMMs,
that they are insensitive to the exact form of $A_{q_j q_i}$,
as long as the dynamics during the interval $t_{n-1} \leq t \leq t_n$
are captured broadly,
e.g. $A_{q_{i+1} q_i} = A_{q_i q_i} = A_{q_{i-1} q_i} = 1/3$
often serves as an adequate model for more complicated Brownian motion
\citep{qui01,suv16,suv17}.

The PDF $p[f(t_n),\dot{f}(t_n) | f(t_{n-1}), \dot{f}(t_{n-1}) ]$ 
at $t=t_n$ describing the ensemble of Langevin trajectories
starting from the state
$[ f(t_{n-1}), \dot{f}(t_{n-1}) ]$ at $t=t_{n-1}$
satisfies the Fokker-Planck equation
\citep{gar94}
\begin{equation}
 \frac{\partial p}{\partial t}
 =
 \frac{\sigma^2}{2}\frac{\partial ^2 p}{dt^2}~.
\label{eq:hmmappb4}
\end{equation}
The coefficients in (\ref{eq:hmmappb4}) are constant,
so the solution is a Gaussian. 
It is defined by the first two moments,
which can be calculated directly from the Langevin trajectory,
\begin{eqnarray}
 f(t) 
 & = &  f(t_{n-1}) 
 + (t-t_{n-1}) \dot{f}(t_{n-1})
 + g(t_{n-1}) [ \Delta f_{\rm p}(t_{n-1}) 
  + (t-t_{n-1}) \Delta \dot{f}_{\rm p}(t_{n-1}) ]
 \nonumber
 \\
 & &
 + \int_{t_{n-1}}^t dt' \int_{t_{n-1}}^{t'} dt'' \, \xi(t'')~.
\label{eq:hmmappb5}
\end{eqnarray}
For any stationary process, we have
\begin{equation}
 \int_0^t dt'' \int_0^{t'} dt''' \, 
 \langle \xi(t'') \xi(t''') \rangle
 =
 \sigma^2 {\rm min}(t,t')~.
\label{eq:hmmappb6}
\end{equation}
Combining (\ref{eq:hmmappb5}) and (\ref{eq:hmmappb6}), we find that
the first moments evolve according to
\begin{eqnarray}
 \langle f(t) \rangle 
 & = &
 f(t_{n-1})
 + (t-t_{n-1}) \dot{f}(t_{n-1})
 \nonumber
 \\
 & & + g(t_{n-1}) [ \Delta f_{\rm p}(t_{n-1}) 
  + (t-t_{n-1}) \Delta \dot{f}_{\rm p}(t_{n-1}) ]~,
\label{eq:hmmappb7}
 \\
 \langle \dot{f}(t) \rangle 
 & = &
 \dot{f}(t_{n-1})
 + g(t_{n-1}) \Delta \dot{f}_{\rm p}(t_{n-1})~,
\label{eq:hmmappb8}
\end{eqnarray}
and the second moments evolve according to
\begin{eqnarray}
 {\rm cov}[f(t),f(t)]
 & = &
 \frac{1}{3} \sigma^2 (t-t_{n-1})^3~,
\label{eq:hmmappb9}
 \\
 {\rm cov}[f(t),\dot{f}(t)]
 & = &
 \frac{1}{2} \sigma^2 (t-t_{n-1})^2~,
\label{eq:hmmappb10}
 \\
 {\rm cov}[\dot{f}(t),\dot{f}(t)]
 & = &
 \sigma^2 (t-t_{n-1})~,
\label{eq:hmmappb11}
\end{eqnarray}
where
${\rm cov}(a,b)
 = \langle (a-\langle a \rangle) (b-\langle b \rangle) \rangle$
denotes the central covariance of $a$ and $b$.
Equations (\ref{eq:hmmappb7})--(\ref{eq:hmmappb11})
together define $A_{q(t_n) q(t_{n-1})}$ through
(\ref{eq:hmm10})--(\ref{eq:hmm13}).

We emphasize that the Wiener process 
(\ref{eq:hmmappb1})--(\ref{eq:hmmappb3})
may not be realistic physically for every pulsar.
Empirically speaking,
pulsar timing noise does not display significant memory
in the torque derivative over typical TOA gaps of days to weeks
\citep{pri12},
so the white noise in (\ref{eq:hmmappb2})
and (\ref{eq:hmmappb3}) represents a fair approximation.
\footnote{
The autocorrelation time-scale of days to weeks 
measured in the phase residuals
\citep{pri12}
arises after integrating the torque twice with respect to time.
}
However, for longer TOA gaps, redder timing noise,
or long-lasting post-glitch recoveries 
(see \S\ref{sec:hmm2d} and \S\ref{sec:hmm7}),
equations (\ref{eq:hmmappb1})--(\ref{eq:hmmappb3})
need to be generalized.
\footnote{
Timing noise is red in all but the ``calmest'' millisecond pulsars, 
when measured over years and many TOAs.
As far as $A_{q(t_n) q(t_{n-1})}$ is concerned, however,
the Wiener process in (\ref{eq:hmmappb1})--(\ref{eq:hmmappb3})
is indifferent to multi-TOA correlations;
it resets at the start of every TOA interval $t_{n-1} \leq t \leq t_n$.
}
In this paper, we test the robustness of
(\ref{eq:hmmappb1})--(\ref{eq:hmmappb3}) in two ways.
First, we deliberately generate synthetic data with a different noise model,
which is white in the torque instead of the torque derivative
[see (\ref{eq:hmm20})--(\ref{eq:hmm22})],
yet still the HMM performs well in the tests in \S\ref{sec:hmm6}.
Second, in \S\ref{sec:hmm7}, the assumption
(\ref{eq:hmmappb1})--(\ref{eq:hmmappb3}) 
does not harm the HMM's ability to locate accurately
the 2016 December 12 glitch in PSR J0835$-$4510
and exclude the existence of a second glitch in its vicinity,
in accord with traditional analyses.
This is comforting,
because in PSR J0835$-$4510 the timing noise is relatively red,
and the post-glitch recoveries are notoriously lengthy
\citep{lyn96}.
Generalizing the calculations in this appendix to redden
(\ref{eq:hmmappb1})--(\ref{eq:hmmappb3}) 
is an interesting avenue for future work,
once the HMM is validated against more real pulsars.
Ultimately glitch detection is an exercise undertaken
conditionally with respect to a phase model;
there is no model-independent answer to the question of whether or not
a stretch of data contains a glitch.
This is equally true of traditional methods, 
whether the model is simple
(e.g.\ step changes in an otherwise smooth Taylor expansion)
or complex 
[e.g.\ phase residuals with a power-law power spectral density
 \citep{sha16}].

\section{Defining the grid and DOI
 \label{sec:hmmappc}}
The DOI encompasses the point 
$(f_{\rm LS},\dot{f}_{\rm LS})$
corresponding to the optimal (least squares), 
constant-coefficient phase model
fitted with $f=f_{\rm LS}$, $\dot{f}=\dot{f}_{\rm LS}$, and $\ddot{f}=0$.
The fit can be generated from the TOAs by running {\sc tempo2}, for example.
At any instant,
the true, unknown $f(t)$ and $\dot{f}(t)$ deviate slightly from
$f_{\rm LS}$ and $\dot{f}_{\rm LS}$ respectively.
One way to estimate the deviations is to
attribute the phase residual $\delta\phi(t_n)$ measured at each HMM step $t_n$ 
($1\leq n \leq N_T$) to a pure frequency fluctuation,
$\varepsilon_f(t_n) = \delta\phi(t_n) / x_n$ 
(with $\dot{f}=\dot{f}_{\rm LS}$),
or a pure frequency derivative fluctuation,
$\varepsilon_{\dot{f}}(t_n) = 2 \delta\phi(t_n) / x_n^2$
(with $f=f_{\rm LS}$).
We then define the DOI to be the rectangular domain
\begin{equation}
 \min_{1\leq n \leq N_T} \varepsilon_f(t_n)
 \leq 
 S^{-1} ( f - f_{\rm LS} )
 \leq
 \max_{1\leq n \leq N_T} \varepsilon_f(t_n)~,
\label{eq:hmm14}
\end{equation}
\begin{equation}
 \min_{1\leq n \leq N_T} \varepsilon_{\dot{f}}(t_n)
 \leq 
 S^{-1} ( \dot{f} - \dot{f}_{\rm LS} )
 \leq
 \max_{1\leq n \leq N_T} \varepsilon_{\dot{f}}(t_n)~,
\label{eq:hmm15}
\end{equation}
where $S \geq 1$ is a dimensionless safety factor chosen by the user.
Equations (\ref{eq:hmm14}) and (\ref{eq:hmm15}) are conservative,
because in reality the fluctuations develop over multiple HMM steps
[thereby reducing $\varepsilon_f(t_n)$ and $\varepsilon_{\dot{f}}(t_n)$]
and occur in tandem
[$\varepsilon_f(t_n) \neq 0 $ and $\varepsilon_{\dot{f}}(t_n) \neq 0$ 
simultaneously].

The continuous physical variables $f(t)$ and $\dot{f}(t)$
are discretized for numerical purposes.
Formally the grid resolution is governed by the curvature
of the likelihood function at its peak
through the Cram\'{e}r-Rao lower bound or related quantities like
the parameter space metric in gravitational wave applications
\citep{lea15,wet16}.
The peaks of $L_{o(t_n)q_i}$ sharpen, as $\kappa \gg 1$ increases.
However, it is unclear how to apply such approaches to the problem at hand, 
because the distribution of the number of pulses
between consecutive TOAs is unknown
\citep{suv18}.
It can be estimated,
say as a Poisson or quasiperiodic process \citep{mel08,ful17,how18},
with the relevant time-scales determined iteratively
(if the data are analysed for the first time) 
or copied from the literature
(if glitches have already been detected)
\citep{car19}.
Alternatively one can approximate the likelihood function
(for the purpose of grid design only)
assuming constant $\dot{f}$ and $\sigma_{\rm TOA} = 0$,
as discussed thoroughly by \citet{suv18}.
In this paper, for simplicity, we set the grid spacing to be
the minimum $\Delta f_{\rm p}$ and $\Delta \dot{f}_{\rm p}$
that we wish to resolve, limited only by computational cost.
Other options which may deliver computational savings,
such as logarithmic gridding,
will be explored in future work.

The set $G$ in (\ref{eq:hmm10}) is constructed as follows.
For the frequency component we allow all jumps with
$\Delta f_{\rm p} > 0$,
such that
$f(t_{n-1}) + \Delta f_{\rm p}$ is a valid state and lies in the DOI.
For the frequency derivative component we allow all jumps of either sign,
such that
$\dot{f}(t_{n-1}) + \Delta \dot{f}_{\rm p}$ is a valid state and lies in the DOI.
Note that $\Delta f_{\rm p} > 0$ does not imply $f(t_n)>f(t_{n-1})$ necessarily,
because the spin down between TOAs may compensate for the glitch.

Gridding modifies the emission probability given by (\ref{eq:hmm7})
and (\ref{eq:hmm8}), as noted in \S\ref{sec:hmm2c},
by changing the effective value of $\kappa$.
This occurs because discretization introduces a state 
and hence a phase uncertainty proportional to the grid spacing,
which adds in quadrature to the phase uncertainty arising from the
intrinsic measurement uncertainty.
Let $\sigma_{{\rm TOA},n}$ be the measurement uncertainty in $t_n$,
and let $\eta_f$ and $\eta_{\dot{f}}$ be the grid spacings in the 
frequency and frequency derivative variables.
Then $\kappa_n$, which equals the inverse square of the phase uncertainty
accumulated over the interval $x_n$, as in (\ref{eq:hmm8}),
depends on $t_n$ and takes the generalized form
\begin{equation}
 \kappa_n
 =
 \{
  [ \sigma_{{\rm TOA},n-1}^2 + \sigma_{{\rm TOA},n}^2 ] f(t_n)^2
  + x_n^2 \eta_f^2
  + x_n^4 \eta_{\dot{f}}^2 / 4
 \}^{-1}~.
\label{eq:hmm8a}
\end{equation}
Equation (\ref{eq:hmm8a}) reduces to (\ref{eq:hmm8}) for
$\sigma_{{\rm TOA},n-1} = \sigma_{{\rm TOA},n} = \sigma_{\rm TOA}$
and $\eta_f = 0 = \eta_{\dot{f}}$.
It preserves the Markovian nature of the HMM,
because $L_{x_n q(t_n)}$ depends only on the state and data at $t_n$,
now expanded to embrace $\sigma_{{\rm TOA},n}$, $\eta_f$, and $\eta_{\dot{f}}$.

\section{Jump Markov model
 \label{sec:hmmappd}}
Instead of relying on Bayesian model selection to detect glitches,
as in \S\ref{sec:hmm3},
one can instruct the HMM to track the hidden Boolean variable $g(t)$ 
introduced in \S\ref{sec:hmm2a}.
Glitches are sparse, so it is needlessly costly to sample all $2^{N_T}$
possible sequences $\{ g(t_0), \dots, g(t_{N_T-1}) \}$.
An approximation, known as a jump Markov model,
involves replacing $g(t_{n-1})$ in (\ref{eq:hmm11}) and (\ref{eq:hmm12})
by the hyperparameter $g=\langle g(t) \rangle$,
i.e.\ the time-averaged glitch probability per TOA.
[A more sophisticated version assumes something about glitch statistics,
e.g.\ a Poisson process, and relates $g(t_{n-1})$ to $x_{n-1}$.]
Typically we have $g\ll 1$.
In the jump Markov model,
$A_{q_i q_j}$ contains a simultaneous mixture of glitch and no-glitch evolution
through (\ref{eq:hmm11}) and (\ref{eq:hmm12}).
Once the HMM generates an optimal sequence,
it is important to check {\em a posteriori} the model evidence;
in effect, $g=\langle g(t) \rangle$ is a uniform prior on every $g(t_n)$,
which must be updated to estimate the posterior of $g(t_n)$,
once the HMM finishes its work.

\citet{suv18} investigated glitch finding with
a jump Markov model and concluded that it does not work as well
as the procedure described in \S\ref{sec:hmm3} for small glitches.
In short, the method finds too many false, small glitches,
which makes sense;
glitches are sparse, so we expect $g(t_n)=0$ for most $1\leq n\leq N_T$.
The reader is referred to the detailed study by \citet{suv18}
for more information.

\section{Generating synthetic data
 \label{sec:hmmappe}}
An infinite family of Langevin equations can generate solutions
of the form (\ref{eq:hmm19}) with a random walk added.
In this paper we solve
\begin{eqnarray}
 \frac{df}{dt}
 & = &
 \frac{f_{\rm s}-f}{2 \tau}
 + \dot{f}(0)
 + \left(\Delta\dot{f}_{\rm p} + \frac{\Delta f_{\rm p}}{\tau} \right) H(t-T)
 + (\Delta f_{\rm p} + \Delta f_1) \delta(t-T)
 + \zeta(t)~,
\label{eq:hmm20}
 \\
 \frac{df_s}{dt}
 & = &
 \frac{f-f_{\rm s}}{2\tau}
 + \left( \Delta\dot{f}_{\rm p} - \frac{\Delta f_{\rm p}}{\tau} \right) H(t-T)
 - (\Delta f_{\rm p} + \Delta f_1) \delta(t-T)~,
\label{eq:hmm21}
\end{eqnarray}
where $\zeta(t)$ is a zero-mean, white-noise torque satisfying
\begin{equation}
 \langle \zeta(t) \zeta(t') \rangle
 = 
 \sigma_{\rm TN}^2 \delta(t-t')~,
\label{eq:hmm22}
\end{equation}
$\sigma_{\rm TN}$ is the timing noise amplitude 
(units: ${\rm Hz\,s^{-1/2}}$),
$\delta(\dots)$ is the Dirac delta function,
and $f_{\rm s}$ is an auxiliary variable,
whose physical interpretation is irrelevant here (see below).
\footnote{
The auxiliary variable is needed,
because behavior of the form (\ref{eq:hmm19})
entails two independent degrees of freedom.
An alternative to (\ref{eq:hmm20}) and (\ref{eq:hmm21}),
also involving two degrees of freedom, is
$df/dt = \dot{f} +
 (\Delta f_{\rm p} + \Delta f_1) \delta(t-T)$
and
$d\dot{f}/dt = 
 -[f-f(0)]/\tau^2 -2\dot{f}/\tau - \dot{f}(0) (2/\tau + t/\tau^2)
 + (\Delta \dot{f}_{\rm p} - \Delta f_1/\tau) \delta(t-T)
 + \tau^{-2}\Delta f_{\rm p} H(t-T)
 + \Delta \dot{f}_{\rm p} [2/\tau + (t-T)/\tau^2] H(t-T)
 + \zeta(t)$.
}
Equations (\ref{eq:hmm20}) and (\ref{eq:hmm21}) are solved subject to
the initial conditions
$f(t=0)=f(0)$ and
$f_{\rm s}(t=0)=f(0)+\dot{f}(0) \tau$.

\begin{figure}
\begin{center}
\includegraphics[width=14cm,angle=0]{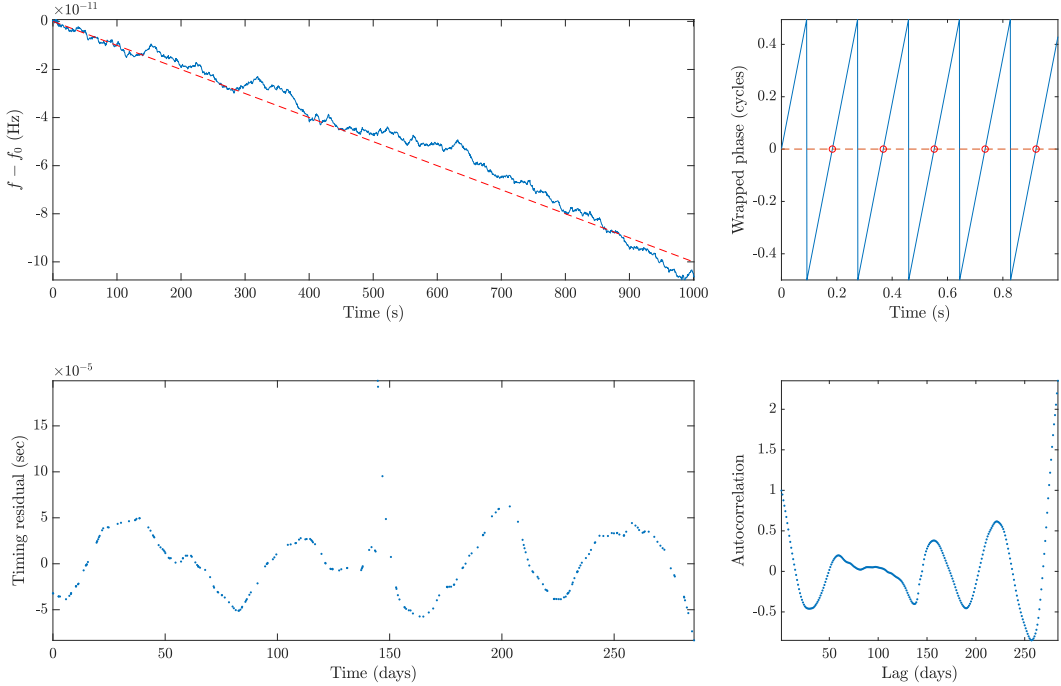}
\end{center}
\caption{
Sample segment of synthetic data comprising
250 TOAs over $286\,{\rm d}$
generated by solving (\ref{eq:hmm20})--(\ref{eq:hmm22})
with a relatively large glitch injected at $t=144.67\,{\rm d}$.
({\em Top left panel.})
Stochastic frequency evolution $f(t)$ (blue solid curve)
and deterministic component $f(0)+\dot{f}(0)t$ (red dashed curve)
within the subinterval $0\leq t/(1\,{\rm s}) \leq 10^3$.
({\em Top right panel.})
Wrapped phase versus time (blue solid curve) and TOA locations (red circles)
within the window $0\leq t/(1 \, {\rm s}) \leq 1$.
({\em Bottom left panel.})
TOA residuals produced by a {\sc tempo2} fit
to the data in the top left panel
for $0\leq t / (1\,{\rm d}) \leq 286$,
which returns the best-fit parameters
$(f_{\rm LS},\dot{f}_{\rm LS})
 = [f(0),\dot{f}(0)]
 + (-3.0\times 10^{-10}\,{\rm Hz},-5.9\times 10^{-18}\,{\rm Hz\,s^{-1}})$,
$\Delta f_{\rm p} = 5.2\times 10^{-8}\,{\rm Hz}$,
$\Delta \dot{f}_{\rm p} = 5.0\times 10^{-14}\,{\rm Hz\,s^{-1}}$,
and glitch epoch $144.9\,{\rm d}$.
({\em Bottom right panel.})
Autocorrelation function of the phase residuals 
produced by the {\sc tempo2} fit in the bottom left panel.
Injection parameters in (\ref{eq:hmm20})--(\ref{eq:hmm22}):
$f(0)=5.435\,{\rm Hz}$,
$\dot{f}(0)=-1\times 10^{-13}\,{\rm Hz\,s^{-1}}$,
$\sigma_{\rm TN} = 5\times 10^{-13}\,{\rm Hz\,s^{-1/2}}$,
$\Delta f_{\rm p} = 5\times 10^{-8}\,{\rm Hz}$,
$\Delta \dot{f}_{\rm p} = 5\times 10^{-14} \,{\rm Hz\,s^{-1}}$,
$\Delta f_1 = 5\times 10^{-8}\,{\rm Hz}$,
$\tau=5\,{\rm d}$.
}
\label{fig:hmm3}
\end{figure}

Figure \ref{fig:hmm3} displays a sample of the synthetic data
generated by the above procedure.
Overall it comprises 250 TOAs sampled according to a Poisson process,
whose waiting times $\Delta t$ are distributed
according to the probability density function
$p(\Delta t) = \lambda_{\Delta t} \exp(-\lambda_{\Delta t} \Delta t)$,
with $\lambda_{\Delta t}=0.864 \,{\rm d^{-1}}$.
By assigning the TOAs randomly,
we ensure that they do not coincide with the glitch in general.
The top left panel displays the time series $f(t)$ sampled with
high temporal resolution over a subinterval lasting $10^3\,{\rm s}$.
The frequency fluctuations generated by the torque noise process in 
(\ref{eq:hmm20})--(\ref{eq:hmm22}) are clearly visible.
Their root-mean-square amplitude, which reaches one part in $\sim 10^{11}$,
is consistent with $\sigma_{\rm TN}=5\times 10^{-13}\,{\rm Hz\,s^{-1/2}}$
over an interval of $10^3\,{\rm s}$.
The top right panel displays the phase evolution (including wrapping)
within a short window lasting $1\,{\rm s}$
and indicates the TOAs of individual pulses within the window. 
The bottom left panel shows the TOA residuals produced by 
a {\sc tempo2} fit to the whole data set, 
$0\leq t/(1\,{\rm d}) \leq 286$,
including a relatively large glitch 
with $\Delta f_{\rm p}=5\times 10^{-8}\,{\rm Hz}$
at $t=144.67\,{\rm d}$.
The secular spin-down parameters $f_{\rm LS}$ and $\dot{f}_{\rm LS}$
inferred from the fit agree well 
with the injected values of $f(0)$ and $\dot{f}(0)$.
The TOA residuals $\sim 10^{-5}\,{\rm s}$
are consistent with $\sigma_{\rm TN}$ over $286\,{\rm d}$
and are consistent visually with the red phase noise produced by
filtered white torque noise in line with 
(\ref{eq:hmm20})--(\ref{eq:hmm22}),
except for a modest spike around the glitch epoch,
because the glitch fit is imperfect.
The bottom right panel shows the autocorrelation function 
of the phase residuals in the bottom left panel.
The half-power point occurs at a lag of $\approx 7\,{\rm d}$.
The oscillations at lags $\gtrsim 50\,{\rm d}$
are characteristic of red phase noise,
as observed in many pulsars
\citep{pri12}.

Equations (\ref{eq:hmm20}) and (\ref{eq:hmm21}) can be extended 
in several ways.
As the system is linear
and obeys the principle of superposition,
it is easy to add more exponential recoveries
with amplitudes $\Delta f_k$ and time-scales $\tau_k$ ($k > 1$)
by lifting the order of the system of differential equations
and adding forcing terms proportional to
$H(t-T)$ and $\delta(t-T)$ 
as in (\ref{eq:hmm20}) and (\ref{eq:hmm21}).
One can also include a secular second derivative 
of the form $\langle \ddot{f} \rangle \propto f^n$.

The reader may notice a similarity between (\ref{eq:hmm20}) 
and (\ref{eq:hmm21}) and the two-component model of a neutron star interior
\citep{bay69},
where $f$ and $f_s$ correspond to the spin frequencies
of the rigid crust and neutron condensate respectively.
The analogy may prove useful in future work,
when interpreting physically the results of HMM-based glitch searches, 
but it is not pertinent to this paper.
Here we merely exploit the mathematical correspondence,
which renders (\ref{eq:hmm20}) and (\ref{eq:hmm21}) with $\zeta(t)=0$
equivalent to (\ref{eq:hmm19}),
in order to generate synthetic data.
We emphasize that the dynamical model for generating synthetic data
differs deliberately from the dynamical model governing the
hidden state evolution in the HMM.
For example, white noise fluctuations 
enter through the torque in (\ref{eq:hmm20}) [i.e.\ $\zeta(t)$]
and through the torque derivative in (\ref{eq:hmm9}) [i.e.\ $\xi(t)$].
This reflects the situation in practice astrophysically,
where the dynamical model for timing noise is unknown.
It also confirms the robustness of the HMM in light of the 
encouraging results in \S\ref{sec:hmm6}.
Likewise, the permanent jumps $\Delta f_{\rm p}$ and $\Delta\dot{f}_{\rm p}$ 
in the HMM transition probability (\ref{eq:hmm10})--(\ref{eq:hmm12})
do not match exactly the eponyomous variables in the synthetic data
generation model (\ref{eq:hmm20}) and (\ref{eq:hmm21}),
although they are related.

\section{Representative worked example: synthetic data
 \label{sec:hmmappf}}
\begin{figure}
\begin{center}
\includegraphics[width=14cm,angle=0]{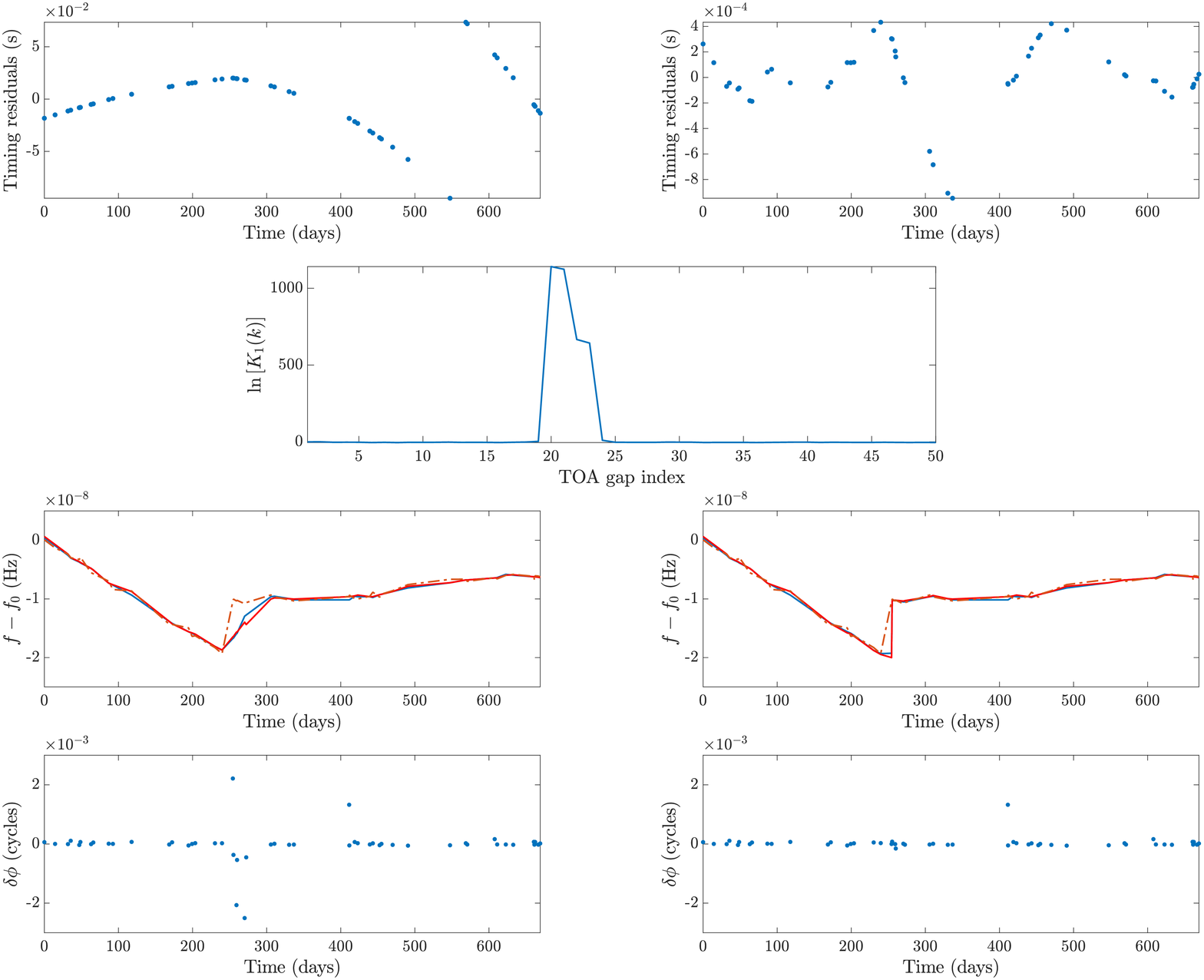}
\end{center}
\caption{
Worked example involving a segment of synthetic data 
generated by solving (\ref{eq:hmm20})--(\ref{eq:hmm22}) 
with the parameters in the penultimate (typical) column 
in Table \ref{tab:hmm1}.
({\em First row.})
Phase residuals $\delta\phi(t_n)$ versus TOA $t_n$
computed with {\sc tempo2}
for no-glitch (left panel) and one-glitch (right panel) models
with the parameters quoted in Appendix \ref{sec:hmmappf}.
({\em Second row.})
Logarithm of the Bayes factor,
$K_1(k)=\Pr[O_{1:N_T}|M_1(k)]/\Pr(O_{1:N_T}|M_0)$,
versus TOA index, $k$,
computed with the HMM using
$\sigma = 1\times 10^{-18}\,{\rm Hz\,s^{-3/2}}$.
The glitch is injected at $k=19$ (epoch $T=253.371\,{\rm d}$)
and recovered at $k=20$.
({\em Third row.})
Recovered frequency $f(t_n)$ versus TOA $t_n$
for no-glitch (left panel) and one-glitch (right panel) models
$M_0$ and $M_1(k=20)$,
showing the injected signal (dash-dotted curve),
HMM forward-backward sequence (blue curve),
and HMM Viterbi sequence (red curve).
({\em Fourth row.})
Unsummed per-gap phase residuals $\delta\phi(t_n)$ versus TOA $t_n$
for the HMM forward-backward sequences in the third row.
The DOI and grid spacing are given by
$-5.9\times 10^{-8} \leq (f - 5.435\,{\rm Hz})/(1\,{\rm Hz})
 \leq 1\times 10^{-8}$,
$-1\times 10^{-15} \leq \dot{f}/(1\,{\rm Hz\,s^{-1}})
 \leq 2\times 10^{-15}$,
$\eta_f = 6.79\times 10^{-11}\,{\rm Hz}$
($10^3$ bins),
and
$\eta_{\dot{f}} = 3\times 10^{-16}\,{\rm Hz\,s^{-1}}$
(11 bins).
}
\label{fig:hmm4}
\end{figure}

\begin{figure}
\begin{center}
\includegraphics[width=14cm,angle=0]{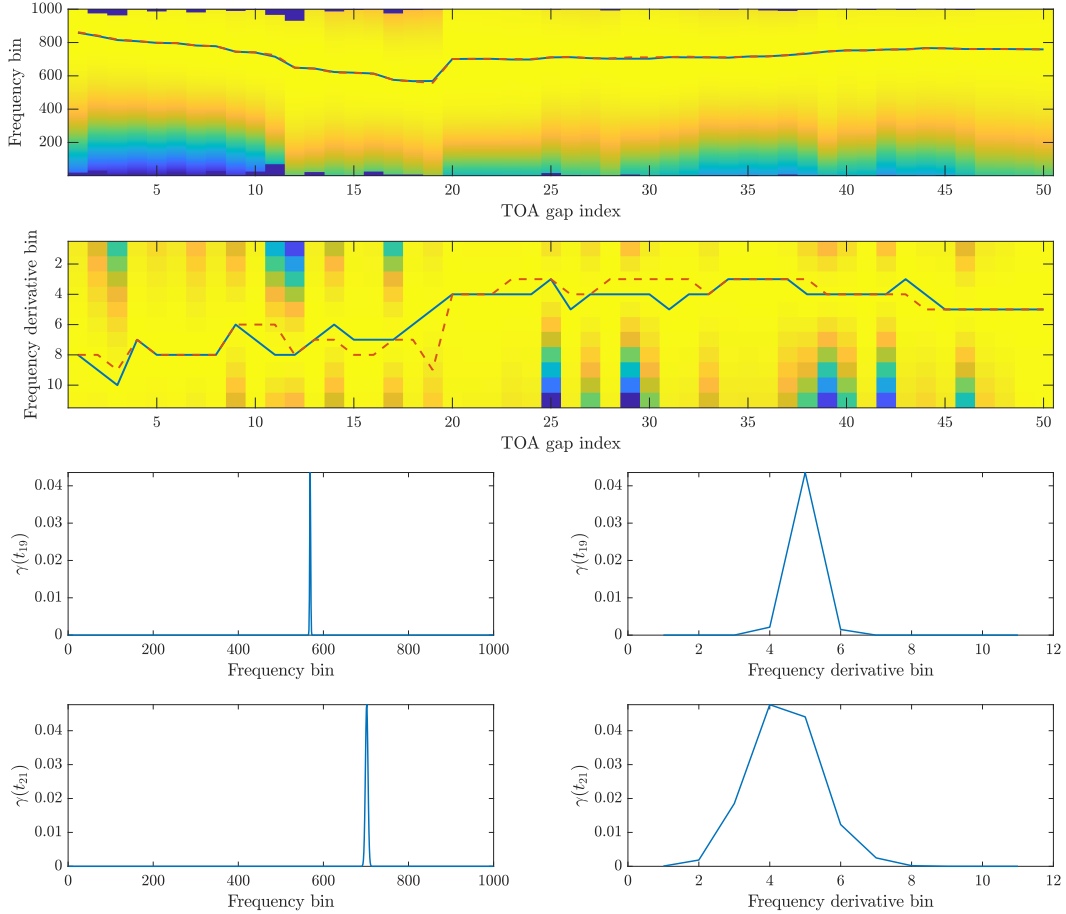}
\end{center}
\caption{
Evolution of the posterior PDF $\gamma_{q_i}(t_n)$ 
defined by (\ref{eq:hmmappa12})
in the vicinity of the optimal values $\hat{f}(t_n)$ and $\hat{\dot{f}}(t_n)$.
({\em First row.})
Contours of $\gamma_{q_i}(t_n)$ marginalized over $\dot{f}$ 
(arbitrary color scale; yellow high, blue low) versus TOA index $n$,
with the point-wise optimal (forward-backward; blue curve) 
and sequence-wise optimal (Viterbi; red curve) state sequences overplotted.
({\em Second row.})
Contours of $\gamma_{q_i}(t_n)$ marginalized over $f$
versus TOA index $n$.
({\em Third row.})
Cross-section of $\gamma_{q_i}(t_n)$ 
marginalized over $\dot{f}$ (left column) and $f$ (right column)
at $n=19$,
i.e.\ at the TOA preceding the recovered glitch.
The horizontal axes display numbers of bins.
({\em Fourth row.})
Cross-section of $\gamma_{q_i}(t_n)$ 
marginalized over $\dot{f}$ (left column) and $f$ (right column)
at $n=21$,
i.e.\ at the TOA following the recovered glitch.
Parameters: see Figure \ref{fig:hmm4}.
}
\label{fig:hmm5}
\end{figure}

In this section, we illustrate how to apply the HMM in \S\ref{sec:hmm2} 
and the model selection procedure in \S\ref{sec:hmm3}
to a sample of synthetic data generated according to the recipe
in \S\ref{sec:hmm4} and the parameters in the penultimate (typical)
column in Table \ref{tab:hmm1}.
The results are plotted in Figures \ref{fig:hmm4} and \ref{fig:hmm5}.
The worked example breaks out the steps in the analysis 
and introduces several useful diagnostics.
It is a training run for the systematic performance tests in \S\ref{sec:hmm6}.

The top two panels of Figure \ref{fig:hmm4} present the raw data
before any analysis with the HMM.
The top left panel graphs the phase residuals $\delta\phi(t_n)$ 
as a function of time 
after subtracting a no-glitch spin-down model with
$f_{\rm LS} - 5.435 \,{\rm Hz} = 1.48\times 10^{-8}\,{\rm Hz}$ 
and $\dot{f}_{\rm LS} = -1.41\times 10^{-15}\,{\rm Hz\,s^{-1}}$,
inferred by fitting the full data set with {\sc tempo2}.
The residuals diverge for $t > T=253.371 \,{\rm d}$ 
quadratically (and the phase wraps at $t\approx 550\,{\rm d}$),
as expected for a glitch with $\Delta f_{\rm p} > 0$
and $\Delta\dot{f}_{\rm p} > 0$.
Phase residuals are also plotted in the top right panel
after subtracting a one-glitch model with 
$f_{\rm LS} -5.435\,{\rm Hz} = -1.2 \times 10^{-10}\,{\rm Hz}$,
$\dot{f}_{\rm LS} = -9.164\times 10^{-16}\,{\rm Hz\,s^{-1}}$,
$T=254.37\,{\rm d}$,
$\Delta f_{\rm p} = 8.535\times 10^{-9}\,{\rm Hz}$,
and
$\Delta \dot{f}_{\rm p} = 1.09\times 10^{-15}\,{\rm Hz\,s^{-1}}$,
again fitted with {\sc tempo2}.
The residuals in the top right panel do not diverge
and have root-mean-square amplitude 
$\sim 10^{-3}\,{\rm rad}$,
consistent with
$\sigma_{\rm TN}=1\times 10^{-12}\,{\rm Hz\,s^{-1/2}}$
integrated over $\sim 10^2\,{\rm d}$
($\sigma_{\rm TN}$ dominates $\sigma_{\rm TOA}$ in this example).
The fitted parameters are close to the injected parameters
quoted in the penultimate column in Table \ref{tab:hmm1}.
Returning to the no-glitch fit,
we convert $\delta\phi(t_n)$ into the DOI and grid spacing
specified in the figure caption following the recipe
in Appendix \ref{sec:hmmappc}.

Model selection is now performed.
The second row of Figure \ref{fig:hmm4} displays the Bayes factor,
$K_1(k)=\Pr[O_{1:N_T}|M_1(k)]/\Pr(O_{1:N_T}|M_0)$,
as a function of the TOA index, $k$.
The Bayes factor peaks at $\ln K_1(k_1^\ast=20)\sim 10^3$,
well above the threshold $K_1(k) > 10^{1/2}$.
In other words, 
a model featuring a glitch near the injected location ($k=19$) 
is preferred categorically over the no-glitch model.
The $\ln K_1(k)$ plateau near $k=k_1^\ast$ is typical
of the HMM output for relatively large glitches,
but the peak stands clearly above neighboring points,
before the logarithm is taken.
The single-TOA mismatch between the injected and recovered epochs 
is also typical.
The HMM cannot say anything about the phase evolution between TOAs, 
so a single-TOA mismatch is always possible,
even when a glitch is injected exactly at a TOA.
Two-glitch models $M_2(k,l)$ are not considered here,
as only one glitch is injected.

Finally, an ephemeris is constructed for the preferred model.
The bottom four panels display $f(t_n)$ (third row) 
and $\delta\phi(t_n)$ (fourth row) as functions of $t_n$
for the point-wise optimal hidden sequence $\hat{q}(t_n)$
found by the HMM forward-backward algorithm (blue curve)
in Appendix \ref{sec:hmmappa}.
For comparison, the sequence-wise optimal sequence $Q^\ast_{1:N_T}$
found by the Viterbi algorithm is also graphed (without residuals)
as a red curve in the third row.
Both HMM sequences lie close to each other and to the true, injected sequence,
plotted as a dash-dotted curve.
The forward-backward sequence $\hat{q}(t_n)$ yields
a root-mean-square error of 
$\approx 1\times 10^{-3} \, {\rm rad}$
for the one-glitch model $M_1(k_1^\ast)$ (right column),
which corresponds to $\approx 0.1 \max_n \delta\phi(t_n)$
for the no-glitch model $M_0$ (left column).
The $M_0$ residuals are highest at $t\approx T$, as expected.

The posterior PDF of the hidden state likelihood in the neighborhood
of the optimal sequence is a useful diagnostic.
It indicates how far the optimal sequence stands out above its
nearest competitors.
It also provides a way of estimating point-by-point confidence intervals
for the optimal ephemeris in practical astrophysical applications,
where the underlying, true ephemeris is unknown.
The top two panels in Figure \ref{fig:hmm5} display heat map contours
of the posterior PDF computed by the forward-backward algorithm,
$\gamma_{q_i}(t_n)$ in (\ref{eq:hmmappa12}),
marginalized over $\dot{f}$ (first row) and $f$ (second row).
The point-wise (forward-backward; blue curve) and 
sequence-wise (Viterbi; red curve) optimal state sequences 
run through the middle of the high-probability (yellow) regions.
The bottom four panels display cross-sections of $\gamma_{q_i}(t_n)$ 
immediately before (third row) and after (fourth row) the recovered glitch.
The optimal state stands out clearly and is localized precisely.
The FWHM of the PDF marginalized over $\dot{f}$ satisfies
$\approx 1.4\times 10^{-10}\,{\rm Hz}$ and 
$\approx 4.8\times 10^{-10}\,{\rm Hz}$ 
before and after the glitch respectively,
while the FWHM of the PDF marginalized over $f$ satisfies
$\approx 3\times 10^{-16}\,{\rm Hz\,s^{-1}}$
and
$\approx 6\times 10^{-16}\,{\rm Hz\,s^{-1}}$
before and after the glitch respectively.
We can compute the jumps in $f$ and $\dot{f}$ during the glitch
by comparing the peaks in the third and fourth rows.
The displacements are clearly visible, once enough time elapses;
we find 
$\hat{f}(t_{21}) - \hat{f}(t_{19}) = 8.5\times 10^{-9}\,{\rm Hz}$ 
and
$\hat{\dot{f}}(t_{21}) - \hat{\dot{f}}(t_{19}) 
 = 3\times 10^{-16}\,{\rm Hz\,s^{-1}}$,
cf.\ the injected values 
$\Delta f_{\rm p} = 1\times 10^{-8}\,{\rm Hz}$ 
and
$\Delta \dot{f}_{\rm p} = 1\times 10^{-15}\,{\rm Hz\,s^{-1}}$.

\section{Schedule of observations: impact on performance
 \label{sec:hmmappg}}
When optimizing an observational campaign aimed at detecting glitches,
it is important to plan how the spacing of observation sessions 
and the number of TOAs affect $P_{\rm fa}$ and $P_{\rm d}$.
A typical observation session may last a few minutes,
with $\sim 10^4$ pulses averaged to produce each TOA.
Sessions are often separated by days to weeks,
although of course there are exceptions;
for instance, PSR J0835$-$4510 is monitored continuously
for extended intervals
\citep{pal16,pal18}.

\begin{figure}
\begin{center}
\includegraphics[width=14cm,angle=0]{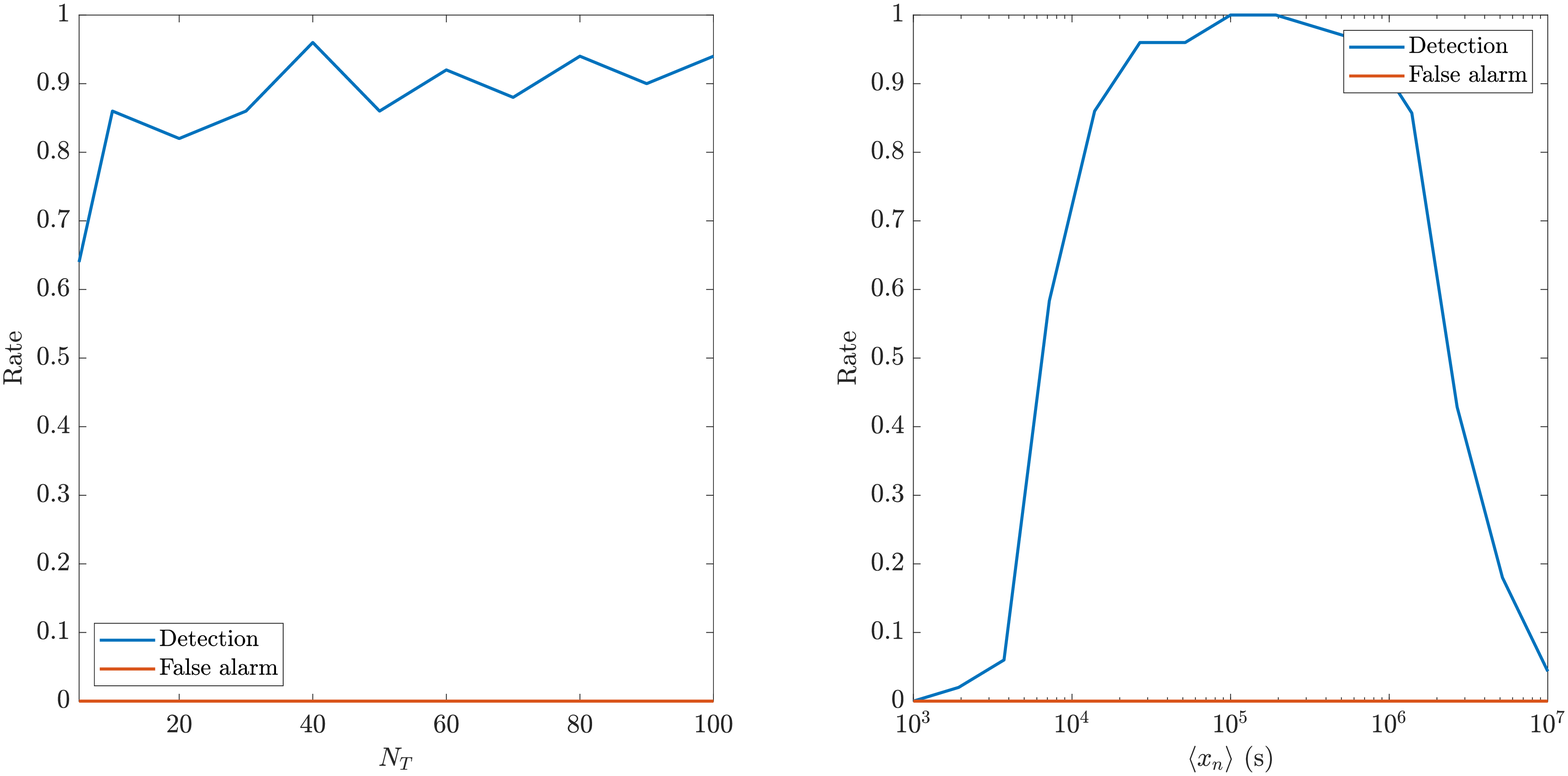}
\end{center}
\caption{
Detection probability (blue curve) and false alarm probability (red curve)
versus the total number of observing sessions ({\em left panel})
and the mean TOA interval averaged over the entire observation
({\em right panel}).
Parameters:
as in Figure \ref{fig:hmm7}.
}
\label{fig:hmm9}
\end{figure}

Figure \ref{fig:hmm9} partially quantifies the above considerations.
The left panel graphs $P_{\rm fa}$ and $P_{\rm d}$ as functions of 
the total number of observation sessions,
after adjusting the Bayes factor threshold to achieve
$P_{\rm fa}=10^{-2}$ on average across the plotted range.
A detection is highly probable in most realistic scenarios;
we obtain $P_{\rm d} \geq 0.9$ for $\gtrsim 35$ sessions.
The right panel graphs $P_{\rm fa}$ and $P_{\rm d}$ as functions of 
the mean interval between sessions averaged over the entire observation
($\sim 1\,{\rm yr}$),
after adjusting the Bayes factor threshold as in the left panel.
We obtain $P_{\rm d} \geq 0.9$ for intervals between $\sim 10^4\,{\rm s}$
and $\sim 10^6\,{\rm s}$,
which are readily achievable with dedicated or multibeam telescopes.
For shorter intervals, the phase error due to $\sigma_{\rm TOA}$,
which is independent of $\langle x_n \rangle$,
impairs the HMM's performance.
For longer intervals, 
the phase error due to binning (see Appendix \ref{sec:hmmappc}) dominates,
because it scales $\propto \langle x_n \rangle$.
The results in Figure \ref{fig:hmm9} are generated for one TOA
per observation session.
Tests show that $P_{\rm fa}$ is roughly constant 
given between one and five TOAs per session
for the parameters in Figure \ref{fig:hmm9}.
A thorough study of multiple TOAs per observation session,
including the related and important matter of pulse jitter
\citep{hel75},
is postponed to future work.

We formulate a useful rule of thumb to predict
how one should space observations
to resolve glitches of a certain size.
During a gap of duration $x_n$, phase deviations
$\Delta f_{\rm p} x_n$ and $\Delta\dot{f}_{\rm p} x_n^2 /2$
develop for frequency and frequency derivative jumps respectively.
Writing their ratio as 
$\approx (\Delta f_{\rm p}/ f_{\rm LS})
 (\Delta\dot{f}_{\rm p}/\dot{f}_{\rm LS})^{-1}
 (x_n \dot{f}_{\rm LS}/f_{\rm LS})^{-1}$,
we see that the two contributions are comparable typically,
e.g.\ for
$\Delta f_{\rm p}/ f_{\rm LS} \sim 10^{-7}$,
$\Delta\dot{f}_{\rm p}/\dot{f}_{\rm LS} \sim 10^{-2}$,
and $x_n \dot{f}_{\rm LS}/f_{\rm LS} \sim 10^{-5}$.
When the glitch-related phase deviations exceed 
those produced by TOA measurement errors ($\sigma_{\rm TOA} f_{\rm LS}$)
and astrophysical timing noise ($\sigma_{\rm TN} x_n^{3/2}$),
the glitch is discerned above the noise.
This occurs for 
$\Delta f_{\rm p} \geq \sigma_{\rm TOA} f_{\rm LS}
 \langle x_n \rangle^{-1}$
and
$\Delta f_{\rm p} \geq \sigma_{\rm TN} \langle x_n \rangle^{1/2}$
for
$\Delta \dot{f}_{\rm p}=0$,
or
$\Delta \dot{f}_{\rm p} \geq 2 \sigma_{\rm TOA} f_{\rm LS}
 \langle x_n \rangle^{-2}$
and
$\Delta \dot{f}_{\rm p} \geq 2 \sigma_{\rm TN} 
 \langle x_n \rangle^{-1/2}$
for 
$\Delta f_{\rm p}=0$.
Both special cases agree with the general expression 
for $\Delta f_{\rm p}\neq 0$ and $\Delta \dot{f}_{\rm p} \neq 0$
presented by \citet{suv18}.

\begin{figure}
\begin{center}
\includegraphics[width=14cm,angle=0]{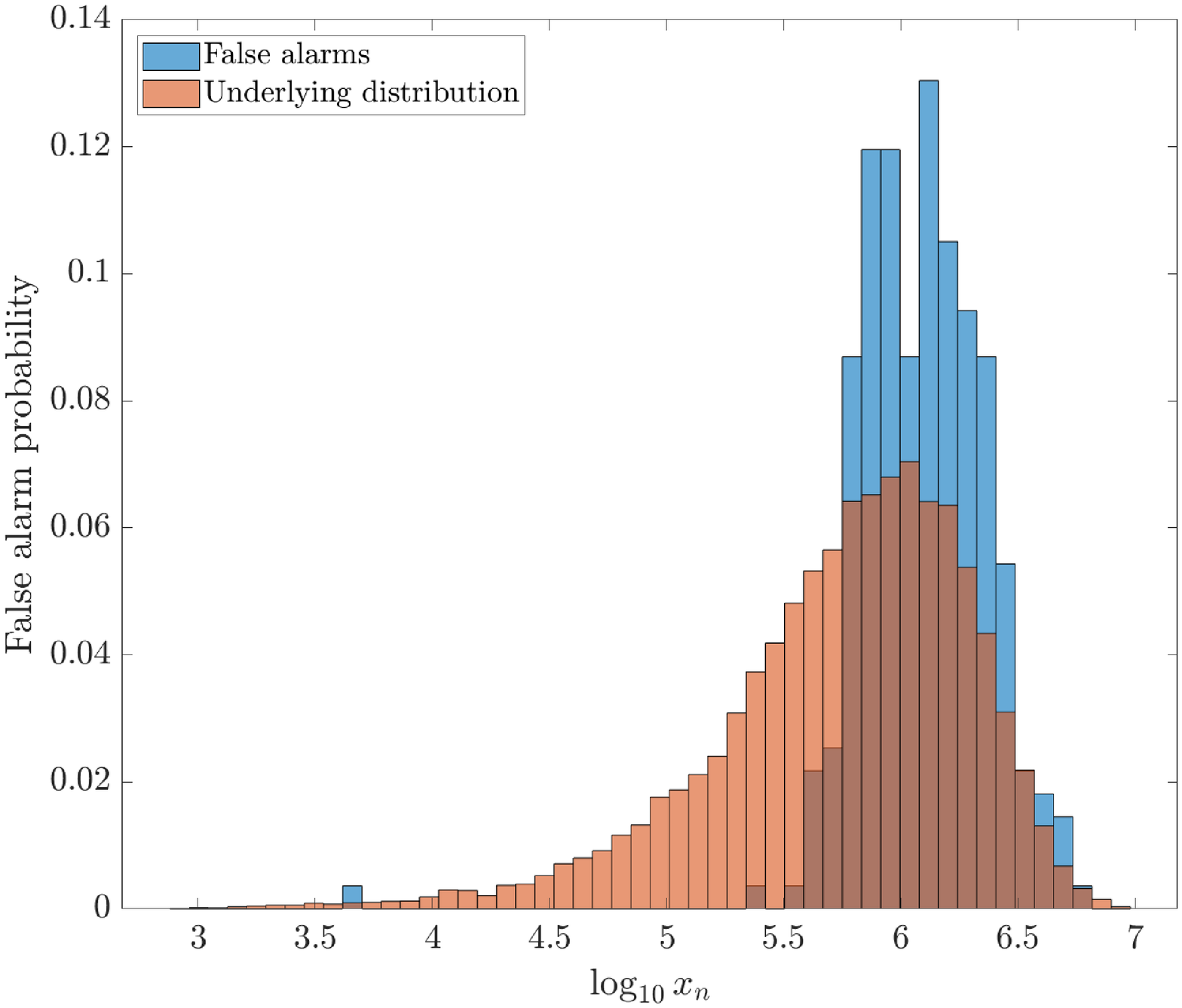}
\end{center}
\caption{
PDF of the logarithm of the TOA gap, $\log x_n$, 
for the subset of gaps adjacent to a false alarm
(blue histogram)
and for all simulated gaps
(orange histogram; Poisson distribution by construction).
Number of realizations: $5\times 10^2$. 
Other parameters:
see penultimate (typical) column in Table \ref{tab:hmm1}.
}
\label{fig:hmm10}
\end{figure}

Another way to gauge the impact of the observational schedule
on the HMM is to note that,
when false alarms occur, they correlate with large TOA gaps.
Figure \ref{fig:hmm10} displays, in orange,
a histogram of $x_n$ values adjacent to false alarms,
along with a blue histogram of all the simulated $x_n$ values,
whether they are adjacent to a false alarm or not.
The simulations are in the regime, where $\ddot{f}$ can be neglected
(see \S\ref{sec:hmm2b}).
As expected, there is a clear trend: 
false alarms occur more frequently near larger gaps,
because the number of peaks in $L_{x_n q(t_n)}$ within the DOI
increases with $x_n$,
even though the peaks sharpen
(see \S\ref{sec:hmm2c}).
The correlation strengthens, 
as $\kappa$ and hence the number of false alarms increase.
The opposite trend applies to false dismissals:
the HMM is more prone to reject a true glitch,
when $x_n$ is relatively short,
because the phase deviation across the gap is relatively small.

\section{Is there a second glitch in PSR J0835$-$4510
 between MJD 57427 and MJD 57810?
 \label{sec:hmmapph}}
In this appendix, we apply the greedy hierarchical algorithm 
introduced in \S\ref{sec:hmm3aa} \citep{suv18} to test for the existence
of a second glitch in PSR J0835$-$4510 
in the interval from MJD 57427 to MJD 57810.
Specifically, we assume that the glitch found at MJD 57734.54 is real
and construct the model $M_2(173,k)$,
which features a glitch at TOA interval
$x_{173}$ and a second glitch at $x_k$.

\begin{figure}
\begin{center}
\includegraphics[width=14cm,angle=0]{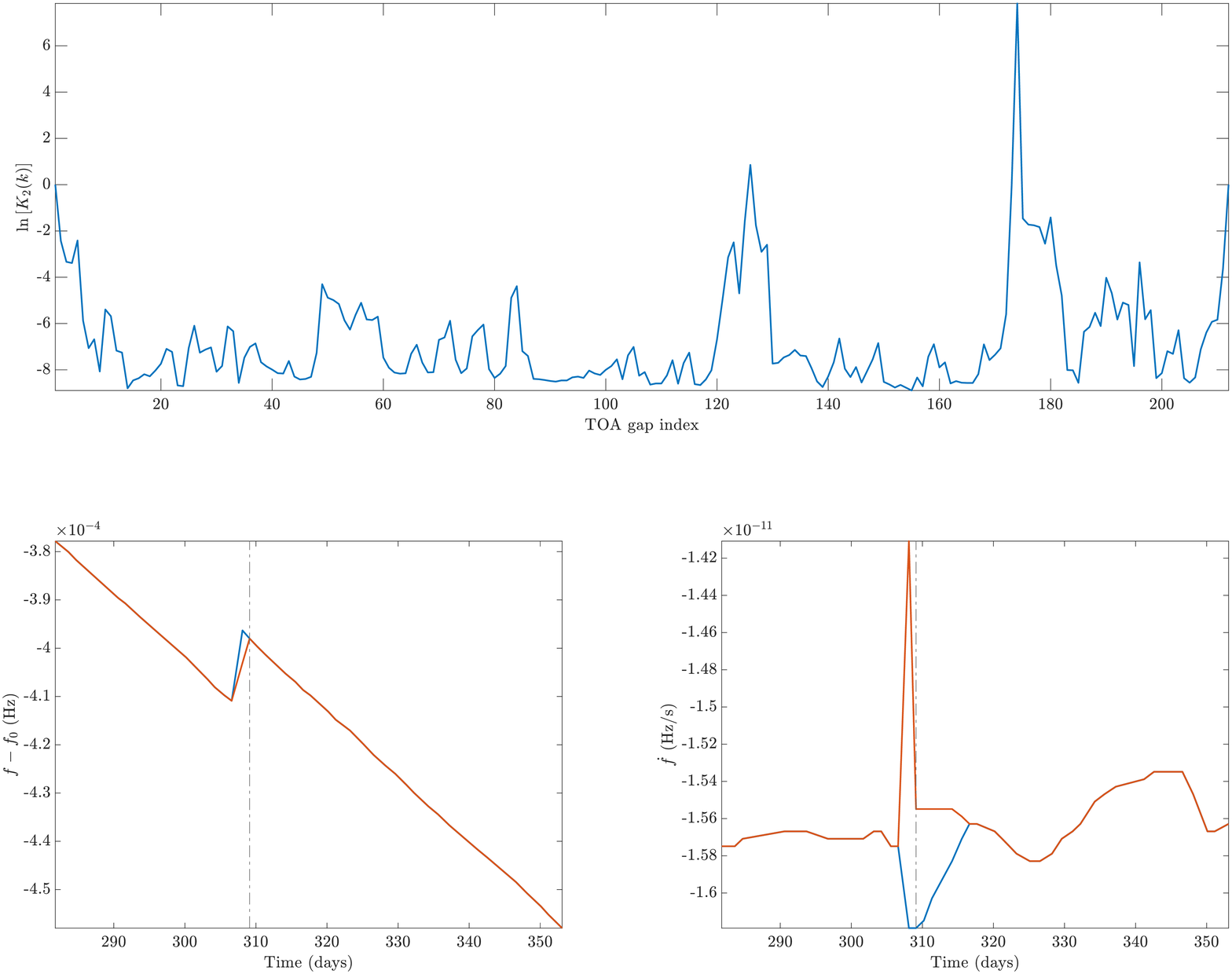}
\end{center}
\caption{
Search for a second glitch in PSR J0835$-$4510
between MJD 57427 and MJD 57810.
({\em Top row.})
Bayes factor $K_2(k)$ versus TOA index $k$.
({\em Bottom row.})
Point-wise optimal state sequence $\hat{f}(t_n)$ (left panel)
and $\hat{\dot{f}}(t_n)$ (right panel) versus $t_n$
for the one-glitch model $M_1(173)$ (blue curves)
and two-glitch model $M_2(173,174)$ (red curves).
Parameters: see Figure \ref{fig:hmm20}.
}
\label{fig:hmm23}
\end{figure}

Figure \ref{fig:hmm23} presents the analysis of $M_2(173,k)$.
The top panel displays the Bayes factor,
$K_2(k) = \Pr[O_{1:N_T} | M_2(173,k) ] / \Pr[O_{1:N_T} | M_1(173) ]$,
as a function of $k$.
A peak is observed at $k=174$,
with $\ln K_2(174) \approx 8$.
Formally this counts as a detection by the criterion in \S\ref{sec:hmm3a}.
(The detection threshold in \S\ref{sec:hmm3a} is roughly consistent
with $P_{\rm fa}\approx 1\times 10^{-2}$ and $P_{\rm d}\approx 0.9$ 
throughout this paper.)
However, it occurs at the TOA immediately following the first glitch
and is likely to be associated with it,
because the introductory version of the HMM in this paper
treats a glitch as an instantaneous step with no quasiexponential recovery,
whereas in reality a recovery with
$\tau = 0.96(17)\,{\rm d}$
is measured independently in the 2016 December 12 event
\citep{pal16b,sar17b}.
The second-highest peak in the top panel of Figure \ref{fig:hmm23},
which occurs at $k=123$
with $\ln K_2(123) \approx 1$,
does not count as a detection by the criterion in \S\ref{sec:hmm3a}.

The bottom left and right panels in Figure \ref{fig:hmm23}
display the point-wise optimal
state sequences $\hat{f}(t_n)$ and $\hat{\dot{f}}(t_n)$ respectively,
calculated by the forward-backward algorithm.
The plots zoom into the neighborhood of the second ``glitch'' at $t_{174}$
(dashed vertical line)
and compare the one-glitch model $M_1(173)$ (blue curve)
with the two-glitch model $M_2(173,174)$ (red curve).
Both models handle the complicated, composite dynamics of the spin up
and quasiexponential recovery with equal dexterity
but in slightly different ways, e.g.\ $\hat{\dot{f}}$ increases
for $t>T$ in $M_2(173,174)$, decreases in $M_1(173)$,
and asymptotes to its long-term, post-glitch value 
over $\sim 5 \,{\rm d}$ in both cases.

We note in closing that the three-glitch Bayes factor $K_3(k)$
does not exceed $10^{1/2}$ for any $k$,
i.e.\ there is no evidence in the above data for a third glitch.

\end{document}